\documentstyle[preprint,12pt,aps,graphicx,tighten,eqsecnum]{revtex}

\newcommand{\beq}{\begin{equation}}
\newcommand{\eeq}{\end{equation}}
\newcommand{\beqa}{\begin{eqnarray}}
\newcommand{\eeqa}{\end{eqnarray}}

\renewcommand{\Re}{\mbox{Re}}
\renewcommand{\Im}{\mbox{Im}}
\newcommand{\no}{\nonumber}
\def\lsim{\mathrel{\rlap{\lower4pt\hbox{\hskip1pt$\sim$}}
    \raise1pt\hbox{$<$}}}         
\def\gsim{\mathrel{\rlap{\lower4pt\hbox{\hskip1pt$\sim$}}
    \raise1pt\hbox{$>$}}}         

\begin{document}
\begin{titlepage}
\preprint{
\vbox{   \hbox{WIS/18/01-Aug-DPP}
         \hbox{}\hbox{} }}

\title{CP Violation - A New Era}
\author{Yosef Nir \thanks{yosef.nir@weizmann.ac.il}} 
\vskip 1cm
\address{Department of Particle Physics, Weizmann Institute of Science\\
 Rehovot 76100, Israel} 
\maketitle
\begin{abstract}
We give a pedagogical review  of the theory of CP violation with emphasis
on the implications of recent experimental results. The review includes:
(i) A detailed description of how CP violation arises in the Standard Model
and in its extension that allows for neutrino masses; (ii) The formalism of
CP violation in meson decays and its application to various $K$ decays
($\varepsilon_K$, $\varepsilon^\prime/\varepsilon$ and $K\to\pi\nu\nu$), 
$D$ decays ($D\to K\pi$ and $D\to KK$) and $B$ decays ($B\to \ell\nu X$,
$B\to\psi K_S$ and $B\to\pi\pi$, including a discussion of the `penguin 
pollution' problem); (iii) Supersymmetry: the CP problems and the use of CP 
violation as a probe of the mechanism of dynamical supersymmetry breaking.
 
\end{abstract}

\bigskip\bigskip\bigskip

\centerline{\it Lectures given at the}
\centerline{\bf 55th Scottish Universities Summer School in Physics}
\centerline{\bf Heavy Flavour Physics} 
\centerline{\it University of St. Andrews, Scotland}
\centerline{\it August 7 $-$ 23 2001}

\end{titlepage}

\tableofcontents
\vfill\eject

\section{Introduction}
The Standard Model predicts that the only way that CP is violated is
through the Kobayashi-Maskawa mechanism \cite{Kobayashi:1973fv}. Specifically,
the source of CP violation is a {\it single} phase in the mixing matrix that
describes the charged current weak interactions of quarks.

In the introductory chapter, we briefly review the present evidence that
supports the Kobayashi-Maskawa picture of CP violation, as well as the various 
arguments against this picture.
 
\subsection{Why Believe the Kobayashi-Maskawa Mechanism?}
Experiments have measured to date three independent CP violating observables:

\noindent 
(i) Indirect CP violation in $K\to\pi\pi$ decays \cite{Christenson:1964fg}
and in $K\to\pi\ell\nu$ decays is given by:
\beq\label{expeps}
\varepsilon_K=(2.28\pm0.02)\times10^{-3}\ e^{i\pi/4}.
\eeq
(ii) Direct CP violation in $K\to\pi\pi$ decays 
\cite{Burkhardt:1988yh,Barr:1993rx,Gibbons:1993zq,ktev,nafe}
is given by
\beq\label{aveepp}
{\varepsilon^\prime\over\varepsilon}=(1.72\pm0.18)\times10^{-3}.
\eeq
(iii) CP violation in $B\to\psi K_S$ decay and other, related modes
has been measured 
\cite{Ackerstaff:1998xz,Affolder:2000gg,Barate:2000tf,Aubert:2001nu,Abe:2001xe}:
\beq\label{aveapk}
a_{\psi K_S}=0.79\pm0.10.
\eeq

All three measurements are consistent with the Kobayashi-Maskawa picture of
CP violation. In particular, the two recent measurements of CP violation in
$B$ decays \cite{Aubert:2001nu,Abe:2001xe} have provided the first precision
test of CP violation in the Standard Model. Since the model has passed this
test successfully, we are able, for the first time, to make the following
statement: {\it The Kobayashi-Maskawa phase is, very likely,
the dominant source of CP violation in low-energy flavor-changing processes.}
 
In contrast, various alternative scenarios of CP violation that have been 
phenomenologically viable for many years are now unambiguously excluded. Two 
important examples are the following:

1. The superweak framework \cite{Wolfenstein:1964ks}, that is, the idea
that CP violation is purely indirect, is excluded by the evidence that
$\varepsilon^\prime/\varepsilon\neq0$.

2. Approximate CP, that is, the idea that all CP violating phases are
small, is excluded by the evidence that $a_{\psi K_S}={\cal O}(1)$.

The experimental result (\ref{aveapk}) and its implications for theory signify
a new era in the study of CP violation. In this series of lectures we will 
explain these recent developements and their significance.

\subsection{Why Doubt the Kobayashi-Maskawa Mechanism?}
\subsubsection{The Baryon Asymmetry of the Universe}
Cosmology shows that the Kobayashi-Maskawa phase cannot be the only source
of CP violation: baryogenesis, that is, the history of matter and antimatter in
the Universe, cannot be accounted for by the Kobayashi-Maskawa mechanism.

To understand this statement, let us provisionally switch off all CP violation.
Then, for every process that occurs in Nature, the corresponding CP conjugate
process proceeds with a precisely equal rate. Let us further assume that the
initial conditions are such that the number density of quarks and the number
density of the matching antiquarks are equal. Then, CP invariance guarantees 
that the number densities remain equal to each other along the history of the 
Universe. In other words, the baryon asymmetry, $\eta\equiv
(n_B-n_{\bar B})/n_\gamma$, is guaranteed to remain zero.
Two particularly significant processes are proton-antiproton annihilation
and production. While the first would happen at any temperature, the latter
is allowed only if the energy of the photons is large enough to produce
a proton-antiproton pair. At high enough termperatures, $T\gsim2m_p$,
annihilation and production will keep the protons and antiprotons in
equilibrium and their number densities would be (precisely equal to each
other and) similar to the photon number density, $n_B=n_{\bar B}\approx
n_\gamma$. But at temperatures well below GeV, proton-antiproton production 
slows down until it practically stops. Since annihilation continues to take 
place, the number densities of protons and antiprotons (remain equal to each 
other but) decrease, and at present there would be practically neither
matter nor antimatter. This is, of course, inconsistent with observations.

Now let us switch on CP violation. That allows a different rate for a process
and its CP conjugate. Such a situation would have relevant consequences if
two more conditions are met \cite{Sakharov:1967dj}: there is a departure from
thermal equilibrium and baryon number can be violated. When all three 
conditions are satisfied, a difference between the number densities of quarks 
and of antiquarks can be induced. We assume that the number of quarks becomes 
slightly larger than the number of antiquarks. This scenario is called 
{\it baryogenesis}. At the electroweak phase transition (temperatures of order 
a few hundred GeV, $t\sim10^{-11}$ seconds) baryon number violating processes 
become  highly suppressed, and the baryon number cannot
change any longer. The history of matter and antimatter in the Universe 
proceeds along the same lines as described in the previous paragraph. In
particular, at temperatures well below GeV the number densities of protons and
antiprotons decrease. There is however an important difference: at some time,
practically all antiprotons would disappear. But the small surplus of protons
have no matching antiprotons to annihilate with. It remains there forever.
The resulting picture of the present Universe is then as follows: there is no
antimatter. There is a small amount of matter, with the present ratio 
$(n_B/n_\gamma)_0$ reflecting the baryon asymmetry, 
$[(n_q-n_{\bar q})/n_\gamma]_{\rm BG}$, induced by baryogenesis.
This picture is qualitatively consistent with observations. Thus we have
good reasons to think that we understand the general mechanism of baryogenesis.

The important point for our purposes is that baryogenesis is a consequence of 
CP violating processes. Therefore the present baryon number, which is
accurately deduced from nucleosynthesis constraints (for a recent analysis, see
\cite{Burles:2000zk}),
\beq\label{barden}
{n_B\over n_\gamma}=(5.5\pm0.5)\times10^{-10},
\eeq
is essentially a CP violating observable! It can be added to the list
of known CP violating observables, eqs. (\ref{expeps}), (\ref{aveepp}) and
(\ref{aveapk}). Within a given model of CP violation, one can check for
consistency between the data from cosmology, eq. (\ref{barden}), and those
from laboratory experiments.

The surprising point is that the Kobayashi-Maskawa mechanism for CP violation
fails to account for (\ref{barden}). It predicts present baryon number density
that is many orders of magnitude below the observed value
\cite{Farrar:1994hn,Huet:1995jb,Gavela:1994ds}. This failure is independent
of other aspects of the Standard Model: the suppression of $n_B/n_\gamma$
from CP violation is much too strong, even if the departure from thermal
equilibrium is induced by mechanisms beyond the Standard Model. This situation
allows us to make the following statement: {\it There must exist sources of 
CP violation beyond the Kobayashi-Maskawa phase}.

Three important examples of viable models of baryogenesis are the following:

1. GUT baryogenesis (for a recent review see \cite{Riotto:1999yt}): the source 
of the baryon asymmetry is in CP violating 
decays of heavy bosons related to grand unified theories. In general, baryon 
number is not a conserved quantity in GUTs. Departure from thermal equilibrium 
is provided if the lifetime of the heavy boson is long enough that it decays 
when the temperature is well below its mass. The relevant CP violating 
parameters are not expected to affect low energy observables.

2. Leptogenesis (for a recent review see \cite{Buchmuller:2000as}): lepton 
asymmetry is induced by CP violating decays of heavy
fermions that are singlets of the Standard Model gauge group (sterile
neutrinos). Departure from thermal equilibrium is provided if the lifetime
of the heavy neutrino is long enough that it decays when the temperature is
below its mass. $B+L$-violating processes are fast before the electroweak
phase transition and convert the lepton asymmetry into a baryon asymmetry.
The CP violating parameters may be related to CP violation in the mixing
matrix for the light neutrinos (but this is a model dependent issue
\cite{Branco:2001pq}).

3. Electroweak baryogenesis (for a review see \cite{Cohen:1993nk}): the source 
of baryon asymmetry is the interactions
of top (anti)quarks with the Higgs field during the electroweak phase
transition. CP violation is induced, for example, by supersymmetric
interactions. Sphaleron configurations provide baryon number violating
interactions. Departure from thermal equilibrium is provided by the
wall between the false vacuum ($\langle\phi\rangle=0$) and the expanding
bubble with the true vacuum, where electroweak symmetry is broken.

\subsubsection{The Strong CP Problem}
Nonperturbative QCD effects induce an additional term in the SM Lagrangian,
\beq\label{ltheta}
{\cal L}_\theta={\theta_{\rm QCD}\over32\pi^2}\epsilon_{\mu\nu\rho\sigma}
F^{\mu\nu a}F^{\rho\sigma a}.
\eeq
This term violates CP. In particular, it induces an electric dipole moment
(EDM) to the neutron. The leading contribution in the chiral limit is
given by \cite{Crewther:1979pi}
\beqa\label{dntheta}
d_N&=&{g_{\pi NN}\bar g_{\pi NN}\over4\pi^2M_N}\ln(M_N/m_\pi)\no\\
&\approx&5\times10^{-16}\ \theta_{\rm QCD}\ e\ {\rm cm},
\eeqa
where $M_N$ is the nucleon mass, and $g_{\pi NN}$ ($\bar g_{\pi NN}$) is the
pseudoscalar coupling (CP-violating scalar coupling) of the pion to the nucleon.
(The leading contribution in the large $N_c$ limit was calculated in the Skyrme
model \cite{Dixon:1991cq} and leads to a similar estimate.) The experimental 
bound on $d_N$ is given by
\beq\label{dnexp}
d_N\leq 6.3\times10^{-26}\ e\ {\rm cm}\ \ \  \cite{Harris:1999jx}.
\eeq
It leads to the following bound on $\theta_{\rm QCD}$:
\beq\label{bouthe}
\theta_{\rm QCD}\lsim10^{-10}.
\eeq

Since $\theta_{\rm QCD}$ arises from nonperturbative QCD effects, it is
impossible to calculate it. Yet, there are good reasons to expect that these
effects should yield $\theta_{\rm QCD}={\cal O}(1)$ (for a clear review of
this subject, see \cite{Dine:2000cj}). Within the SM, a value as small
as (\ref{bouthe}) is unnatural, since setting $\theta_{\rm QCD}$ to zero
does not add symmetry to the model. [In particular, as we will see below, 
CP is violated by $\delta_{\rm KM}={\cal O}(1)$.] Understanding why CP is so
small in the strong interactions is the strong CP problem.

It seems then that the strong CP problem is a clue to new physics. Among
the solutions that have been proposed are a massless $u$-quark (for a review,
see \cite{Banks:1994yg}), the Peccei-Quinn mechanism
\cite{Peccei:1977hh,Peccei:1977ur} and spontaneous CP violation. As concerns
the latter, it is interesting to note that in various string theory 
compactifications, CP is an exact gauge symmetry and must be spontaneously
broken \cite{Dine:1992ya,Choi:1993xp}.

\subsubsection{New Physics}
Another motivation to measure CP violating processes is that almost any
extension of the Standard Model provides new sources of CP violation.
These sources often allow for significant deviations from the Standard
Model predictions. Moreover, various CP violating observables can be calculated
with very small hadronic uncertainties. Consequently, CP violation provides an
excellent probe of new physics.

\subsection{Will New CP Violation Be Observed In Experiments?}
The SM picture of CP violation is testable because the Kobayashi-Maskawa
mechanism is unique and predictive. These features are mainly related to
the fact that there is a single phase that is responsible to all CP violation.
As a consequence of this situation, one finds two classes of tests:

(i) Correlations: many independent CP violating observables are correlated
within the SM. For example, the SM predicts that the CP asymmetries in 
$B\to\psi K_S$ and in $B\to\phi K_S$, which proceed through different quark 
decay processes, are equal to each other. Another important example is the
strong SM correlation between CP violation in $B\to\psi K_S$ and in
$K\to\pi\nu\bar\nu$.

(ii) Zeros: since the KM phase appears in flavor-changing, weak-interaction
couplings of quarks, and only if all three generations are involved, many
CP violating observables are predicted to be negligibly small. For example,
the SM predicts no CP violation in the lepton sector, practically no CP 
violation in flavor-diagonal processes ({\it i.e.} a tiny electric dipole
moment for the neutron) and very small CP violation in tree level $D$ decays.

In addition, several CP violating observables can be calculated with
very small hadronic uncertainties. 

The strongest argument that new sources of CP violation must exist in Nature 
comes from baryogenesis. Whether the CP violation that is responsible for 
baryogenesis would be manifest in measurements of CP asymmetries in $B$ decays 
depends on two issues:

(i) The scale of the new CP violation: if the relevant scale is very high, 
such as in GUT baryogenesis or leptogenesis, the effects cannot be signalled in
these measurements. To estimate the limit on the scale, the following three 
facts are relevant: First, the Standard Model 
contributions to CP asymmetries in $B$ decays are ${\cal O}(1)$. Second,
the expected experimental accuracy would reach in some cases the few percent
level. Third, the contributions from new physics are expected to be suppressed 
by $(\Lambda_{\rm EW}/\Lambda_{\rm NP})^2$. The conclusion is that, if the new
source of CP violation is related to physics at $\Lambda_{\rm NP}\gg1\ TeV$,
it cannot be signalled in $B$ decays. Only if the true mechanism is
electroweak baryogenesis, it can potentially affect $B$ decays.

(ii) The flavor dependence of the new CP violation: if it is flavor diagonal,
its effects on $B$ decays would be highly suppressed. It can still manifest
itself in other, flavor diagonal  CP violating observables, such as electric 
dipole moments.

We conclude that new measurements of CP asymmetries in meson decays are
particularly sensitive to new sources of CP violation that come from physics
at (or below) the few TeV scale and that are related to flavor changing
couplings. This is, for example, the case, in certain supersymmetric models
of baryogenesis \cite{Worah:1997hk,Worah:1997ni}. The search for electric 
dipole moments can reveal the existence of new flavor diagonal CP violation.

Of course, there could be new flavor physics at the TeV scale that is not 
related to the baryon asymmetry and may give signals in $B$ decays. The best
motivated extension of the SM where this situation is likely is that of
supersymmetry. We will discuss supersymmetric CP violation in the last chapter.

\section{The Kobayashi-Maskawa Mechanism}
\subsection{Yukawa Interactions are the Source of CP Violation}
A model of elementary particles and their interactions is defined
by three ingredients:

(i) The symmetries of the Lagrangian;

(ii) The representations of fermions and scalars;

(iii) The pattern of spontaneous symmetry breaking.

The Standard Model (SM) is defined as follows:

(i) The gauge symmetry is 
\beq\label{smsym}
G_{\rm SM}=SU(3)_{\rm C}\times SU(2)_{\rm L}\times U(1)_{\rm Y}.
\eeq

(ii) There are three fermion generations, each consisting of five 
representations of $G_{\rm SM}$:
\beq\label{ferrep}
Q^I_{Li}(3,2)_{+1/6},\ \ U^I_{Ri}(3,1)_{+2/3},\ \ 
D^I_{Ri}(3,1)_{-1/3},\ \ L^I_{Li}(1,2)_{-1/2},\ \ E^I_{Ri}(1,1)_{-1}.
\eeq
Our notations mean that, for example, left-handed quarks, $Q_L^I$, are
triplets of $SU(3)_{\rm C}$, doublets of $SU(2)_{\rm L}$ and carry hypercharge
$Y=+1/6$. The super-index $I$ denotes interaction eigenstates. The sub-index
$i=1,2,3$ is the flavor (or generation) index.

There is a single scalar representation,
\beq\label{scarep}
\phi(1,2)_{+1/2}.
\eeq

(iii) The scalar $\phi$ assumes a VEV,
\beq\label{phivev}
\langle\phi\rangle=\pmatrix{0\cr {v\over\sqrt2}\cr},
\eeq
so that the gauge group is spontaneously broken,
\beq\label{smssb}
G_{\rm SM}\to SU(3)_{\rm C}\times U(1)_{\rm EM}.
\eeq

The Standard Model Lagrangian, ${\cal L}_{\rm SM}$, is the most general
renormalizable Lagrangian that is consistent with the gauge symmetry 
(\ref{smsym}). It can be divided to three parts:
\beq\label{LagSM}
{\cal L}_{\rm SM}={\cal L}_{\rm kinetic}+{\cal L}_{\rm Higgs}
+{\cal L}_{\rm Yukawa}.
\eeq

As concerns the kinetic terms, to maintain gauge invariance, one has 
to replace the derivative with a covariant derivative:
\beq\label{SMDmu}
D^\mu=\partial^\mu+ig_s G^\mu_a L_a+ig W^\mu_b T_b+ig^\prime B^\mu Y.
\eeq
Here $G^\mu_a$ are the eight gluon fields, $W^\mu_b$ the three
weak interaction bosons and $B^\mu$ the single hypercharge boson.
The $L_a$'s are $SU(3)_{\rm C}$ generators (the $3\times3$
Gell-Mann matrices ${1\over2}\lambda_a$ for triplets, $0$ for singlets),
the $T_b$'s are $SU(2)_{\rm L}$ generators (the $2\times2$
Pauli matrices ${1\over2}\tau_b$ for doublets, $0$ for singlets),
and $Y$ are the $U(1)_{\rm Y}$ charges. For example, for the
left-handed quarks $Q_L^I$, we have
\beq\label{DmuQL}
{\cal L}_{\rm kinetic}(Q_L)= i{\overline{Q_{Li}^I}}\gamma_\mu
\left(\partial^\mu+{i\over2}g_s G^\mu_a\lambda_a
+{i\over2}g W^\mu_b\tau_b+{i\over6}g^\prime B^\mu\right)Q_{Li}^I,
\eeq
while for the left-handed leptons $L_L^I$, we have
\beq\label{DmuLL}
{\cal L}_{\rm kinetic}(L_L)=i{\overline{L_{Li}^I}}\gamma_\mu 
\left(\partial^\mu+{i\over2}g W^\mu_b\tau_b-ig^\prime B^\mu\right)L_{Li}^I.
\eeq
These parts of the interaction Lagrangian are always CP conserving.

The Higgs potential, which describes the scalar self interactions, is given by:
\beq\label{HiPo}
{\cal L}_{\rm Higgs}=\mu^2\phi^\dagger\phi-\lambda(\phi^\dagger\phi)^2.
\eeq
For the Standard Model scalar sector, where there is a single doublet,
this part of the Lagrangian is also CP conserving.  For extended scalar
sector, such as that of a two Higgs doublet model, ${\cal L}_{\rm Higgs}$ can
be CP violating. Even in case that it is CP symmetric, it may lead
to spontaneous CP violation.

The quark Yukawa interactions are given by
\beq\label{Hqint}
-{\cal L}_{\rm Yukawa}^{\rm quarks}=Y^d_{ij}{\overline {Q^I_{Li}}}\phi D^I_{Rj}
+Y^u_{ij}{\overline {Q^I_{Li}}}\tilde\phi U^I_{Rj}+{\rm h.c.}.
\eeq
This part of the Lagrangian is, in general, CP violating.
More precisely, CP is violated if and only if \cite{Jarlskog:1985ht}
\beq\label{JarCon}
\Im\left\{\det[Y^d Y^{d\dagger},Y^u Y^{u\dagger}]\right\}\neq0.
\eeq

An intuitive explanation of why CP violation is related to {\it complex} 
Yukawa couplings goes as follows. The hermiticity of the Lagrangian implies
that ${\cal L}_{\rm Yukawa}$ has its terms in pairs of the form
\beq\label{Yukpairs}
Y_{ij}\overline{\psi_{Li}}\phi\psi_{Rj}
+Y_{ij}^*\overline{\psi_{Rj}}\phi^\dagger\psi_{Li}.
\eeq
A CP transformation exchanges the operators 
\beq\label{CPoper}
\overline{\psi_{Li}}\phi\psi_{Rj}\leftrightarrow
\overline{\psi_{Rj}}\phi^\dagger\psi_{Li},
\eeq 
but leaves their coefficients, $Y_{ij}$ and $Y_{ij}^*$, unchanged. This means 
that CP is a symmetry of ${\cal L}_{\rm Yukawa}$ if $Y_{ij}=Y_{ij}^*$.

The lepton Yukawa interactions are given by
\beq\label{Hlint}
-{\cal L}_{\rm Yukawa}^{\rm leptons}=
Y^e_{ij}{\overline {L^I_{Li}}}\phi E^I_{Rj}+{\rm h.c.}.
\eeq
It leads, as we will see in the next section, to charged lepton masses
but predicts massless neutrinos. Recent measurements of the fluxes of 
atmospheric and solar neutrinos provide evidence for neutrino masses.
That means that ${\cal L}_{\rm SM}$ cannot be a complete description of
Nature. The simplest way to allow for neutrino masses is to add 
dimension-five (and, therefore, nonrenormalizable) terms, consistent with the 
SM symmetry and particle content:
\beq\label{Hnint}
-{\cal L}_{\rm Yukawa}^{\rm dim-5}=
{Y_{ij}^\nu\over M}L_iL_j\phi\phi+{\rm h.c.}.
\eeq
The parameter $M$ has dimension of mass. The dimensionless couplings
$Y^\nu_{ij}$ are symmetric ($Y^\nu_{ij}=Y^\nu_{ji}$). We will refer to the SM 
extended to include the terms ${\cal L}_{\rm Yukawa}^{\rm dim-5}$ of eq. 
(\ref{Hnint}) as the ``extended SM" (ESM):
\beq\label{LagESM}
{\cal L}_{\rm ESM}={\cal L}_{\rm kinetic}+{\cal L}_{\rm Higgs}
+{\cal L}_{\rm Yukawa}+{\cal L}_{\rm Yukawa}^{\rm dim-5}.
\eeq
The inclusion of nonrenormalizable terms is equivalent to postulating that the 
SM is only a low energy effective theory, and that new physics appears at the 
scale $M$.

How many independent CP violating parameters are there in 
${\cal L}_{\rm Yukawa}^{\rm quarks}$? Each of the two Yukawa matrices $Y^q$ 
($q=u,d$) is $3\times3$ and complex. Consequently, there are 18 real and 18 
imaginary parameters in these matrices. Not all of them are, however, physical.
One can think of the quark Yukawa couplings as spurions that break a global
symmetry,
\beq\label{Gglobq}
U(3)_Q\times U(3)_{D}\times U(3)_{U}\ \to\ U(1)_B.
\eeq
This means that there is freedom to remove 9 real and 17 imaginary parameters 
[the number of parameters in three $3\times3$ unitary matrices minus the phase
related to $U(1)_B$]. We conclude that there are 10 quark flavor parameters: 9 
real ones and a single phase. This single phase is the source of CP violation
in the quark sector.

How many independent CP violating parameters are there in the lepton Yukawa 
interactions? The matrix $Y^e$ is a general complex $3\times3$ matrix and 
depends, therefore, on 9 real and 9 imaginary parameters. The matrix $Y^\nu$ is 
symmetric and depends on 6 real and 6 imaginary parameters. Not all of these 15
real and 15 imaginary parameters are physical. One can think of the lepton 
Yukawa couplings as spurions that break (completely) a global symmetry,
\beq\label{Gglobl}
U(3)_L\times U(3)_{E}.
\eeq
This means that 6 real and 12 imaginary parameters are not physical. We 
conclude that there are 12 lepton flavor parameters: 9 real ones and three 
phases. These three phases induce CP violation in the lepton sector.

\subsection{CKM Mixing is the (Only!) Source of CP Violation 
in the Quark Sector}
Upon the replacement $\Re(\phi^0)\rightarrow(v+H^0)/\sqrt2$ [see eq. 
(\ref{phivev})], the Yukawa interactions (\ref{Hqint}) give rise to mass terms:
\beq\label{fermasq}
-{\cal L}_M^q=(M_d)_{ij}{\overline {D^I_{Li}}} D^I_{Rj}
+(M_u)_{ij}{\overline {U^I_{Li}}} U^I_{Rj}+{\rm h.c.},
\eeq
where
\beq\label{YtoMq}
M_q={v\over\sqrt2}Y^q,
\eeq
and we decomposed the $SU(2)_{\rm L}$ quark doublets into their components:
\beq\label{doublets}
Q^I_{Li}=\pmatrix{U^I_{Li}\cr D^I_{Li}\cr}.
\eeq

The mass basis corresponds, by definition, to diagonal mass matrices. We can 
always find unitary matrices $V_{qL}$ and $V_{qR}$ such that
\beq\label{diagMq}
V_{qL}M_q V_{qR}^\dagger=M_q^{\rm diag}\ \ \ (q=u,d),
\eeq
with $M_q^{\rm diag}$ diagonal and real. The quark mass eigenstates
are then identified as
\beq\label{masses}
q_{Li}=(V_{qL})_{ij}q_{Lj}^I,\ \ \ q_{Ri}=(V_{qR})_{ij}q_{Rj}^I\ \ \ (q=u,d).
\eeq

The charged current interactions for quarks [that is the interactions of the 
charged $SU(2)_{\rm L}$ gauge bosons $W^\pm_\mu={1\over\sqrt2}
(W^1_\mu\mp iW_\mu^2)$], which in the interaction basis are described 
by (\ref{DmuQL}), have a complicated form in the mass basis:
\beq\label{Wmasq}
-{\cal L}_{W^\pm}^q={g\over\sqrt2}{\overline {u_{Li}}}\gamma^\mu
(V_{uL}V_{dL}^\dagger)_{ij}d_{Lj} W_\mu^++{\rm h.c.}.
\eeq
The unitary $3\times3$ matrix,
\beq\label{VCKM}
V_{\rm CKM}=V_{uL}V_{dL}^\dagger,\ \ \ 
(V_{\rm CKM}V_{\rm CKM}^\dagger=1),
\eeq 
is the Cabibbo-Kobayashi-Maskawa (CKM) {\it mixing matrix} for quarks
\cite{Cabibbo:1963yz,Kobayashi:1973fv}. A unitary $3\times3$ matrix depends on 
nine parameters: three real angles and six phases. 

The form of the matrix is not unique:

$(i)$ There is freedom in defining $V_{\rm CKM}$ in that we can permute between
the various generations. This freedom is fixed by ordering the up quarks and 
the down quarks by their masses, {\it i.e.} $(u_1,u_2,u_3)\to(u,c,t)$ and 
$(d_1,d_2,d_3)\to(d,s,b)$. The elements of $V_{\rm CKM}$ are written as 
follows:
\beq\label{defVij}
V_{\rm CKM}=\pmatrix{V_{ud}&V_{us}&V_{ub}\cr
V_{cd}&V_{cs}&V_{cb}\cr V_{td}&V_{ts}&V_{tb}\cr}.
\eeq

$(ii)$ There is further freedom in the phase structure of $V_{\rm CKM}$. Let us
define $P_q$ ($q=u,d$) to be diagonal unitary (phase) matrices. Then, if 
instead of using $V_{qL}$ and $V_{qR}$ for the rotation (\ref{masses}) to the 
mass basis we use $\tilde V_{qL}$ and $\tilde V_{qR}$, defined by
$\tilde V_{qL}=P_q V_{qL}$ and $\tilde V_{qR}=P_q V_{qR}$,
we still maintain a legitimate mass basis since $M_q^{\rm diag}$ remains
unchanged by such transformations. However, $V_{\rm CKM}$ does change:
\beq\label{eqphase}
V_{\rm CKM}\to P_u V_{\rm CKM}P_d^*.
\eeq 
This freedom is fixed by demanding that $V_{\rm CKM}$ has the minimal number of
phases. In the three generation case $V_{\rm CKM}$ has a single phase. (There 
are five phase differences between the elements of $P_u$ and $P_d$ and, 
therefore, five of the six phases in the CKM matrix can be removed.) This is 
the Kobayashi-Maskawa phase $\delta_{\rm KM}$ which is the single source of 
CP violation in the quark sector of the Standard Model \cite{Kobayashi:1973fv}. 

As a result of the fact that $V_{\rm CKM}$ is not diagonal, the $W^\pm$ gauge 
bosons couple to quark (mass eigenstates) of different generations. Within the 
Standard Model, this is the only source of {\it flavor changing} quark
interactions. 

\subsection{The Three Phases in the MNS Mixing Matrix}
The leptonic Yukawa interactions (\ref{Hlint}) and (\ref{Hnint}) give rise to 
mass terms:
\beq\label{fermasl}
-{\cal L}^\ell_M=(M_e)_{ij}{\overline {e^I_{Li}}}e^I_{Rj}
+(M_\nu)_{ij}\nu^I_{Li}\nu^I_{Lj}+{\rm h.c.},
\eeq
where
\beq\label{YtoMl}
M_e={v\over\sqrt2}Y^e,\ \ \ M_\nu={v^2\over2M}Y^\nu,
\eeq
and we decomposed the $SU(2)_{\rm L}$ lepton doublets into their components:
\beq\label{ldoublets}
L^I_{Li}=\pmatrix{\nu^I_{Li}\cr e^I_{Li}\cr}.
\eeq

We can always find unitary matrices $V_{eL}$ and $V_\nu$ such that
\beq\label{diagMl}
V_{eL}M_eM_e^\dagger V_{eL}^\dagger={\rm diag}(m_e^2,m_\mu^2,m_\tau^2),\ \ \ 
V_\nu M_\nu^\dagger M_\nu V_\nu^\dagger={\rm diag}(m_1^2,m_1^2,m_3^2).
\eeq
The charged current interactions for leptons, which in the interaction basis 
are described by (\ref{DmuLL}), have the following form in the mass basis:
\beq\label{Wmasl}
-{\cal L}_{W^\pm}^\ell={g\over\sqrt2}{\overline {e_{Li}}}\gamma^\mu
(V_{eL}V_{\nu}^\dagger)_{ij}\nu_{Lj} W_\mu^-+{\rm h.c.}.
\eeq
The unitary $3\times3$ matrix,
\beq\label{VMNS}
V_{\rm MNS}=V_{eL}V_{\nu}^\dagger,
\eeq 
is the Maki-Nakagawa-Sakata (MNS) {\it mixing matrix} for leptons 
\cite{Maki:1962mu}. Similarly to the CKM matrix, the form of the MNS matrix is 
not unique. But there are differences in choosing conventions:

$(i)$ We can permute between the various generations. This freedom is usually 
fixed in the following way. We order the charged leptons by their masses, 
{\it i.e.} $(e_1,e_2,e_3)\to(e,\mu,\tau)$. As concerns the neutrinos, one takes
into account that the interpretation of atmospheric and solar neutrino data in 
terms of two-neutrino oscillations implies that $\Delta m^2_{\rm ATM}\gg
\Delta m^2_{\rm SOL}$. It follows that one of the neutrino mass eigenstates is 
separated in its mass from the other two, which have a smaller mass difference.
The convention is to denote this separated state by $\nu_3$. For the remaining 
two neutrinos, $\nu_1$ and $\nu_2$, the convention is to call the heavier state
$\nu_2$. In other words, the three mass eigenstates are defined by the 
following conventions:
\beq\label{conneu}
|\Delta m^2_{3i}|\gg|\Delta m^2_{21}|,\ \ \ \Delta m^2_{21}>0.
\eeq
Note in particular that $\nu_3$ can be either heavier or lighter than 
$\nu_{1,2}$. The elements of $V_{\rm MNS}$ are written as  follows:
\beq\label{defVijl}
V_{\rm MNS}=\pmatrix{V_{e1}&V_{e2}&V_{e3}\cr
V_{\mu1}&V_{\mu2}&V_{\mu3}\cr V_{\tau1}&V_{\tau2}&V_{\tau3}\cr}.
\eeq

$(ii)$ There is further freedom in the phase structure of $V_{\rm MNS}$. 
(In the MNS paper \cite{Maki:1962mu} there is no reference to CP violation.)
One can change the charged lepton mass basis by the transformation
$e_{(L,R)i}\to e^\prime_{(L,R)i}=(P_e)_{ii} e_{(L,R)i}$, where $P_e$ is a phase 
matrix. There is, however, no similar freedom to redefine the neutrino
mass eigenstates: From eq. (\ref{fermasl}) one learns that a transformation
$\nu_{L}\to P_\nu\nu_{L}$ will introduce phases into the diagonal mass matrix. 
This is related to the Majorana nature of neutrino masses, assumed in
eq. (\ref{Hnint}). The allowed transformation modifies $V_{\rm MNS}$:
\beq\label{ephase}
V_{\rm MNS}\to P_e V_{\rm MNS}.
\eeq 
This freedom is fixed by demanding that $V_{\rm MNS}$ will have the minimal 
number of phases. Out of six phases of a generic unitary $3\times3$ matrix,
the multiplication by $P_e$ can be used to remove three. We conclude that the 
three generation $V_{\rm MNS}$ matrix has three
phases. One of these is the analog of the Kobayashi-Maskawa phase. It is 
the only source of CP violation in processes that conserve lepton number,
such as neutrino flavor oscillations. The other two phases
can affect lepton number changing processes.

With $V_{\rm MNS}\neq{\bf 1}$, the $W^\pm$ gauge bosons couple to lepton (mass 
eigenstates) of different generations. Within the ESM, this is the 
only source of {\it flavor changing} lepton interactions. 

\subsection{The Flavor Parameters}
Examining the quark mass basis, one can easily identify the flavor parameters.
In the quark sector, we have six quark masses and four mixing parameters:
three mixing angles and a single phase. 

The fact that there are only three real and one imaginary physical parameters 
in $V_{\rm CKM}$ can be made manifest by choosing an explicit parameterization.
For example, the standard parameterization \cite{Chau:1984fp}, used by the 
particle data group, is given by 
\beq\label{stapar}
V_{\rm CKM}=\pmatrix{c_{12}c_{13}&s_{12}c_{13}&
s_{13}e^{-i\delta}\cr 
-s_{12}c_{23}-c_{12}s_{23}s_{13}e^{i\delta}&
c_{12}c_{23}-s_{12}s_{23}s_{13}e^{i\delta}&s_{23}c_{13}\cr
s_{12}s_{23}-c_{12}c_{23}s_{13}e^{i\delta}&
-c_{12}s_{23}-s_{12}c_{23}s_{13}e^{i\delta}&c_{23}c_{13}\cr},
\eeq
where $c_{ij}\equiv\cos\theta_{ij}$ and $s_{ij}\equiv\sin\theta_{ij}$. The 
three $\sin\theta_{ij}$ are the three real mixing parameters while $\delta$ is 
the Kobayashi-Maskawa phase. Another, very useful, example is the Wolfenstein 
parametrization, where the four mixing parameters are $(\lambda,A,\rho,\eta)$ 
with $\lambda=|V_{us}|=0.22$ playing the role of an expansion parameter
and $\eta$ representing the CP violating phase \cite{Wolfenstein:1983yz}:
\beq\label{WCKM}
V_{\rm CKM}=\pmatrix{1-{\lambda^2\over2}&\lambda&A\lambda^3(\rho-i\eta)\cr
-\lambda&1-{\lambda^2\over2}&A\lambda^2\cr
A\lambda^3(1-\rho-i\eta)&-A\lambda^2&1\cr}+{\cal O}(\lambda^4).
\eeq

Various parametrizations differ in the way that the freedom of phase rotation, 
eq. (\ref{eqphase}), is used to leave a single phase in $V_{\rm CKM}$. One can 
define, however, a CP violating quantity in $V_{\rm CKM}$ that is independent 
of the parametrization \cite{Jarlskog:1985ht}. This quantity, $J_{\rm CKM}$, is 
defined through 
\beq\label{defJ}
\Im[V_{ij}V_{kl}V_{il}^*V_{kj}^*]=
J_{\rm CKM}\sum_{m,n=1}^3\epsilon_{ikm}\epsilon_{jln},\ \ \ (i,j,k,l=1,2,3).
\eeq
In terms of the explicit parametrizations given above, we have
\beq\label{parJ}
J_{\rm CKM}=c_{12}c_{23}c_{13}^2s_{12}s_{23}s_{13}\sin\delta\simeq \lambda^6 
A^2\eta.
\eeq

It is interesting to translate the condition (\ref{JarCon}) to the language
of the flavor parameters in the mass basis. One finds that the following is
a necessary and sufficient condition for CP violation in the quark sector of
the SM:
\beq\label{jarconmas}
\Delta m^2_{tc}\Delta m^2_{tu}\Delta m^2_{cu}\Delta m^2_{bs}\Delta m^2_{bd}
\Delta m^2_{sd}J_{\rm CKM}\neq0.
\eeq
Here
\beq\label{defdelij}
\Delta m^2_{ij}\equiv m_i^2-m_j^2.
\eeq
Equation (\ref{jarconmas}) puts the following requirements on the SM in order
that it violates CP:

(i) Within each quark sector, there should be no mass degeneracy;

(ii) None of the three mixing angles should be zero or $\pi/2$;

(iii) The phase should be neither 0 nor $\pi$.

As concerns the lepton sector of the ESM, the flavor parameters are the six
lepton masses, and six mixing parameters: three mixing angles and three phases.
One can parameterize $V_{\rm MNS}$ in a convenient way by factorizing it into 
$V_{\rm MNS}=VP$. Here $P$ is a diagonal unitary matrix that depends on two 
phases, {\it e.g.} $P={\rm diag}(e^{i\phi_1},e^{i\phi_2},1)$, while $V$ can be 
parametrized in the same way as (\ref{stapar}). The advantage of this 
parametrization is that for the purpose of analyzing lepton number conserving
processes and, in particular, neutrino flavor oscillations, the parameters of 
$P$ are usually irrelevant and one can use the same Chau-Keung parametrization 
as is being used for $V_{\rm CKM}$. (An alternative way to understand these
statements is to use a single-phase mixing matrix and put the extra two phases
in the neutrino mass matrix. Then it is obvious that the effects of these
`Majorana-phases' always appear in conjunction with a factor of the
Majorana mass that is lepton number violating parameter.) On the other hand, 
the Wolfenstein parametrization (\ref{WCKM}) is inappropriate for the lepton 
sector: it assumes $|V_{23}|\ll|V_{12}|\ll1$, which does not hold here.

In order that the CP violating phase $\delta$ in $V$ would be physically
meaningful, {\it i.e.} there would be CP violation that is not related to
lepton number violation, a condition similar to (\ref{jarconmas}) should hold:
\beq\label{jarconlep}
\Delta m^2_{\tau\mu}\Delta m^2_{\tau e}\Delta m^2_{\mu e}
\Delta m^2_{32}\Delta m^2_{31}\Delta m^2_{21}J_{\rm MNS}\neq0.
\eeq

\subsection{The Unitarity Triangles}
A very useful concept is that of the {\it unitarity triangles}. We will focus
on the quark sector, but analogous triangles can be defined in the lepton
sector. The unitarity of the CKM matrix leads to various relations among
the matrix elements, {\it e.g.}
\beqa\label{Unitds}
V_{ud}V_{us}^*+V_{cd}V_{cs}^*+V_{td}V_{ts}^*=0,\\
\label{Unitsb}
V_{us}V_{ub}^*+V_{cs}V_{cb}^*+V_{ts}V_{tb}^*=0,\\
\label{Unitdb}
V_{ud}V_{ub}^*+V_{cd}V_{cb}^*+V_{td}V_{tb}^*=0.
\eeqa
Each of these three relations requires 
the sum of three complex quantities to vanish and so can be geometrically
represented in the complex plane as a triangle. These are
``the unitarity triangles", though the term ``unitarity triangle"
is usually reserved for the relation (\ref{Unitdb}) only. It is a surprising
feature of the CKM matrix that all unitarity triangles are equal in area:
the area of each unitarity triangle equals $|J_{\rm CKM}|/2$ while the sign of 
$J_{\rm CKM}$ gives the direction of the complex vectors around the triangles. 

The rescaled unitarity triangle  is derived from (\ref{Unitdb})
by (a) choosing a phase convention such that $(V_{cd}V_{cb}^*)$
is real, and (b) dividing the lengths of all sides by $|V_{cd}V_{cb}^*|$.
Step (a) aligns one side of the triangle with the real axis, and
step (b) makes the length of this side 1. The form of the triangle
is unchanged. Two vertices of the rescaled unitarity triangle are
thus fixed at (0,0) and (1,0). The coordinates of the remaining
vertex correspond to the Wolfenstein parameters $(\rho,\eta)$.
The area of the rescaled unitarity triangle is $|\eta|/2$.

Depicting the rescaled unitarity triangle in the
$(\rho,\eta)$ plane, the lengths of the two complex sides are
\beq\label{RbRt}
R_u\equiv\left|{V_{ud}V_{ub}\over V_{cd}V_{cb}}\right|
=\sqrt{\rho^2+\eta^2},\ \ \
R_t\equiv\left|{V_{td}V_{tb}\over V_{cd}V_{cb}}\right|
=\sqrt{(1-\rho)^2+\eta^2}.
\eeq
The three angles of the unitarity triangle are defined as follows 
\cite{Dib:1990uz,Rosner:1988nx}:
\beq\label{abcangles}
\alpha\equiv\arg\left[-{V_{td}V_{tb}^*\over V_{ud}V_{ub}^*}\right],\ \ \
\beta\equiv\arg\left[-{V_{cd}V_{cb}^*\over V_{td}V_{tb}^*}\right],\ \ \
\gamma\equiv\arg\left[-{V_{ud}V_{ub}^*\over V_{cd}V_{cb}^*}\right].
\eeq
They are physical quantities and can be independently 
measured by CP asymmetries in $B$ decays 
\cite{Carter:1980hr,Carter:1981tk,Bigi:1981qs,Dunietz:1986vi,Bigi:1987vr}. 
It is also useful to define the two 
small angles of the unitarity triangles (\ref{Unitsb}) and (\ref{Unitds}):
\beq\label{bbangles}
\beta_s\equiv\arg\left[-{V_{ts}V_{tb}^*\over V_{cs}V_{cb}^*}\right],\ \ \
\beta_K\equiv\arg\left[-{V_{cs}V_{cd}^*\over V_{us}V_{ud}^*}\right].
\eeq

To make predictions for future measurements of CP violating observables, we 
need to find the allowed ranges for the CKM phases. There are three ways to 
determine the CKM parameters (see {\it e.g.} \cite{Harari:1987ex}):

(i) {\bf Direct measurements} are related to SM tree level  processes. 
At present, we have direct measurements of $|V_{ud}|$, $|V_{us}|$, $|V_{ub}|$,
$|V_{cd}|$, $|V_{cs}|$, $|V_{cb}|$ and $|V_{tb}|$. 

(ii) {\bf CKM Unitarity}  ($V_{\rm CKM}^\dagger V_{\rm CKM}={\bf 1}$) relates 
the various matrix elements. At present, these relations are useful to 
constrain $|V_{td}|$, $|V_{ts}|$, $|V_{tb}|$ and $|V_{cs}|$.
 
(iii) {\bf Indirect measurements} are related to SM loop processes. 
At present, we constrain in this way $|V_{tb}V_{td}|$ (from $\Delta m_B$ 
and  $\Delta m_{B_s}$) and $\delta_{\rm KM}$ or, equivalently, $\eta$ or 
$\beta$ (from $\varepsilon_K$ and $a_{\psi K_S}$).

\begin{figure}[h]   
  \begin{center}
    \includegraphics[width=0.8\textwidth]{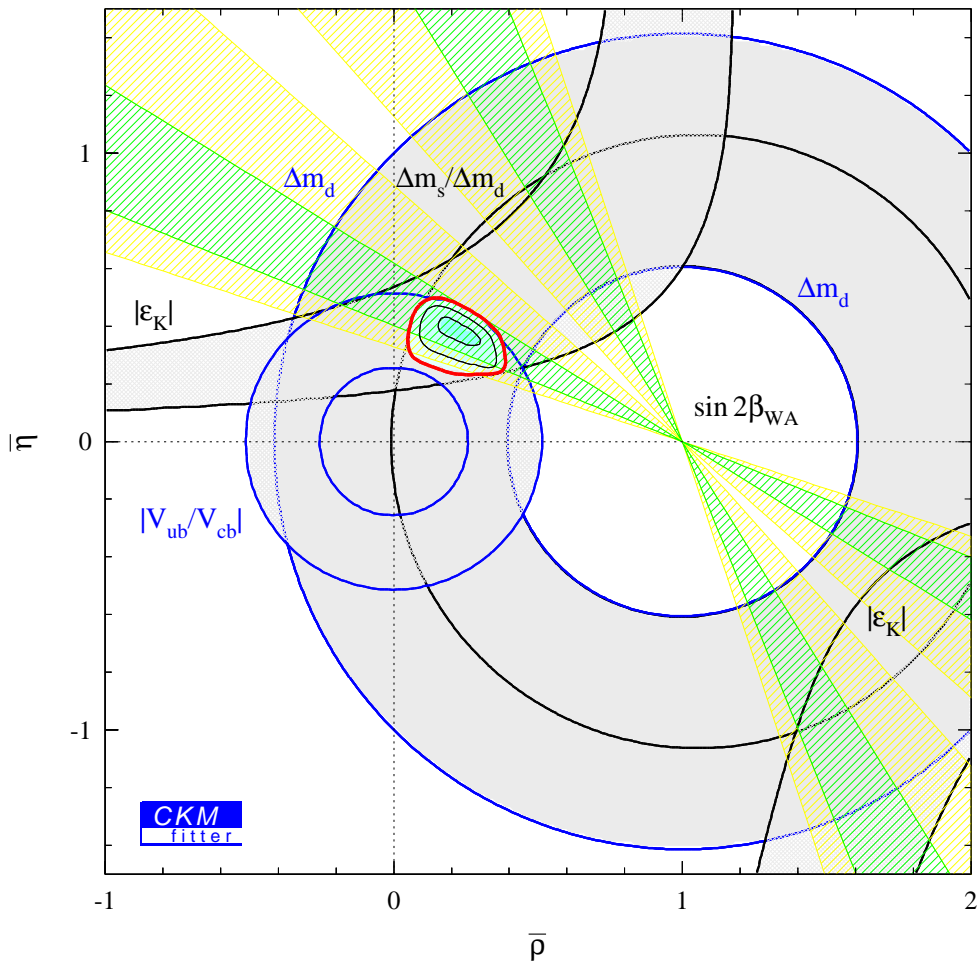}
    \caption{Present Standard Model constraints and the result from
     the global CKM fit.}
    \label{fig:fig1}
  \end{center}
\end{figure}

When all available data are taken into account, one finds \cite{Hocker:2001xe}:
\beqa\label{lacon}
\lambda&=&0.2221\pm0.0021,\ \ \ A=0.827\pm0.058,\\
\label{recon}
\rho&=&0.23\pm0.11,\ \ \ \eta=0.37\pm0.08,\\
\label{abccon}
\sin2\beta&=&0.77\pm0.08,\ \ \ \sin2\alpha=-0.21\pm0.56,\ \ \ 
0.43\lsim\sin^2\gamma\leq0.91.
\eeqa
Of course, there are correlations between the various parameters. The full 
information in the $(\rho,\eta)$ plane is given in fig. \ref{fig:fig1} 
\cite{Hocker:2001xe}.

\subsection{The Uniqueness of the Standard Model Picture of CP Violation}
In the previous subsections, we have learnt several features of CP violation 
as explained by the Standard Model. It is important to understand that various 
reasonable (and often well-motivated) extensions of the SM provide examples 
where some or all of these features do not hold. Furthermore, until a few years
ago, none of the special features of the Kobayashi-Maskawa mechanism of CP 
violation has been experimentally tested. This situation has dramatically 
changed recently. Let us survey some of the SM features, how they can be
modified with new physics, and whether experiment has shed light on
these questions.

(i) {\it $\delta_{\rm KM}$ is the only source of CP violation in meson decays.}
This is arguably the most unique feature of the SM and gives the model a strong
predicitive power. It is violated in almost any low-energy extension.
For example, in the supersymmetric extension of the SM there are
44 physical CP violating phases, many of which affect meson decays.
The measured value of $a_{\psi K_S}$ is consistent with the correlation 
between $K$ and $B$ decays that is predicted by the SM. It is therefore very
likely that $\delta_{\rm KM}$ is indeed the dominant source of CP violation in
meson decays.

(ii) {\it CP violation is small in $K\to\pi\pi$ decays because of flavor
suppression and not because CP is an approximate symmetry.} In many
(though certainly not all) supersymmetric models, the flavor suppression
is too mild, or entirely ineffective, requiring approximate CP to hold.
The measurement of $a_{\psi K_S}={\cal O}(1)$ confirms that not all
CP violating phases are small.
 
(iii) {\it CP violation appears in both $\Delta F=1$ (decay) and
$\Delta F=2$ (mixing) amplitudes.} Superweak models suggest that CP is
violated only in mixing amplitudes.
The measurement of $\varepsilon^\prime/\varepsilon$ confirms that there is
CP violation in $\Delta S=1$ processes.

(iv) {\it CP is not violated in the lepton sector.} Models that allow for
neutrino masses, such as the ESM framework presented above, predict
CP violation in leptonic charged current interactions.
The data from neutrino oscillation experiments makes it very likely that 
charged current weak interactions violate CP also in the lepton sector.

(v) {\it CP violation appears only in the charged current weak interactions
and in conjunction with flavor changing processes.} Here both various 
extensions of the SM (such as supersymmetry) and non-perturbative effects 
within the SM ($\theta_{\rm QCD}$) allow for CP violation in other types of 
interactions and in flavor diagonal processes. In particular, it is difficult 
to avoid flavor-diagonal phases in the supersymmetric framework.
The fact that no electric dipole moment has been measured yet poses 
difficulties to many models with diagonal CP violation (and, of course, is
responsible to the strong CP problem within the SM).

(vi) {\it CP is explicitly broken.} In various extensions of the scalar
sector, it is possible to achieve spontaneous CP violation.
It will be very difficult to test this question experimentally.

This situation, where the Standard Model has a very unique and predictive
description of CP violation and the number of experimentally measured CP
violating observables is very limited ($\varepsilon_K$, 
$\varepsilon^\prime/\varepsilon$ and $a_{\psi K_S}$), is the basis for the 
strong interest, experimental and theoretical, in CP violation. There are two 
types of unambiguous tests concerning CP violation in the Standard Model:
First, since there is a single source of CP violation, all observables
are correlated with each other. For example, the CP asymmetries in
$B\to\psi K_S$ and in $K\to\pi\nu\bar\nu$ are strongly correlated
\cite{Buchalla:1994tr,Buchalla:1996fp,Bergmann:2000ak}.
Second, since CP violation is restricted to flavor changing fermion
processes, it is predicted to be highly suppressed in the lepton sector
and practically vanish in flavor diagonal processes. For example, the 
transverse lepton polarization in semileptonic meson decays, CP violation
in $t\bar t$ production, and (assuming $\theta_{\rm QCD}=0$) the electric 
dipole moment of the neutron are all predicted to be orders of magnitude
below the (present and near future) experimental sensitivity. We conclude that
it is highly important to search for CP violation in many different systems.

\section{Meson Decays}
In the previous section, we explained how CP violation arises in the Standard 
Model. In the next three sections, we would like to understand the implications
of this theory for the phenomenology of CP violation in $K$, $D$ and $B$ 
decays. To do so, we first present a model independent analysis of CP violation
in meson decays.

We distinguish between three different types of CP violation in meson decays:

(i) CP violation in mixing, which occurs when the two neutral mass
eigenstate admixtures cannot be chosen to be CP-eigenstates;

(ii) CP violation in decay, which occurs in both charged and neutral decays, 
when the amplitude for a decay and its CP-conjugate process have different 
magnitudes;

(iii) CP violation in the interference of decays with and without mixing, which
occurs in decays into final states that are common to $B^0$ and $\bar B^0$. 

\subsection{Notations and Formalism}
To define these three types and to discuss their theoretical calculation and 
experimental measurement, we first introduce some notations and formalism. We 
refer specifically to $B$ meson mixing and decays, but most of our discussion 
applies equally well to $K$, $B_s$ and $D$ mesons.

A $B^0$ meson is made from a $b$-type antiquark and an $d$-type quark, while
the $\bar B^0$ meson is made from a $b$-type quark and an $d$-type antiquark.
Our phase convention for the CP transformation law of the neutral $B$ mesons is
defined by
\beq\label{phacon}
{\rm CP}|{B^0}\rangle=\omega_B|{\bar B^0}\rangle,\ \ \
{\rm CP}|{\bar B^0}\rangle=\omega_B^*|{B^0}\rangle,\ \ \ (|\omega_B|=1).
\eeq
Physical observables do not depend on the phase factor $\omega_B$. 

The light, $B_L$, and heavy, $B_H$, mass eigenstates can be written as linear
combinations of $B^0$ and $\bar B^0$:
\beqa\label{defqp}
|B_L\rangle&=&p|{B^0}\rangle+q|{\bar B^0}\rangle,\no\\
|B_H\rangle&=&p|{B^0}\rangle-q|{\bar B^0}\rangle,
\eeqa
with
\beq\label{norqp}
|q|^2+|p|^2=1.
\eeq  
The mass difference $\Delta m_B$ and the width difference $\Delta\Gamma_B$
are defined as follows:
\beq\label{DelmG}
\Delta m\equiv M_H-M_L,\ \ \ \Delta\Gamma\equiv\Gamma_H-\Gamma_L.
\eeq
The average mass and width are given by
\beq\label{aveMG}
m_B\equiv{M_H+M_L\over2},\ \ \ \Gamma_B\equiv{\Gamma_H+\Gamma_L\over2}.
\eeq
It is useful to define dimensionless ratios $x$ and $y$:
\beq\label{defxy}
x\equiv{\Delta m\over\Gamma},\ \ \ y\equiv{\Delta\Gamma\over2\Gamma}.
\eeq

The time evolution of the mass eigenstates is simple:
\beqa\label{tievmes}
|B_H(t)\rangle&=& e^{-iM_Ht}e^{-\Gamma_Ht/2}|B_H\rangle,\no\\
|B_L(t)\rangle&=& e^{-iM_Lt}e^{-\Gamma_Lt/2}|B_L\rangle.
\eeqa
The time evolution of the strong interaction eigenstates is complicated and 
obeys a Schr\"odinger-like equation,
\beq\label{Schro}
i{d\over dt}\pmatrix{B\cr \bar B\cr}=
\left(M-{i\over2}\Gamma\right)\pmatrix{B\cr \bar B\cr},
\eeq
where $M$ and $\Gamma$ are $2\times2$ Hermitian matrices.

The off-diagonal terms in these matrices, $M_{12}$ and $\Gamma_{12}$, are 
particularly important in the discussion of mixing and CP violation. $M_{12}$ 
is the dispersive part of the transition amplitude from $B^0$ to $\bar B^0$, 
while $\Gamma_{12}$ is the absorptive part of that amplitude.

Solving the eigenvalue equation gives
\beqa\label{eveq}
(\Delta m)^2-{1\over4}(\Delta\Gamma)^2&=&(4|M_{12}|^2-|\Gamma_{12}|^2),\ \ \ \ 
\Delta m\Delta\Gamma=4\Re(M_{12}\Gamma_{12}^*),\\
\label{solveqp}
{q\over p}&=&-{2M_{12}^*-i\Gamma_{12}^*\over\Delta m-{i\over2}\Delta\Gamma}=
-{\Delta m-{i\over2}\Delta\Gamma\over 2M_{12}-i\Gamma_{12}}.
\eeqa
In the $B$ system, $|\Gamma_{12}|\ll|M_{12}|$ (see discussion below), and then,
to leading order in $|\Gamma_{12}/M_{12}|$, eqs. (\ref{eveq}) and 
(\ref{solveqp}) can be written as
\beqa\label{eveqB}
\Delta m_B&=&2|M_{12}|,\ \ \ 
\Delta\Gamma_B=\ 2\Re(M_{12}\Gamma_{12}^*)/|M_{12}|,
\\ \label{solveqpB}
{q/p}&=&-{M_{12}^*/|M_{12}|}.
\eeqa

To discuss CP violation in mixing, it is useful to write eq. (\ref{solveqp}) 
to first order in $|\Gamma_{12}/M_{12}|$ [rather than to zeroth order as in
(\ref{solveqpB})]:
\beq\label{solveqpC}
{q\over p}=-{M_{12}^*\over|M_{12}|}\left[1-{1\over2}
\Im\left({\Gamma_{12}\over M_{12}}\right)\right].
\eeq

To discuss CP violation in decay, we need to consider decay amplitudes. The CP 
transformation law for a final state $f$ is
\beq\label{phaconf}
{\rm CP}|{f}\rangle=\omega_f|{\bar f}\rangle,\ \ \
{\rm CP}|{\bar f}\rangle=\omega_f^*|{f}\rangle,\ \ \ (|\omega_f|=1).
\eeq
For a final CP eigenstate $f=\bar f=f_{\rm CP}$, the phase factor $\omega_f$ is 
replaced by $\eta_{f_{\rm CP}}=\pm1$, the CP eigenvalue of the final state.
We define the decay amplitudes $A_f$ and $\bar A_f$ according to
\beq\label{defAf}
A_f=\langle f|{\cal H}_d|B^0\rangle,\ \ \ 
\bar A_f=\langle f|{\cal H}_d|\bar B^0\rangle,
\eeq
where ${\cal H}_d$ is the decay Hamiltonian. 

CP relates $A_f$ and $\bar A_{\bar f}$. There are two types of phases that 
may appear in $A_f$ and $\bar A_{\bar f}$. Complex parameters in any Lagrangian
term that contributes to the amplitude will appear in complex conjugate form in
the CP-conjugate amplitude. Thus their phases appear in $A_f$ and 
$\bar A_{\bar f}$ with opposite signs. In the SM these phases occur only in the
mixing matrices that parameterize the charged current weak interactions, hence 
these are often called ``weak phases''. The weak phase of any single term is
convention dependent. However the difference between the weak phases in two 
different terms in $A_f$ is convention independent because the phase rotations 
of the initial and final states are the same for every term. A second type of 
phase can appear in scattering or decay amplitudes even when the Lagrangian is 
real. Such phases do not violate CP and they appear in $A_f$ and 
$\bar A_{\bar f}$ with the same sign. Their origin is the possible contribution
from intermediate on-shell states in the decay process, that is an absorptive 
part of an amplitude that has contributions from coupled channels. Usually the 
dominant rescattering is due to strong interactions and hence the designation
``strong phases'' for the phase shifts so induced. Again only the relative 
strong phases of different terms in a scattering amplitude have physical 
content, an overall phase rotation of the entire amplitude has no physical 
consequences. Thus it is useful to write each contribution to $A$ in three 
parts: its magnitude $A_i$; its weak phase term $e^{i\phi_i}$; and its strong
phase term $e^{i\delta_i}$. Then, if several amplitudes contribute to
$B\to f$, we have
\beq\label{defAtoA}
\left|{\bar A_{\bar f}\over A_f}\right|=\left|{\sum_i A_i 
e^{i(\delta_i-\phi_i)}\over  \sum_i A_i e^{i(\delta_i+\phi_i)}}\right|.
\eeq

To discuss CP violation in the interference of decays with and without 
mixing, we introduce a complex quantity $\lambda_f$ defined by
\beq\label{deflam}
\lambda_f\ =\ {q\over p}\ {\bar A_f\over A_f}.
\eeq

We further define the CP transformation law for the quark fields in the 
Hamiltonian (a careful treatment of CP conventions can be found in 
\cite{Branco:1999fs}):
\beq\label{CPofq}
q\ \rightarrow\ \omega_q\bar q,\ \ \ 
\bar q\ \rightarrow\ \omega_q^*q,\ \ \ (|\omega_q|=1).
\eeq
The effective Hamiltonian that is relevant to $M_{12}$ is of the form
\beq\label{Hbtwo}
H^{\Delta b=2}_{\rm eff}\propto e^{+2i\phi_B}\left[\bar d\gamma^\mu(1-\gamma_5)
b\right]^2+e^{-2i\phi_B}\left[\bar b\gamma^\mu(1-\gamma_5)d\right]^2,
\eeq
where $2\phi_B$ is a CP violating (weak) phase. (We use the SM $V-A$ amplitude,
but the results can be generalized to any Dirac structure.) For the $B$ system,
where $|\Gamma_{12}|\ll |M_{12}|$, this leads to
\beq\label{qpforB}
q/p=\omega_B\omega_b^*\omega_d e^{-2i\phi_B}.
\eeq
(We implicitly assumed that the vacuum insertion approximation gives the 
correct sign for $M_{12}$. In general, there is a sign($B_B$) factor on the 
right hand side of eq. (\ref{qpforB}) \cite{Grossman:1997xn}.) To understand 
the phase structure of decay amplitudes, we take as an example the $b\to 
q\bar qd$ decay ($q=u$ or $c$). The decay Hamiltonian is of the form
\beq\label{Hdecay}
H_d\propto e^{+i\phi_f}\left[\bar q\gamma^\mu(1-\gamma_5)d\right]
\left[\bar b\gamma_\mu(1-\gamma_5)q\right]
+e^{-i\phi_f}\left[\bar q\gamma^\mu(1-\gamma_5)b\right]
\left[\bar d\gamma_\mu(1-\gamma_5)q\right],
\eeq
where $\phi_f$ is the appropriate weak phase. (Again, for simplicity we use a 
$V-A$ structure, but the results hold for any Dirac structure.) Then
\beq\label{AbarA}
\bar A_{\bar f}/A_f=\omega_f\omega_B^*\omega_b\omega_d^* e^{-2i\phi_f}.
\eeq
Eqs. (\ref{qpforB}) and (\ref{AbarA}) together imply that for a final CP 
eigenstate,
\beq\label{lamfCP}
\lambda_{f_{\rm CP}}=\eta_{f_{\rm CP}}e^{-2i(\phi_B+\phi_f)}.
\eeq

\subsection{The Three Types of CP Violation in Meson Decays}
\subsubsection{CP violation in mixing}
\beq\label{inmixin}
|q/p|\neq1.
\eeq
This type of CP violation results from the mass eigenstates being different 
from the CP eigenstates, and requires a relative phase between $M_{12}$
and $\Gamma_{12}$. For the neutral $B$ system, this effect could
be observed through the asymmetries in semileptonic decays:
\beq\label{mixexa}
a_{\rm SL}={\Gamma(\bar B^0_{\rm phys}(t)\to\ell^+\nu X)-
\Gamma(B^0_{\rm phys}(t)\to\ell^-\nu X)\over
\Gamma(\bar B^0_{\rm phys}(t)\to\ell^+\nu X)+
\Gamma(B^0_{\rm phys}(t)\to\ell^-\nu X)}.
\eeq
In terms of $q$ and $p$,
\beq\label{mixter}
a_{\rm SL}={1-|q/p|^4\over1+|q/p|^4}.
\eeq
CP violation in mixing has been observed in the neutral $K$ system
($\Re\ \varepsilon_K\neq0$).

In the neutral $B$ system, the effect is expected to be small, 
$\lsim{\cal O}(10^{-2})$. The reason is that, model independently, the effect 
cannot be larger than ${\cal O}(\Delta\Gamma_B/\Delta m_B)$. The difference in 
width is produced by decay channels common to $B^0$ and $\bar B^0$. The 
branching ratios for such channels are at or below the level of $10^{-3}$.
Since various channels contribute with differing signs, one expects that their 
sum does not exceed the individual level. Hence, we can safely assume that 
$\Delta\Gamma_B/\Gamma_B={\cal O}(10^{-2})$. On the other hand, it is
experimentaly known that $\Delta m_B/\Gamma_B\approx0.7$.

To calculate $a_{\rm SL}$, we use (\ref{mixter}) and (\ref{solveqpC}), and get:
\beq\label{Absqp}
a_{\rm SL}=\Im(\Gamma_{12}/M_{12}).
\eeq
To predict it in a given model, one needs to calculate $M_{12}$ and 
$\Gamma_{12}$. This involves large hadronic uncertainties, in particular in the
hadronization models for $\Gamma_{12}$.

\subsubsection{CP violation in decay}
\beq\label{indecay}
|\bar A_{\bar f}/A_f|\neq1.
\eeq
This appears as a result of interference among various terms in the decay 
amplitude, and will not occur unless at least two terms have different weak 
phases and different strong phases. CP asymmetries in charged $B$ decays,
\beq\label{decexa}
a_{f^\pm}={\Gamma(B^+\to f^+)-\Gamma(B^-\to f^-)\over
\Gamma(B^+\to f^+)+\Gamma(B^-\to f^-)},
\eeq
are purely an effect of CP violation in decay. In terms of the decay amplitudes,
\beq\label{decter}
a_{f^\pm}={1-|\bar A_{f^-}/A_{f^+}|^2\over1+|\bar A_{f^-}/A_{f^+}|^2}.
\eeq
CP violation in decay has been observed in the neutral $K$ system 
($\Re\ \varepsilon_K^\prime\neq0$).

To calculate $a_{f^\pm}$, we use (\ref{decter}) and (\ref{defAtoA}). For 
simplicity, we consider decays with contributions from two weak phases and with
$A_2\ll A_1$. We get:
\beq\label{AbsAA}
a_{f^\pm}=-2(A_2/A_1)\sin(\delta_2-\delta_1)\sin(\phi_2-\phi_1).
\eeq
The magnitude and strong phase of any amplitude involve long distance strong 
interaction physics, and our ability to calculate these from first principles 
is limited. Thus quantities that depend only on the weak phases are much 
cleaner than those that require knowledge of the relative magnitudes or strong 
phases of various amplitude contributions, such as CP violation in decay.

\subsubsection{CP violation in the interference between decays
 with and without mixing}
\beq\label{ininter}
\Im\ \lambda_{f_{\rm CP}}\neq0.
\eeq
This effect is the result of interference between a direct decay amplitude and 
a first-mix-then-decay path to the same final state. For the neutral $B$ 
system, the effect can be observed by comparing decays into final CP 
eigenstates of a time-evolving neutral $B$ state that begins at time zero as 
$B^0$ to those of the state that begins as $\bar B^0$:
\beq\label{intexa}
a_{f_{\rm CP}}(t)={\Gamma(\bar B^0_{\rm phys}(t)\to f_{\rm CP})-
\Gamma(B^0_{\rm phys}(t)\to f_{\rm CP})\over
\Gamma(\bar B^0_{\rm phys}(t)\to f_{\rm CP})+
\Gamma(B^0_{\rm phys}(t)\to f_{\rm CP})}.
\eeq
This time dependent asymmetry is given, in general, by
\beq\label{intgene}
a_{f_{\rm CP}}(t)=-{1-|\lambda_{f_{\rm CP}}|^2\over1+|\lambda_{f_{\rm CP}}|^2}
\cos(\Delta m_B t)
+{2\Im\lambda_{f_{\rm CP}}\over1+|\lambda_{f_{\rm CP}}|^2}\sin(\Delta m_B t).
\eeq
In decays with $|\lambda_{f_{\rm CP}}|=1$, (\ref{ininter}) is the only 
contributing effect:
\beq\label{intters}
a_{f_{\rm CP}}(t)=\Im\lambda_{f_{\rm CP}}\sin(\Delta m_B t).
\eeq
We often use
\beq\label{defafcp}
a_{f_{\rm CP}}\equiv{2\Im\lambda_{f_{\rm CP}}\over1+|\lambda_{f_{\rm CP}}|^2}.
\eeq

CP violation in the interference of decays with and without mixing has been 
observed for the neutral $K$ system ($\Im\ \varepsilon_K\neq0$) and for the 
neutral $B$ system ($a_{\psi K_S}\neq0$). In the latter, it is an effect of 
${\cal O}(1)$. For such cases, the contribution from CP violation in mixing is 
clearly negligible. For decays that are dominated by a single CP violating 
phase (for example, $B\to\psi K_S$ and $K_L\to\pi^0\nu\bar\nu$), so that the 
contribution from CP violation in decay is also negligible, $a_{f_{\rm CP}}$ is
cleanly interpreted in terms of purely electroweak parameters. Explicitly,
$\Im\lambda_{f_{\rm CP}}$ gives the relative phase between the $B-\bar B$ 
mixing amplitude and the relevant decay amplitudes [see  eq. (\ref{lamfCP})]:
\beq\label{intCKM}
\Im\lambda_{f_{\rm CP}}=-\eta_{f_{\rm CP}}\sin[2(\phi_B+\phi_f)].
\eeq

\subsubsection{Direct and Indirect CP Violation} 
The terms indirect CP violation and direct CP violation are commonly used in 
the literature. While various authors use these terms with different meanings, 
the most useful definition is the following:

{\bf Indirect CP violation} refers to CP violation in meson decays where the CP
violating phases can all be chosen to appear in $\Delta F=2$ (mixing) 
amplitudes.

{\bf Direct CP violation} refers to CP violation in meson decays where some CP
violating phases necessarily appear in $\Delta F=1$ (decay) amplitudes.

Examining eqs. (\ref{inmixin}) and (\ref{solveqp}), we learn that CP violation 
in mixing is a manifestation of indirect CP violation. Examining eqs. 
(\ref{indecay}) and (\ref{defAf}), we learn that CP violation in decay is a 
manifestation of direct CP violation. Examining eqs. (\ref{ininter}) and 
(\ref{deflam}), we learn that the situation concerning CP violation in the 
interference of decays with and without mixing is more subtle. For any single
measurement of $\Im\lambda_f\neq0$, the relevant CP violating phase can be 
chosen by convention to reside in the $\Delta F=2$ amplitude [$\phi_f=0$, 
$\phi_B\neq0$ in the notation of eq. (\ref{lamfCP})], and then we would call it
indirect CP violation. Consider, however, the CP asymmetries for two different 
final CP eigenstates (for the same decaying meson), $f_a$ and $f_b$. Then, a 
non-zero difference between $\Im\lambda_{f_a}$ and $\Im\lambda_{f_b}$ requires
that there exists CP violation in $\Delta F=1$ processes ($\phi_{f_a}-
\phi_{f_b}\neq0$), namely direct CP violation.

Experimentally, both direct and indirect CP violation have been established.
Below we will see that $\varepsilon_K$ signifies indirect CP violation
while $\varepsilon^\prime_K$ signifies direct CP violation. 

Theoretically, most models of CP violation (including the Standard Model)
have predicted that both types of CP violation exist. There is, however,
one class of models, that is {\it superweak models}, that predict
only indirect CP violation. The measurement of $\varepsilon^\prime_K\neq0$ has
excluded this class of models.

\section{$K$ Decays}
Measurements of CP violation have played an enormous role in particle physics.
First, the measurement of $\varepsilon_K$ in 1964 provided the first evidence 
that CP is not a symmetry of Nature. This discovery revolutionized the thinking
of particle physicists and was essential for understanding baryogenesis.
Second, the measurement in 1988 of $\varepsilon^\prime_K$ provided the first
evidence for direct CP violation and excluded the superweak scenario.
In the future, the search for CP violation in $K\to\pi\nu\bar\nu$ decays
will add significantly to our understanding of CP violation.
 
\subsection{$\varepsilon_K$ and $\varepsilon^\prime_K$}
Historically, a different language from the one used by us has been employed to
describe CP violation in $K\to\pi\pi$ and $K\to\pi\ell\nu$ decays. In this 
section we `translate' the language of $\varepsilon_K$ and 
$\varepsilon_K^\prime$ to our notations. Doing so will make it easy to 
understand which type of CP violation is related to each quantity.

The two CP violating quantities measured in neutral $K$ decays are
\beq\label{defetaij}
\eta_{00}={\langle\pi^0\pi^0|{\cal H}|K_L\rangle\over\langle\pi^0\pi^0|
{\cal H}|K_S\rangle},\ \ \ \eta_{+-}={\langle\pi^+\pi^-|{\cal H}|K_L\rangle
\over\langle\pi^+\pi^-|{\cal H}|K_S\rangle}.
\eeq
Define, for $(ij)=(00)$ or $(+-)$,
\beq\label{epsamp}
A_{ij}=\langle\pi^i\pi^j|{\cal H}|K^0\rangle,\ \ \ 
\bar A_{ij}=\langle\pi^i\pi^j|{\cal H}|\bar K^0\rangle,\ \ \ 
\lambda_{ij}=\left({q\over p}\right)_K{\bar A_{ij}\over A_{ij}}.
\eeq
Then
\beq\label{etapqA}
\eta_{00}={1-\lambda_{00}\over1+\lambda_{00}},\ \ \ 
\eta_{+-}={1-\lambda_{+-}\over1+\lambda_{+-}}.
\eeq
The $\eta_{00}$ and $\eta_{+-}$ parameters get contributions from CP violation 
in mixing ($|(q/p)|_K\neq1$) and from the interference of decays with and 
without mixing ($\Im\lambda_{ij}\neq0$) at ${\cal O}(10^{-3})$ and from CP 
violation in decay ($|\bar A_{ij}/A_{ij}|\neq1$) at ${\cal O}(10^{-6})$.

There are two isospin channels in $K\to\pi\pi$ leading to final $(2\pi)_{I=0}$ 
and $(2\pi)_{I=2}$ states:
\beqa\label{twois}
\langle\pi^0\pi^0| &=&\ \sqrt{1/3}\langle(\pi\pi)_{I=0}|-
\sqrt{2/3}\langle(\pi\pi)_{I=2}|,\no\\
\langle\pi^+\pi^-| &=&\ \sqrt{2/3}\langle(\pi\pi)_{I=0}|+
\sqrt{1/3}\langle(\pi\pi)_{I=2}|.
\eeqa
The fact that there are two strong phases allows for CP violation in decay. The
possible effects are, however, small (on top of the smallness of the relevant 
CP violating phases) because the final $I=0$ state is dominant (this is the 
$\Delta I=1/2$ rule). Define
\beq\label{defAI}
A_I=\langle(\pi\pi)_I|{\cal H}|K^0\rangle,\ \ \ 
\bar A_I=\langle(\pi\pi)_I|{\cal H}|\bar K^0\rangle,\ \ \ 
\lambda_I=\left({q\over p}\right)_K\left({\bar A_I\over A_I}\right).
\eeq
Experimentally, $|A_2/A_0|\approx1/20$. Instead of $\eta_{00}$ and $\eta_{+-}$ 
we may define two combinations, $\varepsilon_K$ and $\varepsilon^\prime_K$, in 
such a way that the possible effects of direct (indirect) CP violation are 
isolated into $\varepsilon^\prime_K$  ($\varepsilon_K$).

The experimental definition of the $\varepsilon_K$ parameter is 
\beq\label{defepsex}
\varepsilon_K\equiv{1\over3}(\eta_{00}+2\eta_{+-}).
\eeq
The experimental value is given by eq. (\ref{expeps}). To zeroth order in 
$A_2/A_0$, we have $\eta_{00}=\eta_{+-}=\varepsilon_K$. However, the specific 
combination (\ref{defepsex}) is chosen in such a way that  the following 
relation holds to {\it first} order in $A_2/A_0$:
\beq\label{defepsth}
\varepsilon_K={1-\lambda_0\over1+\lambda_0}.
\eeq
Since, by definition, only one strong channel contributes to $\lambda_0$, there
is indeed no CP violation in decay in (\ref{defepsth}). It is simple to show 
that $\Re\ \varepsilon_K\neq0$ is a manifestation of CP violation in mixing 
while $\Im\ \varepsilon_K\neq0$ is a manifestation of CP violation in the 
interference between decays with and without mixing. Since experimentally 
$\arg\varepsilon_K\approx\pi/4$, the two contributions are comparable. It is 
also clear that $\varepsilon_K\neq0$ is a manifestation of indirect CP 
violation: it could be described entirely in terms of a CP violating phase in 
the $M_{12}$ amplitude.

The experimental definition of the $\varepsilon^\prime_K$ parameter is 
\beq\label{defepspex}
\varepsilon^\prime_K\equiv{1\over3}(\eta_{+-}-\eta_{00}).
\eeq
The quantity that is actually measured in experiment is
\beq\label{epsexpe}
1-\left|{\eta_{00}\over\eta_{+-}}\right|^2=6\Re(\varepsilon^\prime/\varepsilon).
\eeq
The world average is given in eq. (\ref{aveepp}). The theoretical expression is
\beq\label{defepspth}
\varepsilon^\prime_K\approx{1\over6}(\lambda_{00}-\lambda_{+-}).
\eeq
Obviously, any type of CP violation which is independent of the final state 
does not contribute to $\varepsilon^\prime_K$. Consequently, there is no 
contribution from CP violation in mixing to (\ref{defepspth}). It is simple to 
show that $\Re\ \varepsilon^\prime_K\neq0$ is a manifestation of CP violation 
in decay while $\Im\ \varepsilon^\prime_K\neq0$ is a manifestation of CP 
violation in the interference between decays with and without mixing. Following
our explanations in the previous section, we learn that $\varepsilon^\prime_K
\neq0$ is a manifestation of direct CP violation: it requires $\phi_2-\phi_0
\neq0$ [where $\phi_I$ is the CP violating phase in the $A_I$ amplitude defined
in (\ref{defAI})].

\subsubsection{The $\varepsilon_K$ Parameter in the Standard Model}
An approximate expression for $\varepsilon_K$, that is convenient for 
calculating it, is given by
\beq\label{appeps}
\varepsilon_K={e^{i\pi/4}\over\sqrt2}{\Im M_{12}\over\Delta m_K}.
\eeq
A few points concerning this expression are worth emphasizing:

(i) Eq. (\ref{appeps}) is given in a specific phase convention, where $A_2$ is 
real. Within the SM, this is a phase convention where $V_{ud}V_{us}^*$ is real,
a condition fulfilled in both the standard parametrization of eq. 
(\ref{stapar}) and the Wolfenstein parametrization of eq. (\ref{WCKM}).

(ii) The phase of $\pi/4$ is approximate. It is determined by hadronic
parameters and therefore is independent of the electroweak model. Specifically,
\beq\label{argeps}
\arg(\varepsilon_K)\approx{\rm arctan}(-2\Delta m_K/\Delta\Gamma_K)\approx\pi/4.
\eeq

(iii) A term of order $2{\Im\ A_0\over\Re\ A_0}{\Re\ M_{12}\over\Im\ M_{12}}
\lsim0.02$ is neglected when (\ref{appeps}) is derived.

(iv) There is a large hadronic uncertainty in the calculation of $M_{12}$ 
coming from long distance contributions. There are, however, good reasons to 
believe that the long distance contributions are important in $\Re\ M_{12}$ 
(where they could be even comparable to the short distance contributions), but 
negligible in $\Im\ M_{12}$. To avoid this uncertainty, one uses 
$\Im M_{12}/\Delta m_K$ with the experimentally measured value of 
$\Delta m_K$, instead of $\Im M_{12}/2\Re\ M_{12}$ with the theoretically
calculated value of $\Re\ M_{12}$.

(v) The matrix element $\langle\bar K^0|(\bar sd)_{V-A}(\bar sd)_{V-A}|K^0
\rangle$ is yet another source of hadronic uncertainty. If both $\Im\ M_{12}$ 
and $\Re\ M_{12}$ were dominated by short distance contributions, one would 
use the ratio $\Im M_{12}/\Re\ M_{12}$ where the matrix element cancels out. 
However, as explained above, this is not the case.

Within the Standard Model, $\Im\ M_{12}$ is accounted for by box diagrams. 
We follow here the notations of ref. \cite{Buras:2001pn}, where precise
definitions, numerical values and appropriate references are given. One 
obtains:
\beq\label{epsCKM}
\varepsilon_K=e^{i\pi/4}C_\varepsilon B_K\Im(V_{ts}^*V_{td})\left\{\Re
(V_{cs}^*V_{cd})[\eta_1 S_0(x_c)-\eta_3 S_0(x_c,x_t)]-\Re(V_{ts}^*V_{td})
\eta_2S_0(x_t)\right\},
\eeq
where $C_\varepsilon\equiv{G_F^2 f_K^2 m_K m_W^2\over6\sqrt{2}\pi^2\Delta m_K}$
is a well known parameter, the $\eta_i$ are QCD correction factors, $S_0$ is a 
kinematic factor, and $B_K$ is the ratio between the matrix element of the four
quark operator and its value in the vacuum insertion approximation.

We would like to emphasize the following points:

(i) CP violation was discovered through the measurement of $\varepsilon_K$.
Hence this measurement played a significant role in the history of particle
physics. 

(ii) For a long time, $\varepsilon_K$ has been the only measured CP violating
parameter. Roughly speaking, this measurement set the value of 
$\delta_{\rm KM}$ (and, by requiring  $\delta_{\rm KM}={\cal O}(1)$, made
the KM mechanism plausible) but could not serve as a test of the KM mechanism.
(More precisely, a value of $|\varepsilon_K|\gg10^{-3}$ would have invalidated
the KM mechanism, but any value $|\varepsilon_K|\lsim10^{-3}$ was acceptable.)
It is only the combination of the new measurement of $a_{\psi K_S}$ with 
$\varepsilon_K$ that provides the first precision test of the KM mechanism.

(iii) Within the SM, the smallness of $\varepsilon_K$ is not related to 
suppression of CP violation but rather to suppression of flavor violation.
Specifically, it is the smallness of the ratio $|(V_{td}V_{ts})/(V_{ud}V_{us})|
\sim\lambda^4$ that explains $|\varepsilon_K|\sim10^{-3}$.

(iv) Until recently, the measured value of $\varepsilon_K$ provided a unique
type of information on the CKM phase. For example, the 
measurement of sign(Re $\varepsilon_K)>0$ tells us that $\eta>0$ and
excludes the lower half of the $\rho-\eta$ plane. Such information cannot be
obtained from any CP conserving observable.

(v) The $\varepsilon_K$ constraint gives hyperbolae in the $\rho-\eta$ plane. 
It is shown in fig. \ref{fig:fig1}. The measured value is consistent with all 
other CKM-related measurements and further narrows the allowed region.

(vi) The main sources of uncertainty are in the $B_K$ parameter, 
$B_K=0.85\pm0.15$, and in the $|V_{cb}|^4$ dependence.

(vii) $\varepsilon_K$ is an extremely powerful probe of new physics. Its
small value poses a problem to any model of new physics where the flavor
suppression is less efficient than the GIM mechanism \cite{Glashow:1970gm}
of the SM. For example, the construction of viable supersymmetric models is
highly constrained by the requirement that they do not give contributions
that are orders of magnitude higher than the experimental value.

\subsubsection{The $\varepsilon^\prime_K$ Parameter in the Standard Model}
Direct CP violation in $K\to\pi\pi$ decays was first measured in 1988
\cite{Burkhardt:1988yh}. Two recent measurements achieved impressive accuracy:
\beq\label{expepp}
{\varepsilon^\prime\over\varepsilon}=\cases{
(20.7\pm2.8)\times10^{-4}&KTeV \cite{ktev},\cr
(15.3\pm2.6)\times10^{-4}&NA48 \cite{nafe}.\cr}
\eeq
In combination with previous results \cite{Barr:1993rx,Gibbons:1993zq}, 
the present world average has an accuracy of order 10\% [see eq. 
(\ref{aveepp})]. 

A convenient approximate expression for $\varepsilon^\prime_K$ is given by:
\beq\label{appepsp}
\varepsilon^\prime_K={i\over\sqrt2}\left|{A_2\over A_0}\right|
e^{i(\delta_2-\delta_0)}\sin(\phi_2-\phi_0).
\eeq
We would like to emphasize a few points:

(i) The approximations used in (\ref{appepsp}) are 
$|q/p|=1$ and $|A_2/A_0|\ll1$.

(ii) The phase of $\varepsilon^\prime_K$ is determined by hadronic parameters
and is, therefore, model independent: $\arg(\varepsilon^\prime_K)=\pi/2+
\delta_2-\delta_0\approx\pi/4$. The fact that, accidentally, 
$\arg(\varepsilon_K)\approx\arg(\varepsilon^\prime_K)$, means that
\beq\label{epReAbs}\Re(\varepsilon^\prime/\varepsilon)\approx
\varepsilon^\prime/\varepsilon.
\eeq

(iii) $\Re\ \varepsilon^\prime_K\neq0$ requires $\delta_2-\delta_0\neq0$, 
consistent with our statement that it is a manifestation of CP violation in 
decay. $\varepsilon^\prime_K\neq0$ requires $\phi_2-\phi_0\neq0$, consistent 
with our statement that it is a manifestation of direct CP violation.

The calculation of $\varepsilon^\prime/\varepsilon$ within the Standard Model 
suffers from large hadronic uncertainties. A very naive order of magnitude 
estimate gives $\varepsilon^\prime/\varepsilon\sim(A_2/A_0)
(A_0^{\rm penguin}/A_0^{\rm tree})\sim10^{-3}$. Note that 
$\varepsilon^\prime/\varepsilon$ is not small because of small CP violating 
parameters but because of hadronic parameters. 

The value of the phase $\beta_K$ cancels in the ratio $\varepsilon^\prime/
\varepsilon$ and therefore did not affect our estimate. In actual calculations,
one usually uses the experimental value of $\varepsilon_K$ and the theoretical 
expression for $\varepsilon^\prime_K$. Then the expression for 
$\varepsilon^\prime/\varepsilon$ depends on the CP violating phase.
 
The detailed calculation of $\varepsilon^\prime/\varepsilon$ is complicated. 
There are several comparable contributions with differing signs. The final 
result can be written in the form (for details and references, see 
\cite{Buras:2001pn}):
\beqa\label{SMepap}
\varepsilon^\prime/\varepsilon&=&\Im(V_{td}V_{ts}^*)\left[P^{(1/2)}-
P^{(3/2)}\right]\no\\
&\approx&13\ \Im(V_{td}V_{ts}^*)\left(110\ MeV\over m_s(2\ GeV)\right)^2
\left({\Lambda^{(4)}_{\overline{\rm MS}}\over340\ MeV}\right)\no\\
&\times&\left[B_6^{(1/2)}(1-\Omega_{\eta+\eta^\prime})
-0.4B_8^{(3/2)}\left({m_t\over165\ GeV}\right)^{2.5}\right].
\eeqa
We omitted here a phase factor using the approximation $\arg(\varepsilon_K)=
\arg(\varepsilon^\prime_K)$. Here $P^{(1/2)}$, which is dominated by QCD 
penguins, gives the contributions from $\Delta I=1/2$ transitions, while 
$P^{(3/2)}$, which is dominated by electroweak penguins, gives the 
contributions from $\Delta I=3/2$ transitions. The $B_6^{(1/2)}$ and 
$B_8^{(3/2)}$ factors parameterize the corresponding hadronic matrix elements. 
The QCD penguin contributions are suppressed by isospin breaking effects 
($m_u\neq m_d$), parametrized by $\Omega_{\eta+\eta^\prime}$.
The resulting estimates vary in the range \cite{Buras:2001pn}:
\beq\label{SMepeval}
\Re(\varepsilon^\prime/\varepsilon)^{\rm SM}=(0.5-4)\times10^{-3}.
\eeq

We would like to emphasize the following points:

(i) Direct CP violation was discovered through the measurement of
$\varepsilon^\prime$.

(ii) The SM range (\ref{SMepeval}) is consistent with the experimental result 
(\ref{aveepp}).

(iii) The main sources of uncertainties lie then in the parameters $m_s$, 
$B_6^{(1/2)}$, $B_8^{(3/2)}$, $\Omega_{\eta+\eta^\prime}$ and
$\Lambda^{(4)}_{\overline{\rm MS}}$. The importance of these uncertainties is 
increased because of the cancellation between the two contributions in 
(\ref{SMepap}). 

(iv) The large hadronic uncertainties make it difficult to use the experimental
value of $\varepsilon^\prime/\varepsilon$ to constrain the CKM parameters.
Still, a negative value or a value much smaller than $10^{-4}$ would have been 
very puzzling in the context of the SM.

(v) The experimental result is useful in probing and constraining new physics.

\subsection{CP violation in $K\to\pi\nu\bar\nu$}
Observing CP violation in the rare $K\to\pi\nu\bar\nu$ decays would be 
experimentally very challenging and theoretically very rewarding. It is 
very different from the CP violation that has been observed in $K\to\pi\pi$ 
decays which is small and involves theoretical uncertainties. Similar to the CP
asymmetry in $B\to\psi K_S$, it is predicted to be large and can be cleanly 
interpreted. Furthermore, observation of the $K_L\to\pi^0\nu\bar\nu$ decay at 
the rate predicted by the Standard Model will provide further evidence that CP 
violation cannot be attributed to mixing ($\Delta S=2$) processes only, as in 
superweak models. 
 
Define 
\beq\label{defApnn}
A_{\pi^0\nu\bar\nu}=\langle\pi^0\nu\bar\nu|{\cal H}|K^0\rangle,\ \ \
\bar A_{\pi^0\nu\bar\nu}=\langle\pi^0\nu\bar\nu|{\cal H}|\bar K^0\rangle,\ \ \ 
\lambda_{\pi\nu\bar\nu}=\left({q\over p}\right)_K
{\bar A_{\pi^0\nu\bar\nu}\over A_{\pi^0\nu\bar\nu}}.
\eeq
The ratio between the neutral $K$ decay rates is then
\beq\label{KLKSpnn}
{\Gamma(K_L\to \pi^0\nu\bar\nu)\over \Gamma(K_S\to \pi^0\nu\bar\nu)}=
{1+|\lambda_{\pi\nu\bar\nu}|^2-2\Re\lambda_{\pi\nu\bar\nu}\over
1+|\lambda_{\pi\nu\bar\nu}|^2+2\Re\lambda_{\pi\nu\bar\nu}}.
\eeq
We learn that the $K_L\to \pi^0\nu\bar\nu$ decay rate vanishes in the CP limit 
($\lambda_{\pi\nu\bar\nu}=1$), as expected on general grounds 
\cite{Littenberg:1989ix}. (The CP conserving contributions were explicitly 
calculated within the Standard Model \cite{Buchalla:1998ux} and within its 
extensions with massive neutrinos \cite{Perez:1999kw} and with extra scalars 
\cite{Perez:2000qa} and found to be negligible.)

CP violation in decay and in mixing are expected to be negligibly small, of 
order $10^{-5}$ and $10^{-3}$, respectively. Consequently,
$\lambda_{\pi\nu\bar\nu}$ is, to an excellent approximation, 
a pure phase. Defining $2\theta_K$ to be the relative phase between the 
$K-\bar K$ mixing amplitude and twice the $s\to d\nu\bar\nu$ decay amplitude, 
namely $\lambda_{\pi\nu\bar\nu}=e^{2i\theta_K}$, we get from (\ref{KLKSpnn}):
\beq\label{KLKSpn}
{\Gamma(K_L\to \pi^0\nu\bar\nu)\over\Gamma(K_S\to \pi^0\nu\bar\nu)}=
\tan^2\theta_K.
\eeq
Using the isospin relation $A(K^0\to\pi^0\nu\bar\nu)/A(K^+\to\pi^+\nu\bar\nu)=
1/\sqrt2$, we get
\beq\label{defapnn}
a_{\pi\nu\bar\nu}\equiv{\Gamma(K_L\to \pi^0\nu\bar\nu)\over 
\Gamma(K^+\to \pi^+\nu\bar\nu)}=\sin^2\theta_K.
\eeq

The present experimental searches give
\beqa\label{expkpnn}
{\cal B}(K^+\to\pi^+\nu\bar\nu)&=&(1.5^{+3.4}_{-1.2})\times10^{-10}\ \ \ 
\cite{Adler:2000by},\no\\
{\cal B}(K_L\to\pi^0\nu\bar\nu)&<&5.9\times10^{-7}\ \ \
\cite{Alavi-Harati:2000hd}.
\eeqa

Eq. (\ref{defapnn}) implies that $a_{\pi\nu\bar\nu}\leq1$. This inequality is 
based on isospin considerations only. Consequently a measurement of 
$\Gamma(K^+\to \pi^+\nu\bar\nu)$ can be used to set a model independent upper 
limit on $\Gamma(K_L\to \pi^0\nu\bar\nu)$ \cite{Grossman:1997sk}:
\beq\label{miubkpnn}
{\cal B}(K_L\to\pi^0\nu\bar\nu)\ <\ 4.4\ {\cal B}(K^+\to\pi^+\nu\bar\nu).
\eeq
From the range in (\ref{expkpnn}) of the $K^+$ decay, the isospin bound on
the $K_L$ decay is ${\cal B}(K_L\to\pi^0\nu\bar\nu)<2.6\times10^{-9}$, 
which is more than two orders of magnitude below the direct bound.

Within the Standard Model, the $K\to\pi\nu\bar\nu$ decays are dominated by 
short distance $Z$-penguins and box diagrams and can be expressed in terms of 
$\rho$ and $\eta$ (see \cite{Buras:2001pn} for details and references)
\beqa\label{BrKppnnc}
{\cal B}(K^+\to\pi^+\nu\bar\nu)&=&4.11\times10^{-11}[X(x_t)]^2A^4
\left[\eta^2+\left(\rho_0-\rho\right)^2\right],\no\\
{\cal B}(K_L\to\pi^0\nu\bar\nu)&=&1.80\times10^{-10}[X(x_t)]^2A^4\eta^2.
\eeqa
Here $\rho_0=1+{P_0(X)\over A^2X(x_t)}$, and $X(x_t)$ and $P_0(X)$ represent 
the electroweak loop contributions in NLO for the top quark and for the charm 
quark, respectively. 

We would like to emphasize the following points:

(i) The $K\to\pi\nu\bar\nu$ decays are theoretically clean. The main 
theoretical uncertainty in the $K^+$ decay is related to the strong dependence 
of the charm contribution on the renormalization scale and the QCD scale, 
$P_0(X)=0.42\pm0.06$. The $K_L$ decay has hadronic uncertainties smaller
than a percent.

(ii) In the future, these decays will provide excellent $\rho-\eta$
constraints. 

(iii) Present constraints on the CKM parameters give the SM predictions
\cite{Hocker:2001xe}: 
\beqa\label{BrKsm}
{\cal B}(K^+\to\pi^+\nu\bar\nu)&=&(7.0\pm1.9)\times10^{-11},\no\\
{\cal B}(K_L\to\pi^0\nu\bar\nu)&=&(2.9\pm1.1)\times10^{-11}.
\eeqa
The experimental range for the $K^+$ decay (\ref{expkpnn}) is then consistent 
with the SM but not yet accurate enough to constrain it, while the experimental 
bound on the $K_L$ decay is still four orders of magnitude above the SM range.

(iv) The CP violations in $K\to\pi\nu\bar\nu$ and in $B\to\psi K_S$ are
strongly correlated and can provide the most stringent test of the
Kobayashi-Maskawa mechanism.

(v) The $K\to\pi\nu\bar\nu$ decays are interesting probes of CP violation 
related to new physics.

\section{$D$ Decays}
Within the Standard Model, $D-\bar D$ mixing is expected to be well below the 
experimental bound. Furthermore, effects related to CP violation in $D-\bar D$ 
mixing are expected to be negligibly small since this mixing is described to an 
excellent approximation by physics of the first two generations. An 
experimental observation of $D-\bar D$ mixing close to the present bound or, 
more strongly, of related CP violation, will then be evidence for New Physics.

To explain how $D-\bar D$ mixing is searched for and how CP violation can be
signalled, we use notations similar to those of the $B$ system. We thus use 
eq. (\ref{defqp}) to define the two mass eigenstates $|D_{1,2}\rangle$,
eq. (\ref{aveMG}) to define the average width $\Gamma$,
eq. (\ref{defxy}) to define the width and mass differences $y$ and $x$,
eq. (\ref{defAf}) to define the decay amplitudes $A_f$ and $\bar A_f$ and
eq. (\ref{deflam}) to define $\lambda_f$.

\subsection{$D\to K\pi$ and $D\to KK$ Decays}
The processes that are relevant to the most sensitive measurements at present
are the doubly-Cabibbo-suppressed $D^0\to K^+\pi^-$ decay, the 
singly-Cabibbo-suppressed $D^0\to K^+K^-$ decay, the Cabibbo-favored 
$D^0\to K^-\pi^+$ decay, and the three CP-conjugate decay processes. We follow
here the analysis presented in ref.  \cite{Bergmann:2000id}. We write down 
approximate expressions for the time-dependent decay rates that are valid for 
times $t\lsim1/\Gamma$. We take into account the experimental information that 
$x$, $y$ and $\tan\theta_c$ are small. In particular, the smallness of 
$\tan\theta_c$ implies that
\beq\label{Dexp}
|\lambda_{K^+\pi^-}^{-1}|\ll1;\ \ \ |\lambda_{K^-\pi^+}|\ll1.
\eeq
We expand each of the rates only to the order that is relevant to present
measurements:
\beqa\label{dpikt}
\Gamma[D^0(t)&\to&\ K^+\pi^-]\ =\ e^{-\Gamma t}|\bar A_{K^+\pi^-}|^2
|q/p|^2\no\\ 
&\times&\left\{|\lambda_{K^+\pi^-}^{-1}|^2+[\Re(\lambda^{-1}_{K^+\pi^-})y
+\Im(\lambda^{-1}_{K^+\pi^-})x]\Gamma t+{1\over4}(y^2+x^2)(\Gamma t)^2\right\},
\no\\
\Gamma[\overline{D^0}(t)&\to&\ K^-\pi^+]\ =\ e^{-\Gamma t}|A_{K^-\pi^+}|^2
|p/q|^2\no\\ 
&\times&\left\{|\lambda_{K^-\pi^+}|^2+[\Re(\lambda_{K^-\pi^+})y
+\Im(\lambda_{K^-\pi^+})x]\Gamma t+{1\over4}(y^2+x^2)(\Gamma t)^2\right\},\\
\label{dkkt}
\Gamma[D^0(t)&\to& K^+K^-]\ =\ e^{-\Gamma t}|A_{K^+K^-}|^2
\left\{1+[\Re(\lambda_{K^+K^-})y-\Im(\lambda_{K^+K^-})x]\Gamma t\right\},\no\\
\Gamma[\overline{D^0}(t)&\to& K^+K^-]\ =\ e^{-\Gamma t}|\bar A_{K^+K^-}|^2
\left\{1+[\Re(\lambda^{-1}_{K^+K^-})y-\Im(\lambda^{-1}_{K^+K^-})x]
\Gamma t\right\},\\
\label{dkpit}
\Gamma[D^0(t)&\to&\ K^-\pi^+]\ =\ e^{-\Gamma t}|A_{K^-\pi^+}|^2,\no\\
\Gamma[\overline{D^0}(t)&\to&\ K^+\pi^-]\ =\ 
e^{-\Gamma t}|\bar A_{K^+\pi^-}|^2.
\eeqa
Within the Standard Model, the physics of $D^0-\overline{D^0}$ mixing and of 
the tree level decays is dominated by the first two generations and,
consequently, CP violation can be safely neglected. In almost all `reasonable'
extensions of the SM, the six decay modes of eqs. (\ref{dpikt}), (\ref{dkkt}) 
and (\ref{dkpit}) are still dominated by the SM CP conserving contributions 
\cite{Bergmann:1999pm,D'Ambrosio:2001wg}. On the other hand, there could be new
short distance, possibly CP violating contributions to the mixing amplitude 
$M_{12}$. Allowing for only such effects of new physics, the picture of CP 
violation is simplified since there is no direct CP violation. The effects of 
indirect CP violation can be parameterized in the following way 
\cite{Nir:1999mg}:
\beqa\label{parlkpi}
|q/p|&=&R_m,\no\\
\lambda^{-1}_{K^+\pi^-}&=&\sqrt{R}\ R_m^{-1}\ e^{-i(\delta+\phi_D)},\no\\
\lambda_{K^-\pi^+}&=&\sqrt{R}\ R_m\ e^{-i(\delta-\phi_D)},\no\\
\lambda_{K^+K^-}&=&-R_m\ e^{i\phi_D}.
\eeqa
Here $R$ and $R_m$ are real and positive dimensionless numbers. CP violation in
mixing is related to $R_m\neq1$ while CP violation in the interference of 
decays with and without mixing is related to $\sin\phi_D\neq0$. The choice of 
phases and signs in (\ref{parlkpi}) is consistent with having $\phi_D=0$ in the 
SM and $\delta=0$ in the $SU(3)$ limit. We further define
\beqa\label{defxpyp}
x^\prime&\equiv&x\cos\delta+y\sin\delta,\no\\
y^\prime&\equiv&y\cos\delta-x\sin\delta.
\eeqa

With our assumption that there is no direct CP violation in the processes that 
we study, and using the parametrizations (\ref{parlkpi}) and (\ref{defxy}), we 
can rewrite eqs. (\ref{dpikt}), (\ref{dkkt}) and (\ref{dkpit}) as follows:
\beqa\label{dpikcpv}
\Gamma[D^0(t)&\to&\ K^+\pi^-]\ =\ e^{-\Gamma t}|A_{K^-\pi^+}|^2\no\\ 
 &\times&\left[R+\sqrt{R}R_m(y^\prime\cos\phi_D-x^\prime\sin\phi_D)\Gamma t
 +{R_m^2\over4}(y^2+x^2)(\Gamma t)^2\right],\no\\
\Gamma[\overline{D^0}(t)&\to&\ K^-\pi^+]\ =\ e^{-\Gamma t}|A_{K^-\pi^+}|^2\no\\
 &\times&\left[R+\sqrt{R}R_m^{-1}(y^\prime\cos\phi_D+x^\prime\sin\phi_D)
 \Gamma t+ {R_m^{-2}\over4}(y^2+x^2)(\Gamma t)^2\right],\\
\label{dkkcpv} 
\Gamma[D^0(t)&\to&\ K^+K^-]\ =\ e^{-\Gamma t}|A_{K^+K^-}|^2
\left[1-R_m(y\cos\phi_D-x\sin\phi_D)\Gamma t\right],\no\\
\Gamma[\overline{D^0}(t)&\to&\ K^+K^-]\ =\ e^{-\Gamma t}|A_{K^+K^-}|^2
\left[1-R_m^{-1}(y\cos\phi_D+x\sin\phi_D)\Gamma t\right],\\
\label{dkpicpv}
\Gamma[D^0(t)&\to& K^-\pi^+]\ =\ 
\Gamma[\overline{D^0}(t)\to K^+\pi^-]\ =\ e^{-\Gamma t}|A_{K^-\pi^+}|^2.
\eeqa

Of particular interest is the linear term  in eq. (\ref{dpikcpv}) which is 
potentially CP violating \cite{Blaylock:1995ay,Wolfenstein:1995kv}. It is
useful to define a CP violating quantity $a_{D\to K\pi}$ which depends on the
six measurable coefficients in (\ref{dpikcpv}):
\beqa\label{defadkpi}
a_{D\to K\pi}&=&
{\Re(\lambda_{K^-\pi^+})y+\Im(\lambda_{K^-\pi^+})x\over
2|\lambda_{K^-\pi^+}|\sqrt{x^2+y^2}}-
{\Re(\lambda^{-1}_{K^+\pi^-})y+\Im(\lambda^{-1}_{K^+\pi^-})x\over
2|\lambda^{-1}_{K^+\pi^-}|\sqrt{x^2+y^2}}\no\\
&=&{x^\prime\over\sqrt{x^2+y^2}}\sin\phi_D.
\eeqa
Observing $a_{D\to K\pi}\neq0$ would be the most convincing evidence for
new physics in $D-\bar D$ mixing.

The CLEO measurement \cite{Godang:2000yd} gives the coefficient of each of the 
three terms [$1$, $\Gamma t$ and $(\Gamma t)^2$] in the doubly-Cabibbo 
suppressed decays (\ref{dpikcpv}). Such measurements allow a fit to the 
parameters $R$, $R_m$, $x^\prime\sin\phi$, $y^\prime\cos\phi$, and $x^2+y^2$. 
Fit A of ref.  \cite{Godang:2000yd} quotes the following one sigma ranges:
\beqa\label{cleoyx}
R&=&(0.48\pm0.13)\times10^{-2},\no\\
y^\prime\cos\phi_D&=&(-2.5^{+1.4}_{-1.6})\times10^{-2},\no\\
x^\prime&=&(0.0\pm1.5)\times10^{-2},\no\\
A_m&=&0.23^{+0.63}_{-0.80},\no\\
\sin\phi_D&=&0.0\pm0.6.
\eeqa
It is assumed here that $R_m$ is not very different from one and can
be parameterized by a small parameter $A_m$,
\beq\label{defAm}
R_m^{\pm2}=1\pm A_m.
\eeq

We would like to make two further comments in this regard:

(i) The experimental results in eq. (\ref{cleoyx}) do not show any
signal of CP violation, that is, both $\sin\phi_D$ and $A_m$ are consistent
with zero. Consequently, there is no hint of new physics in the present results.

(ii) To test models of new physics, it would be useful to know the value of
the strong phase $\delta$. Such an estimate is a difficult theoretical task
\cite{Chau:1994ec,Browder:1996ay,Falk:1999ts} but experimental data on
related channels would be useful \cite{Golowich:2001hb,Gronau:2001nr}.

As concerns the singly-Cabibbo suppressed modes (\ref{dkkcpv}), several
experiments fit the time dependent decay rates to pure exponentials. We define 
$\hat\Gamma$ to be the parameter that is extracted in this way. More 
explicitly, for a time dependent decay rate with $\Gamma[D(t)\to f]\propto 
e^{-\Gamma t}(1-z\Gamma t+\cdots)$, where $|z|\ll1$, we have 
$\hat\Gamma(D\to f)=\Gamma(1+z)$.
The above equations imply the following relations:
\beqa\label{afitexp}
\hat\Gamma(D^0\to K^+K^-)&=& \Gamma\ [1+R_m(y\cos\phi_D-x\sin\phi_D)],\no\\
\hat\Gamma(\overline{D^0}\to K^+K^-)&=&
\Gamma\ [1+R_m^{-1}(y\cos\phi_D+x\sin\phi_D)],\no\\
\hat\Gamma(D^0\to K^-\pi^+)&=&\hat\Gamma(\overline{D^0}\to K^+\pi^-)=\Gamma.
\eeqa
Note that deviations of $\hat\Gamma(D\to K^+K^-)$ from $\Gamma$ do not require 
that $y\neq0$. They can be accounted for by $x\neq0$ and $\sin\phi_D\neq0$, but 
then they have a different sign in the $D^0$ and $\overline{D^0}$ decays. 
Combining the two $D\to K^+K^-$ modes, one obtains the CP conserving 
quantity $y_{\rm CP}$:
\beqa\label{FkkGcpv}
y_{\rm CP}&\equiv&\ {\hat\Gamma(D\to K^+K^-)\over
\hat\Gamma(D^0\to K^-\pi^+)}-1\no\\
&=&\ y\cos\phi_D-{A_m\over2}x\sin\phi,
\eeqa
where we made the approximations of zero production asymmetry and small $A_m$
\cite{Bergmann:2000id}.
The one sigma ranges measured by various experiments are given by
\beq\label{ycpexp}
y_{\rm CP}=\cases{(3.4\pm1.6)\times10^{-2}&FOCUS \cite{Link:2000cu}\cr
(0.8\pm3.1)\times10^{-2}&E791 \cite{Aitala:1999dt}\cr
(-1.1\pm2.9)\times10^{-2}&CLEO \cite{Smith:2001ej}\cr
(0.5\pm1.3)\times10^{-2}&BELLE \cite{Yabsley}\cr}
\eeq
giving a world average of
\beq\label{ycpnum}
y_{\rm CP}=(1.3\pm0.9)\times10^{-2}.
\eeq

Finally, we note that direct CP violation has been searched for in the
Cabibbo-favored \cite{Bartelt:1995vr}, singly-Cabibbo-suppressed 
\cite{Aitala:1998ff,Link:2000aw,Bonvicini:2001qm}
and doubly-Cabibbo-suppressed \cite{Godang:2000yd} decays with all
results consistent with zero.

We conclude that at present there is no evidence for mixing and certainly
not for CP violation in the neutral $D$ system. These results are consistent
with the SM and constrain models of new physics. If evidence is found in the
future, the $D\to K\pi$ and $D\to KK$ decays will provide rich enough 
information that we will be able to point out the origin of the signals in much
detail.

\section{$B$ Decays}
\subsection{CP Violation in Mixing}
CP violation in mixing is related to a non-zero value for the following
quantity [see eq. (\ref{solveqpC})]:
\beq\label{solveqpD}
1-\left|{q\over p}\right|\simeq
{1\over2}\Im\left({\Gamma_{12}\over M_{12}}\right).
\eeq
The effect can be isolated by measuring the asymmetry in semileptonic
decays [see eq. (\ref{mixter})]:
\beq\label{mixterB}
a_{\rm SL}\simeq 2(1-|q/p|)\simeq \Im(\Gamma_{12}/M_{12}).
\eeq
This has been searched for in several experiments, with sensitivity at the
level of $10^{-2}$:
\beq\label{aslexp}
a_{\rm SL}=\cases{(1.4\pm4.2)\times10^{-2}&CLEO \cite{Jaffe:2001hz}\cr
(0.4\pm5.7)\times10^{-2}&OPAL \cite{Abbiendi:2000av}\cr
(-1.2\pm2.8)\times10^{-2}&ALEPH \cite{Barate:2001uk}\cr
(0.48\pm1.85)\times10^{-2}&BABAR \cite{Aubert:2001xc}\cr}
\eeq
giving a world average of
\beq\label{aveasl}
a_{\rm SL}=(0.2\pm1.4)\times10^{-2}.
\eeq

As explained above, in the $B_d$ system we expect model 
independently that $|\Gamma_{12}/M_{12}|\ll1$. Within any given model we can 
actually calculate the two quantities from quark diagrams. Within the SM, 
$M_{12}$ is given by box diagrams. For both the $B_d$ and $B_s$ systems, the 
long distance contributions are expected to be negligible and the calculation 
of these diagrams with a high loop momentum is a very good approximation. 
$\Gamma_{12}$ is calculated from a cut of box diagrams \cite{Bigi:1987in}. 
Since the cut of a diagram always involves on-shell particles and thus long 
distance physics, the calculation is, at best, a reasonable 
approximation to $\Gamma_{12}$. (For $\Gamma_{12}(B_s)$ it has been shown that 
local quark-hadron duality holds exactly in the simultaneous limit of small
velocity and large number of colors. We thus expect an uncertainty of ${\cal O}
(1/N_C)\sim30\%$ \cite{Aleksan:1993qp,Beneke:1999sy}. For $\Gamma_{12}(B_d)$
the small velocity limit is not as good an approximation but an uncertainty of 
order 50\% still seems a reasonable estimate \cite{Wolfenstein:1998qz}.)

Within the Standard Model, $M_{12}$ is dominated by top-mediated box diagrams
(see \cite{Buras:2001pn} for details and references):
\beq\label{MontwB}
M_{12}={G_F^2\over12\pi^2}m_Bm_W^2\eta_BB_Bf_B^2(V_{tb}V_{td}^*)^2S_0(x_t),
\eeq
where $S_0(x_t)$ is a kinematic factor, $\eta_B$ is a QCD correction, and 
$B_Bf_B^2$ parametrizes the hadronic matrix element.
For $\Gamma_{12}$, we have \cite{Hagelin:1981zk,Buras:1984pq,Beneke:1996gn}
\beqa\label{GontwB}
\Gamma_{12}\ &=&\ -{G_F^2\over24\pi}m_Bm_b^2B_Bf_B^2
(V_{tb}V_{td}^*)^2\no\\ 
&\times&\left[{5\over3}{m_B^2\over(m_b+m_d)^2}{B_S\over B_B}(K_2-K_1)
+{4\over3}(2K_1+K_2)+8(K_1+K_2){m_c^2\over m_b^2}{V_{cb}V_{cd}^*\over 
V_{tb}V_{td}^*}\right],
\eeqa
where $K_1=-0.39$ and $K_2=1.25$ \cite{Beneke:1996gn} are combinations of 
Wilson coefficients and $B_S$ parametrizes the $(S-P)^2$ matrix element. New 
physics usually takes place at a high energy scale and is relevant to the short 
distance part only. Therefore, the SM estimate in eq. (\ref{GontwB}) remains
valid model independently. Combining (\ref{MontwB}) and (\ref{GontwB}), 
we learn that $|\Gamma_{12}/M_{12}|={\cal O}(m_b^2/m_t^2)$, which confirms our 
model independent order of magnitude estimate, $|\Gamma_{12}/M_{12}|
\lsim10^{-2}$. As concerns the imaginary part of this ratio, we have
\beq\label{aSLSM}
a_{\rm SL}=\Im{\Gamma_{12}\over M_{12}}\approx-1.4\times10^{-3}{\eta\over
(1-\rho)^2+\eta^2}.
\eeq
The suppression by a factor of ${\cal O}(10)$ of $a_{\rm SL}$ compared to 
$|\Gamma_{12}/M_{12}|$ comes from the fact that the leading contribution to 
$\Gamma_{12}$ has the same phase as $M_{12}$. Consequently, $a_{\rm SL}=
{\cal O}(m_c^2/m_t^2)$. The CKM factor does not give any further significant
suppression, $\Im{V_{cb}V_{cd}^*\over V_{tb}V_{td}^*}={\cal O}(1)$. In 
contrast, for the $B_s$ system, where the same expressions holds except that 
$V_{cd}/V_{td}$ is replaced by $V_{cs}/V_{ts}$, there is an additional 
CKM suppression from $\Im{V_{cb}V_{cs}^*\over V_{tb}V_{ts}^*}={\cal O}
(\lambda^2)$.

In the SM and in most of its reasonable extensions, both $\Gamma_{12}$ and
$b\to c\bar cs$ transitions are dominated by SM tree level decays. Consequently,
new physics affects $a_{\rm SL}$ and $a_{\psi K_S}$ only through its 
contributions to $M_{12}$. This leads to interesting correlations between
$a_{\rm SL}$ and $a_{\psi K_S}$ that can be used to probe flavor parameters
\cite{Randall:1999te,Cahn:1999gx}. Conversely, one can use the measured value
of $a_{\psi K_S}$ to give model independent predictions for $a_{\rm SL}$
\cite{Barenboim:1999in,Eyal:1999ii}.

\subsection{Penguin Pollution}
In purely hadronic $B$ decays, CP violation in decay and in the interference of
decays with and without mixing is $\geq {\cal O}(10^{-2})$. We can therefore
safely neglect CP violation in mixing in the following discussion and use
\beq
\label{qpSM}
{q\over p}={V_{tb}^*V_{td}\over V_{tb}V_{td}^*}\omega_B.
\eeq
(From here on we omit the convention-dependent quark phases $\omega_q$
defined in eq. (\ref{CPofq}). Our final expressions for physical
quantities are of course unaffected by such omission.)

A crucial aspect of our discussion is the number of relevant weak phases
for a given decay process:

(i) If there is a single weak phase that dominates the decay, CP violation
in decay will be small and difficult to observe. On the other hand, CP 
asymmetries in neutral $B$ decays into final CP eigenstates are subject to 
clean theoretical interpretation: we will either have precise measurements of 
CKM parameters or be provided with unambiguous evidence for new physics.

(ii) If there are two (or more) weak phases that contribute comparably,
hadronic uncertainties will appear in the theoretical interpretation of
CP violation in the interference of decays with and without mixing. On the
other hand, if there are also large strong phase differences, CP violation
in decay can be observed in the corresponding charged and neutral $B$ decays.

In many cases of interest, different weak phases are carried by tree and 
penguin contributions. The difficulties arising from hadronic uncertainties 
related to comparable tree and penguin contributions became known as
``penguin pollution."

To illustrate the problem, we will consider two relevant CP asymmetries. First,
the CP asymmetry in $B\to\psi K_S$ is an example of a case where the penguin 
pollution is negligibly small and a theoretically very clean interpretation of 
the experimental measurement is possible. Second, the CP asymmetry in 
$B\to\pi\pi$ is an example of a case where penguin pollution cannot be a-priori
ignored. We also list various ways in which the problem might be overcome.
 
\subsection{$B\to\psi K_S$}
The first evidence for CP violation outside $K$ decays has been provided by the
recent BaBar and Belle measurements of the CP asymmetry in $B\to\psi K_S$,
\beq\label{expapk}
a_{\psi K_S}=\cases{0.59\pm0.15&Babar \cite{Aubert:2001nu}\cr
0.99\pm0.15&Belle \cite{Abe:2001xe}\cr}
\eeq
These results in combination with previous ones 
\cite{Ackerstaff:1998xz,Affolder:2000gg,Barate:2000tf} 
give the world average quoted in eq. (\ref{aveapk}). The process $B\to\psi K_S$
is one where the penguin contribution is harmless and the CP asymmetry is 
subject to an impressingly clean theoretical interpretation. 

The decay is mediated by the quark transition $\bar b\to\bar cc\bar s$. It gets 
contributions from a tree level diagram and from penguin diagrams with
intermediate $u$, $c$ and $t$ quarks. Using the unitarity relation
(\ref{Unitsb}), we can write the various contributions in terms of two
CKM combinations: 
\beq\label{ccstype}
A(\bar b\to\bar cc\bar s)=(T_{c\bar cs}+P^c_s-P^t_s)V_{cb}^*V_{cs}
+(P^u_s-P^t_s)V_{ub}^*V_{us}.
\eeq

The second term is suppressed by two factors. First, there is the ratio
between penguin and tree contributions,
\beqa\label{pengtree}
r_{PT}^{\psi K}&\equiv& {P_{\psi K}\over T_{\psi K}}\equiv
{P^u_s-P^t_s\over T_{c\bar cs}+P^c_s-P^t_s}\no\\
&\approx&\left[{\alpha_s\over12\pi}\ln{m_t^2\over m_b^2}\right] {
\langle\psi K_S|\bar b\gamma^\mu T^a s\bar c\gamma_\mu T^a c|B^0\rangle\over
\langle\pi^+\pi^-|\bar b_L\gamma^\mu c_L\bar c_L\gamma_\mu s_L|B^0\rangle}.
\eeqa
The term is brackets is $ {\cal O}(0.03)$ but the ratio of matrix elements 
may partially compensate for this suppresion. Second, there is the ratio of 
CKM elements, $|(V_{ub}^*V_{us})/(V_{cb}^*V_{cs})|\sim\lambda^2$. We conclude 
that the second term is suppressed by $r_{PT}^{\psi K}\lambda^2\lsim10^{-2}$ 
and we can safely neglect $P_{\psi K_S}$. Thus the $B\to\psi K$ decay is 
dominated by a single weak phase, that is, $\arg(V_{cb}^*V_{cs})$.

Nelecting $P_{\psi K_S}$ means that, to a very good approximation, we
have $|\lambda_{\psi K_S}|=1$,
\beq\label{acppsik}
a_{\psi K_S}=\Im\lambda_{\psi K_S},
\eeq
and that the experimental value of $a_{\psi K_S}$ [eq. (\ref{aveapk})]
can be cleanly interpreted in terms of a CP violating phase.

A new ingredient in the analysis is the effect of $K-\bar K$ mixing. For decays
with a single $K_S$ in the final state, $K-\bar K$ mixing is essential because 
$B^0\to K^0$ and $\bar B^0\to\bar K^0$, and interference is possible only due 
to $K-\bar K$ mixing. This adds a factor of
\beq\label{qpK}
\left({p\over q}\right)_K={V_{cs}V_{cd}^*\over V_{cs}^*V_{cd}}\omega_K^*
\eeq
into $(\bar A/A)$:
\beq\label{ApsiK}
{\bar A_{\psi K_S}\over A_{\psi K_S}}= \eta_{\psi K_S}
\left({V_{cb}V_{cs}^*\over V_{cb}^*V_{cs}}\right)
\left({V_{cs}V_{cd}^*\over V_{cs}^*V_{cd}}\right)\omega_B^*.
\eeq
The CP-eigenvalue of the state is $\eta_{\psi K_S} = -1$. Combining eqs. 
(\ref{qpSM}) and (\ref{ApsiK}), we find
\beq\label{lampsiKa}
\lambda(B\to\psi K_S)=-\left({V_{tb}^*V_{td}\over
V_{tb}V_{td}^*}\right)\left({V_{cb}V_{cs}^*\over V_{cb}^*V_{cs}}\right)
\left({V_{cd}^*V_{cs}\over V_{cd}V_{cs}^*}\right),
\eeq
which leads to
\beq\label{lampsiK}
a_{\psi K_S}=\sin2\beta.
\eeq

What we have learnt above is that eq. (\ref{lampsiK}) is clean of hadronic 
uncertainties to ${\cal O}(r_{PT}^{\psi K}\lambda^2)\lsim10^{-2}$. This means 
that the measuremnet of $a_{\psi K_S}$ can give the theoretically cleanest 
determination of a CKM parameter, even cleaner than the determination of 
$|V_{us}|$ from $K\to\pi\ell\nu$. [If BR($K_L\to\pi\nu\bar\nu$) is measured, 
it will give a comparably clean determination of $\eta$.]

Taking into account all the constraints on the CKM parameters {\it except}
for the $a_{\psi K_S}$ measurements, the SM prediction is \cite{Hocker:2001xe}
\beq\label{smapk}
\sin2\beta=0.68\pm0.18,
\eeq
consistent with the experimental result (\ref{aveapk}). This consistency
has important implications. In particular,

(i) The Kobayashi-Maskawa mechanism has successfully passed its first
precision test;

(ii) Models of approximate CP which, by definition, predict 
$|a_{\psi K_S}|\ll1$, are excluded.

\subsection{$B\to\pi\pi$}
The CP asymmetry in the $B\to\pi^+\pi^-$ mode has the form
\beq\label{acppipi}
a_{\pi\pi}(t)=-{1-|\lambda_{\pi\pi}|^2\over1+|\lambda_{\pi\pi}|^2}\cos\Delta mt
+{2\Im\lambda_{\pi\pi}\over1+|\lambda_{\pi\pi}|^2}\sin\Delta mt.
\eeq
Recently, the BaBar collaboration presented the first constraints on this
asymmetry \cite{Aubert:2001qj}:
\beqa\label{expapipi}
{2\Im\lambda_{\pi\pi}\over1+|\lambda_{\pi\pi}|^2}&=&0.03^{+0.54}_{-0.57},\no\\
{1-|\lambda_{\pi\pi}|^2\over1+|\lambda_{\pi\pi}|^2}&=&-0.25^{+0.47}_{-0.49}.
\eeqa
The results are not yet precise enough to give useful constraints. But we 
discuss this mode to show how penguin pollution arises and how it complicates
the analysis.

The decay is mediated by the quark transition $\bar b\to\bar uu\bar d$. It gets 
contributions from a tree level diagram and from penguin diagrams with
intermediate $u$, $c$ and $t$ quarks. Using the unitarity relation
(\ref{Unitdb}), we can write the various contributions in terms of two
CKM combinations: 
\beq\label{uudtype}
A(\bar b\to\bar uu\bar d)=(T_{u\bar ud}+P^u_d-P^c_d)V_{ub}^*V_{ud}
+(P^t_d-P^c_d)V_{tb}^*V_{td}.
\eeq

The ratio between the magnitudes of the second and first terms is given by
$r_{PT}^{\pi\pi}\left|{V_{tb}^*V_{td}\over V_{ub}^*V_{ud}}\right|$. Since both 
$|V_{ub}V_{ud}^*|$ and $|V_{tb}^*V_{td}|$ are of ${\cal O}(\lambda^3)$, 
the second term is suppressed only by the factor $r_{PT}^{\pi\pi}$, where
\beq\label{pentre}
r_{PT}^{\pi\pi}\equiv{P_{\pi\pi}\over T_{\pi\pi}}\equiv
{P^t_d-P^c_d\over T_{u\bar ud}+P^u_d-P^c_d}.
\eeq
One may make a rough estimate of $|P_{\pi\pi}/T_{\pi\pi}|$ from the decay
$B\to K\pi$, which can be parameterized as follows:
\beq\label{ampbkpi}
A(B^0\to K^+\pi^-)=T_{K\pi}V_{ub}^*V_{us}+P_{K\pi}V_{tb}V_{ts}^*.
\eeq
In this case $|P_{K\pi}/T_{K\pi}|={\cal O}(r_{PT}^{K\pi}/\lambda^2)$. If QCD 
enhances the penguin contribution to $B\to\pi\pi$ by a significant amount, 
that is, $r_{PT}\gg\lambda^2$, then $B\to K\pi$ would be dominated by the 
penguin process. Let us provisionally make the following assumptions: (i) 
flavor SU(3) symmetry in the QCD matrix elements; (ii) electroweak penguins and
``color suppressed" processes are negligible; (iii) penguins dominate 
$B\to K\pi$, so $T_{K\pi}$ may be ignored in BR$(B^0\to K^+\pi^-)$; (iv) 
penguins make a small enough contribution to $B\to\pi\pi$ that $P_{\pi\pi}$ may
be ignored in BR$(B^0\to \pi^+\pi^-)$. Then
\beq\label{kpipipi}
\left|{P_{\pi\pi}\over T_{\pi\pi}}\right|=
\left|{P_{\pi\pi}\over P_{K\pi}}\right|\left|{P_{K\pi}\over T_{\pi\pi}}\right|=
\left|{V_{ub}V_{ud}\over V_{ts}V_{tb}}\right|
\sqrt{{\rm BR}(B^0\to K^+\pi^-)\over{\rm BR}(B^0\to \pi^+\pi^-)}.
\eeq
Recent measurements \cite{Cronin-Hennessy:2000hw,Abe:2001nq,Aubert:2001hs} give
world averages ${\rm BR}(B^0\to \pi^+\pi^-)=(4.4\pm0.9)\times10^{-6}$ and 
${\rm BR}(B^0\to K^+\pi^-)=(17.3\pm1.5)\times10^{-6}$. We thus find 
${\rm BR}(B^0\to K^+\pi^-)/{\rm BR}(B^0\to \pi^+\pi^-)\approx3.9$ and
obtain the rough estimate
\beq\label{kppp}
\left|r_{PT}^{\pi\pi}\right|\sim0.2-0.3.
\eeq
It is clear that penguin effects are unlikely to  be negligible in 
$B\to\pi\pi$. 

Combining eqs. (\ref{qpSM}) and (\ref{uudtype}), we find
\beq\label{lampipia}
\lambda(B\to\pi\pi)=\left({V_{tb}^*V_{td}\over V_{tb}V_{td}^*}\right)
\left({V_{ub}V_{ud}^*\over V_{ub}^*V_{ud}}\right)
\left[{1+r_{PT}^{\pi\pi}(V_{tb}V_{td}^*)/(V_{ub}V_{ud}^*)\over
1+r_{PT}^{\pi\pi}(V_{tb}^*V_{td})/(V_{ub}^*V_{ud})}\right].
\eeq

If the last factor could be approximated by unity, that is, $r_{PT}^{\pi\pi}=0$,
we would obtain $|\lambda_{\pi\pi}|=1$ and
\beq\label{lampipi}
a_{\pi\pi}=\sin2\alpha.
\eeq
This approximation is however unjustified. To get an idea of the effects of 
$P_{\pi\pi}\neq0$, we give the leading corrections due to a small $|r_{PT}|$:
\beqa\label{alphaeff}
|\lambda_{\pi\pi}|&=&1-2(R_t/R_u)\Im(r_{PT}^{\pi\pi})\sin\alpha,\no\\
\Im\lambda_{\pi\pi}/|\lambda_{\pi\pi}|&=&\sin2\alpha+2(R_t/R_u)
\Re(r_{PT}^{\pi\pi})\cos2\alpha\sin\alpha.
\eeqa
(For a more detailed discussion, see \cite{Charles:1999qx}.)
Note that if strong phases can be neglected, $r_{PT}$ is real and
$|\lambda_{\pi\pi}|=1$ would be a good approximation. But it is not clear 
whether the strong phases are indeed small. In any case, one 
needs to know $r_{PT}^{\pi\pi}$ to extract $\alpha$ from $a_{\pi\pi}(t)$.
This is the problem of the penguin pollution.

A variety of solutions to this problem have been proposed, falling roughly
into two classes. The first type of approach is to convert the estimate given 
above into an actual measurement of $|P_{K\pi}|$. (The list of papers on this
subject is long. Early works include 
\cite{Nir:1991cu,Silva:1994sv,Hernandez:1994rh}. For a much more comprehensive 
list of references, see \cite{Charles:1999qx}.) Once $|P_{K\pi}|$ is known,
flavor $SU(3)$ is used to relate $|P_{K\pi}|$ to $|P_{\pi\pi}|$. One must then
include a number of additional effects:

(i) Electroweak penguins. The effects are calculable \cite{Neubert:1998pt}.

(ii) Color suppressed and rescattering processes. These must be bounded or
estimated using data and some further assumptions.

(iii) $SU(3)$ corrections. Some, such as $f_K/f_\pi$, can be included, but 
$SU(3)$ corrections generally remain a source of irreducible uncertainty.

The second type of approach is to exploit the fact that the penguin 
contribution to $P_{\pi\pi}$ is pure $\Delta I={1\over2}$, while the tree
contribution to $T_{\pi\pi}$ contains a piece which is $\Delta I={3\over2}$.
(This is not true of the electroweak penguins \cite{Fleischer:1997bv}, but
these are expected to be small.) Isospin symmetry allows one to form a
relation among the amplitudes $B^0\to\pi^+\pi^-$, $B^0\to\pi^0\pi^0$, and
$B^+\to\pi^+\pi^0$,
\beq\label{isopipi}
{1\over\sqrt2}A(B^0\to\pi^+\pi^-)+A(B^0\to\pi^0\pi^0)=A(B^+\to\pi^+\pi^0).
\eeq
There is also a relation for the charge conjugate processes. A simple
geometric construction then allows one to disentangle the unpolluted
$\Delta I={3\over2}$ amplitudes, from which $\sin2\alpha$ may be extracted
cleanly \cite{Gronau:1990ka}.

The key experimental difficulty is that one must measure accurately the
flavor-tagged rate for $B^0\to\pi^0\pi^0$. Since the final state consists of
only four photons, and the branching fraction is expected to be of
${\cal O}(10^{-6})$, this is very hard. It has been noted that an upper bound 
on this rate, if sufficiently strong, would also allow one to bound
$P_{\pi\pi}$ usefully \cite{Grossman:1998jr,Charles:1999qx,Gronau:2001ff}.

An alternative is to perform an isospin analysis of the process $B^0\to\rho\pi
\to\pi^+\pi^-\pi^0$ 
\cite{Lipkin:1991st,Gronau:1991dq,Snyder:1993mx,Quinn:2000by}. Here one must 
study the time-dependent asymmetry over the entire Dalitz plot, probing 
variously the intermediate states $\rho^\pm\pi^\mp$ and $\rho^0\pi^0$. The 
advantage here is that the final states with two $\pi^0$'s need not be 
considered. On the other hand, thousands of cleanly reconstructed events would 
be needed.

Finally, one might attempt to calculate the penguin matrix elements.
Model-dependent analyses are not really adequate for this purpose, since the
goal is the extraction of fundamental parameters. Precise calculations of such
matrix elements from lattice QCD are far in the future, given the large
energies of the $\pi$'s and the need for an unquenched treatment. Recently,
a new QCD-based analysis of the $B\to\pi\pi$ matrix elements has been
proposed \cite{Beneke:1999br,Beneke:2000ry,Beneke:2001ev,Keum:2001ph}.
For details, see \cite{susbuc}.

\section{CP Violation in Supersymmetry}
\subsection{CP Violation as a Probe of New Physics}
We have argued that the Standard Model picture of CP violation is rather unique
and highly predictive. We have also stated that reasonable extensions of the 
Standard Model have a very different picture of CP violation. Experimental 
results are now starting to decide between the various possibilities. Our 
discussion of CP violation in the presence of new physics is aimed to 
demonstrate that, indeed, models of new physics can significantly modify the 
Standard Model predictions and that the near future measurements will therefore
have a strong impact on the theoretical understanding of CP violation.

To understand how the Standard Model predictions could be modified by New 
Physics, we focus on CP violation in the interference between decays with and 
without mixing. As explained above, this type of CP violation may give, due to 
its theoretical cleanliness, unambiguous evidence for New Physics most easily.
We now demonstrate what type of questions can be answered when many such 
observables are measured.

Consider $a_{\psi K_S}$, the CP asymmetry in $B\rightarrow\psi K_S$. This 
measurement will cleanly determine the relative phase between the $B-\bar B$ 
mixing amplitude and the $b\to c\bar cs$ decay amplitude ($\sin2\beta$ in the 
SM). The $b\to c\bar cs$ decay has Standard Model tree contributions and
therefore is very unlikely to be significantly affected by new physics. On the 
other hand, the mixing amplitude can be easily modified by new physics. We 
parametrize such a modification by a phase $\theta_d$:
\beq\label{derthed}
2\theta_d=\arg(M_{12}/M_{12}^{\rm SM}).
\eeq
Then
\beq\label{apksNP}
a_{\psi K_S}=\sin[2(\beta+\theta_d)].
\eeq
Examining whether $a_{\psi K_S}$ fits the SM prediction, that is, whether
$\theta_d\neq0$, we can answer the following question 
(see {\it e.g.} \cite{Grossman:1997dd}):

(i) {\it Is there new physics in $B-\bar B$ mixing}?

It is interesting to note that already now the measured value of $a_{\psi K_S}$
(\ref{aveapk}), which is consistent with the SM range, excludes many models
that require a modification of CP violation in $B-\bar B$ mixing due to
new physics. Among these are various models of soft CP violation
\cite{Georgi:1999wt,Frampton:2001dn} aimed to solve the strong CP problem,
models of geometric CP violation due to extra dimensions \cite{Chang:2001uk},
models of spontaneous CP violation in the left-right symmetric framework
\cite{Ball:2000mb,Bergmann:2001pm}, and several models that aim to solve the
supersymmetric CP problems \cite{Pomarol:1993uu,Babu:1994ai,Eyal:1998bk}.
 
Next consider $a_{\phi K_S}$, the CP asymmetry in $B\to\phi K_S$. This 
measurement will cleanly determine the relative phase between the $B-\bar B$ 
mixing amplitude and the $b\to s\bar ss$ decay amplitude ($\sin2\beta$ in the 
SM). The $b\to s\bar ss$ decay has only Standard Model penguin contributions 
and therefore is sensitive to new physics. We parametrize the modification of 
the decay amplitude by a phase $\theta_A$ \cite{Grossman:1997ke}:
\beq\label{dertheA}
\theta_A=\arg(\bar A_{\phi K_S}/\bar A_{\phi K_S}^{\rm SM}).
\eeq
Then
\beq\label{aphksNP}
a_{\phi K_S}=\sin[2(\beta+\theta_d+\theta_A)].
\eeq
Comparing $a_{\phi K_S}$ to $a_{\psi K_S}$, that is, examining whether 
$\theta_A\neq0$, we can answer the following  question:

(ii) {\it Is the new physics related to $\Delta B=1$ processes? $\Delta B=2$? 
both?}
 
Consider $a_{\pi\nu\bar\nu}$, the CP violating ratio of $K\to\pi\nu\bar\nu$ 
decays, defined in (\ref{defapnn}). This measurement will cleanly determine the
relative phase between the $K-\bar K$ mixing amplitude and the 
$s\to d\nu\bar\nu$ decay amplitude (of order $\sin^2\beta$ in the SM). The
experimentally measured small value of $\varepsilon_K$ requires that the phase 
of the $K-\bar K$ mixing amplitude is not modified from the Standard Model 
prediction. (More precisely, it requires that the phase of the mixing amplitude
is very close to twice the phase of the $s\to d\bar uu$ decay amplitude 
\cite{Nir:1990hj}.) On the other hand, the decay, which in the SM is a loop 
process with small mixing angles, can be easily modified by new physics.
Examining whether the SM correlation between $a_{\pi\nu\bar\nu}$ and 
$a_{\psi K_S}$ is fulfilled, we can answer the following  question:

(iii) {\it Is the new physics related to the third generation? to all 
generations?}

Consider $a_{D\to K\pi}$, the CP violating quantity in $D\to K^\pm\pi^\mp$ 
decays defined in (\ref{defadkpi}). It depends on $\phi_D$, the relative phase 
between the $D-\bar D$ mixing amplitude and the $c\to d\bar su$ and 
$c\to s\bar du$ decay amplitudes. Within the Standard Model, the two decay 
channels are tree level. It is unlikely that they are affected by new physics. 
On the other hand, the  mixing amplitude can be easily modified by new physics.
Examining whether $a_{D\to K\pi}=0$, that is, whether $\phi_D$ (and/or 
$\theta_d$)$\neq0$, we can answer the following question:

(iv) {\it Is the new physics related to the down sector? the up sector? both?}

Consider $d_N$, the electric dipole moment of the neutron. We did not discuss 
this quantity so far because, unlike CP violation in meson decays, flavor 
changing couplings are not necessary for $d_N$. In other words, the CP 
violation that induces $d_N$ is {\it flavor diagonal}. It does in general get 
contributions from flavor changing physics, but it could be induced by sectors 
that are flavor blind. Within the SM (and ignoring $\theta_{\rm QCD}$), the 
contribution from $\delta_{\rm KM}$ arises at the three loop level and is at 
least six orders of magnitude below the experimental bound (\ref{dnexp}).
If the bound is further improved (or a signal observed), we can answer
the following question:

(v) {\it Are the new sources of CP violation flavor changing? flavor diagonal? 
both?}

It is no wonder then that with such rich information, flavor and CP violation 
provide an excellent probe of new physics. We will now demostrate this 
situation more concretely by discussing CP violation in supersymmetry.

\subsection{The Supersymmetric Framework}
Supersymmetry solves the fine-tuning problem of the Standard Model and has
many other virtues. But at the same time, it leads to new problems:
baryon number violation, lepton number violation, large flavor changing
neutral current processes and large CP violation. The first two problems
can be solved by imposing $R$-parity on supersymmetric models. There is no
such simple, symmetry-related solution to the problems of flavor and CP 
violation. Instead, suppression of the relevant couplings can be achieved
by demanding very constrained strcutures of the soft supersymmetry breaking
terms. There are two important questions here: First, can theories of
dynamical supersymmetry breaking naturally induce such structures? (For an 
excellent review of dynamical supersymmetry breaking, see \cite{Shadmi:2000jy}.)
Second, can measurements of flavor changing and/or CP violating processes
shed light on the structure of the soft supersymmetry breaking terms?
Since the answer to both questions is in the affirmative, we conclude that
flavor changing neutral current processes and, in particular, CP violating
observables will provide clues to the crucial question of how supersymmetry
breaks.

\subsubsection{CP Violating Parameters}
A generic supersymmetric extension of the Standard Model contains a host of new
flavor and CP violating parameters. (For a review of CP violation in 
supersymmetry see \cite{Grossman:1997pa,Dine:2001ne}.) It is an amusing 
exercise to count the number of parameters \cite{Haber:1998if}. The 
supersymmetric part of the Lagrangian depends, in addition to the three gauge 
couplings of $G_{\rm SM}$, on the parameters of the superpotential $W$:
\beq\label{superp}
W=\sum_{i,j}\left(Y^u_{ij}H_u Q_{Li} \overline{U}_{Lj}
+Y^d_{ij}H_d Q_{Li} \overline{D}_{Lj}
+Y^\ell_{ij}H_d L_{Li} \overline{E}_{Lj}\right)+\mu H_u H_d.
\eeq
In addition, we have to add soft supersymmetry breaking terms:
\beqa\label{Lsoft}
{\cal L}_{\rm soft}=&-&\left(A^u_{ij}H_u\tilde Q_{Li}\tilde{\overline{U}}_{Lj}
+A^d_{ij}H_d\tilde Q_{Li}\tilde{\overline{D}}_{Lj}+A^\ell_{ij}H_d\tilde L_{Li}
\tilde{\overline{E}}_{Lj}+B H_u H_d+{\rm h.c.}\right)\no\\
&-&\sum_{\rm all\ scalars}(m^2_S)_{ij}A_i\bar A_j-{1\over2}\sum_{(a)=1}^3
\left(\tilde m_{(a)}(\lambda\lambda)_{(a)}+{\rm h.c.}\right).
\eeqa
where $S=Q_L,\overline{D}_L,\overline{U}_L,L_L,\overline{E}_L$. The three 
Yukawa matrices $Y^f$ depend on 27 real and 27 imaginary parameters. Similarly,
the three $A^f$-matrices depend on 27 real and 27 imaginary parameters. The 
five $m^2_S$ hermitian $3\times3$ mass-squared matrices for sfermions have 30 
real parameters and 15 phases. The gauge and Higgs sectors depend on 
\beq\label{GaHi}
\theta_{\rm QCD},\tilde m_{(1)},\tilde m_{(2)},\tilde m_{(3)},
g_1,g_2,g_3,\mu,B,m_{h_u}^2,m_{h_d}^2,
\eeq
that is 11 real and 5 imaginary parameters. Summing over all sectors, we get
95 real and 74 imaginary parameters. The various couplings (other than the
gauge couplings) can be thought of as spurions that break a global symmetry,
\beq\label{MSBro}
U(3)^5\times U(1)_{\rm PQ}\times U(1)_R\ \to\ U(1)_B\times U(1)_L.
\eeq
The $U(1)_{\rm PQ}\times U(1)_R$ charge assignments are:
\beq\label{PQRcharges}
\matrix{&H_u&H_d&Q\overline{U}&Q\overline{D}&L\overline{E}\cr
U(1)_{\rm PQ}&1&1&-1&-1&-1\cr U(1)_{\rm R}&1&1&1&1&1\cr}.
\eeq
Consequently, we can remove 15 real and 30 imaginary parameters, which leaves
\beq\label{MSpar}
124\ =\ \cases{80&real\cr 44&imaginary}\ {\rm physical\ parameters}.
\eeq 
In particular, there are 43 new CP violating phases! In addition to the single
Kobayashi-Maskawa of the SM, we can put 3 phases in $M_1,M_2,\mu$ (we used the 
$U(1)_{\rm PQ}$ and $U(1)_R$ to remove the phases from $\mu B^*$ and $M_3$, 
respectively) and the other 40 phases appear in the mixing matrices of the 
fermion-sfermion-gaugino couplings. (Of the 80 real parameters, there are 11 
absolute values of the parameters in (\ref{GaHi}), 9 fermion masses, 21 
sfermion masses, 3 CKM angles and 36 SCKM angles.) Supersymmetry provides a 
nice example to our statement that reasonable extensions of the Standard Model 
may have more than one source of CP violation.

The requirement of consistency with experimental data provides strong 
constraints on many of these parameters. For this reason, the physics of flavor
and CP violation has had a profound impact on supersymmetric model building. 
A discussion of CP violation in this context can hardly avoid addressing the 
flavor problem itself.  Indeed, many of the supersymmetric models that we
analyze below were originally aimed at solving flavor problems.
For details on the supersymmetric flavor problem, see \cite{susabe}.
 
As concerns CP violation, one can distinguish two classes of experimental
constraints. First, bounds on nuclear and atomic electric dipole moments
determine what is usually called the {\it supersymmetric CP problem}. Second, 
the physics of neutral mesons and, most importantly, the small experimental 
value of $\varepsilon_K$ pose the {\it supersymmetric $\varepsilon_K$ problem}.
In the next two subsections we describe the two problems.
 
\subsubsection{The Supersymmetric CP Problem}
One aspect of supersymmetric CP violation involves effects that are flavor 
preserving. Then, for simplicity, we describe this aspect in a supersymmetric 
model without additional flavor mixings, {\it i.e.} the minimal supersymmetric 
standard model (MSSM) with universal sfermion masses and with the trilinear 
SUSY-breaking scalar couplings proportional to the corresponding Yukawa 
couplings. (The generalization to the case of non-universal soft terms is 
straightforward.)  In such a constrained framework, there are four new phases 
beyond the two phases of the SM ($\delta_{\rm KM}$ and $\theta_{\rm QCD}$). One
arises in the bilinear $\mu$-term of the superpotential (\ref{superp}), while 
the other three arise in the soft supersymmetry breaking parameters of 
(\ref{Lsoft}): $\tilde m$ (the gaugino mass), $A$ (the trilinear scalar 
coupling) and $B$ (the bilinear scalar coupling). Only two combinations of the 
four phases are physical \cite{Dugan:1985qf,Dimopoulos:1996kn}:
\beq\label{phiAB}
\phi_A=\arg(A^* \tilde m),\ \ \ \phi_B=\arg(\tilde m\mu B^*).
\eeq
In the more general case of non-universal soft terms there is one independent 
phase $\phi_{A_{i}}$ for each quark and lepton flavor. Moreover, complex 
off-diagonal entries in the sfermion mass-squared matrices represent 
additional sources of CP violation.
 
The most significant effect of $\phi_A$ and $\phi_B$ is their contribution to 
electric dipole moments (EDMs). For example, the contribution from one-loop 
gluino diagrams to the down quark EDM is given by 
\cite{Buchmuller:1983ye,Polchinski:1983zd}:
\beq\label{ddsusy}
d_d=m_d{e\alpha_3\over 18\pi\tilde m^3}\left(
|A|\sin\phi_A+\tan\beta|\mu|\sin\phi_B\right),
\eeq
where we have taken $m^2_Q\sim m^2_D\sim m^2_{\tilde g}\sim\tilde m^2$, for 
left- and right-handed squark and gluino masses. We define, as usual,
$\tan\beta=\langle{H_u}\rangle/\langle{H_d}\rangle$. Similar one-loop diagrams 
give rise to chromoelectric dipole moments. The electric and chromoelectric 
dipole moments of the light quarks $(u,d,s)$ are the main source of $d_N$ (the 
EDM of the neutron), giving \cite{Fischler:1992ha}
\beq\label{dipole}
d_N\sim 2\, \left({100\, GeV\over \tilde m}\right )^2\sin \phi_{A,B}
\times10^{-23}\ e\, {\rm cm},
\eeq
where, as above, $\tilde m$ represents the overall SUSY scale. In a generic 
supersymmetric framework, we expect $\tilde m={\cal O}(m_Z)$ and 
$\sin\phi_{A,B}={\cal O}(1)$. Then the constraint (\ref{dnexp}) is generically 
violated by about two orders of magnitude. This is {\it the Supersymmetric CP 
Problem}.

Eq. (\ref{dipole}) shows two possible ways to solve the 
supersymmetric CP problem:

(i) Heavy squarks: $\tilde m\gsim1\ TeV$;

(ii) Approximate CP: $\sin\phi_{A,B}\ll1$.

Recently, a third way has been investigated, that is cancellations between 
various contributions to the electric dipole moments. However, there seems to 
be no symmetry that can guarantee such a cancellation. This is in contrast to 
the other two mechanisms mentioned above that were shown to arise naturally in 
specific models. We therefore do not discuss any further this third mechanism.

The electric dipole moment of the electron is also a 
sensitive probe of flavor diagonal CP phases. The present experimental bound,
\beq\label{deexp}
|d_e|\leq4\times10^{-27}\ e\ {\rm cm} \cite{Commins:1994gv},
\eeq
is also violated by about two orders of magnitue for `natural' values of 
supersymmetric parameters. A new experiment \cite{Semertzidis:1999kv} has been 
proposed to search for the electric dipole moment of the muon at a level 
smaller by five orders of magnitude than present bounds; such improvement will
make $d_\mu$ another sensitive probe of supersymmetry \cite{Feng:2001sq}.

\subsubsection{The Supersymmetric $\varepsilon_K$ Problem}
The supersymmetric contribution to the $\varepsilon_K$ parameter is dominated 
by diagrams involving $Q$ and $\bar d$ squarks in the same loop. For $\tilde m=
m_{\tilde g}\simeq m_Q \simeq m_D$ (our results depend only weakly on this 
assumption) and focusing on the contribution from the first two squark 
families, one gets (see, for example, \cite{Gabbiani:1996hi}):
\beq\label{epsKSusy}
\varepsilon_K={5\ \alpha_3^2  \over 162\sqrt2}{f_K^2m_K\over\tilde
m^2\Delta m_K}\left [\left({m_K\over m_s+m_d}\right)^2+{3\over 25}\right]
\Im\left[(\delta_{12}^d)_{LL}(\delta_{12}^d)_{RR}\right].
\eeq
Here
\beqa\label{defdsusy}
(\delta_{12}^d)_{LL}&=&\left({m^2_{\tilde Q_2}-m^2_{\tilde Q_1}\over
m^2_{\tilde Q}}\right)\left|K^{dL}_{12}\right|,\no\\
(\delta_{12}^d)_{RR}&=&\left({m^2_{\tilde D_2}-m^2_{\tilde D_1}\over
m^2_{\tilde D}}\right)\left|K^{dR}_{12}\right|,
\eeqa
where $K^{dL}_{12}$ ($K^{dR}_{12}$) are the mixing angles in the gluino
couplings to left-handed (right-handed) down quarks and their scalar partners.
Note that CP would be violated even if there were two families only 
\cite{Nir:1986te}. Using the experimental value of $\varepsilon_K$, we get
\beq\label{epsKScon}
{(\Delta m_K\varepsilon_K)^{\rm SUSY}\over(\Delta m_K\varepsilon_K)^{\rm EXP}}
\sim10^7\left ({300 \ GeV\over\tilde m}\right)^2
\left({m^2_{\tilde Q_2}-m^2_{\tilde Q_1}\over m_{\tilde Q}^2}\right)
\left({m^2_{\tilde D_2}-m^2_{\tilde D_1}\over m_{\tilde D}^2}\right)
|K_{12}^{dL}K_{12}^{dR}|\sin\phi,
\eeq
where $\phi$ is the CP violating phase. In a generic supersymmetric framework, 
we expect $\tilde m={\cal O}(m_Z)$, $\delta m_{Q,D}^2/m_{Q,D}^2={\cal O}(1)$, 
$K_{ij}^{Q,D}={\cal O}(1)$ and $\sin\phi={\cal O}(1)$. Then the constraint 
(\ref{epsKScon}) is generically violated by about seven orders of magnitude. 

The $\Delta m_K$ constraint on $\Re\left[(\delta_{12}^d)_{LL}
(\delta_{12}^d)_{RR}\right]$ is about two orders of magnitude weaker.
One can distinguish then three interesting regions for $\langle{\delta_{12}^d}
\rangle=\sqrt{(\delta_{12}^d)_{LL}(\delta_{12}^d)_{RR}}$\, :
\beqa\label{ranmot}
0.003\ll &\langle&{\delta_{12}^d}\rangle  \ \ \ \ \ \ \ \ \ \ \ \ \ \
\ \ {\rm excluded},\nonumber\\
0.0002\ll&\langle&{\delta_{12}^d}\rangle \lsim0.003\ \ \ \ \ 
{\rm viable\ with\ small\ phases},\\
&\langle&{\delta_{12}^d}\rangle
\ll 0.0002 \ \ \ {\rm viable\ with}\ {\cal O}(1)\ {\rm phases}.\nonumber
\eeqa
The first bound comes from the $\Delta m_K$ constraint (assuming that the
relevant phase is not particularly close to $\pi/2$). The bounds here apply to
squark masses of order 500~GeV and scale like $\tilde m$. There is also
dependence on $m_{\tilde g}/\tilde m$, which is here taken to be one.

Eq. (\ref{epsKScon}) also shows what are the possible ways to solve
the supersymmetric $\varepsilon_K$ problem:

(i) Heavy squarks: $\tilde m\gg300\ GeV$;

(ii) Universality: $\delta m_{Q,D}^2\ll m_{Q,D}^2$;

(iii) Alignment: $|K_{12}^d|\ll1$;

(iv) Approximate CP: $\sin\phi\ll1$.

\subsubsection{A Supersymmetric $\varepsilon^\prime/\varepsilon$?}
In this section we discuss the question of whether supersymmetric contributions
to $\varepsilon^\prime/\varepsilon$ can be dominant. A typical supersymmetric 
contribution to $\varepsilon^\prime/\varepsilon$ is given by 
\cite{Buras:2000da}
\beqa\label{susyepe}
|\varepsilon^\prime/\varepsilon|=58B_G&\ &\ \left[{\alpha_s(m_{\tilde g})\over
\alpha_s(500\ {\rm GeV})}\right]^{23/21}\left({158\ MeV\over m_s+m_d}\right)
\nonumber\\
&\times&\ \left({500\ {\rm GeV}\over m_{\tilde g}}\right)\left|\Im\left[
(\delta_{LR}^d)_{12}-(\delta_{LR}^d)_{21}^*\right]\right|\ ,
\eeqa
where $B_G$ parameterizes the matrix element of the relevant four-quark
operator. Consequently, the supersymmetric contribution saturates 
$\varepsilon^\prime/\varepsilon$ for
\beq\label{satepe}
\Im\left[(\delta_{LR}^d)_{12}-(\delta_{LR}^d)_{21}^*\right]\
\sim\ \lambda^7\left({m_{\tilde g}\over500\ {\rm GeV}}\right)
\eeq
where, motivated by flavor symmetries (see below), we parameterize the 
suppression by powers of $\lambda\sim0.2$.

Without proportionality, a naive guess would give
\beqa\label{naiepe}
(\delta_{LR}^d)_{12}\ &\sim&\ {m_s|V_{us}|\over\tilde m}\sim\lambda^{5-6}
{m_t\over\tilde m},\nonumber\\
(\delta_{LR}^d)_{21}\ &\sim&\ {m_d\over|V_{us}|\tilde m}\ \sim\lambda^{5-6}
{m_t\over\tilde m}\ .
\eeqa
This is not far from the value required to account for 
$\varepsilon^\prime/\varepsilon$ \cite{Masiero:1999ub}. Thus, it is certainly
{\it possible} that supersymmetry accounts for, at least, a large part of
$\varepsilon^\prime/\varepsilon$ (see, for example, the models of refs.
\cite{Baek:2000jq,Baek:2001kc,Khalil:1999zn,Babu:2000xf,Khalil:1999ym,Kagan:1999iq,Khalil:2000ci}). 
Yet, it has been argued that such a situation is not {\it generic} 
\cite{Eyal:1999gk}. The problem is that eq. (\ref{naiepe}) gives an 
overestimate of the supersymmetric contribution in most viable models of 
supersymmetry breaking that have appeared in the literature. We will encounter 
concrete examples to this statement when we survey the various supersymmetric 
flavor models.

\subsection{Supersymmetry Breaking and Flavor Models}
Before turning to a detailed discussion, we define two scales that play an 
important role in supersymmetry: $\Lambda_S$, where the soft supersymmetry 
breaking terms are generated, and $\Lambda_F$, where flavor dynamics takes 
place. When $\Lambda_F\gg\Lambda_S$, it is possible that there are no genuinely
new sources of flavor and CP violation. This leads to models with exact 
universality. When $\Lambda_F\lsim\Lambda_S$, we do not expect, in general, 
that flavor and CP violation are limited to the Yukawa matrices. One way to 
suppress CP violation would be to assume that, similarly to the Standard Model,
CP violating phases are large, but their effects are screened, possibly by the 
same physics that explains the various flavor puzzles, such as models with 
Abelian or non-Abelian horizontal symmetries. It is also possible that CP 
violating effects are suppressed because squarks are heavy. Another option,
which is now excluded, was to assume that CP is an approximate symmetry 
of the full theory (namely, CP violating phases are all small).

\subsubsection{Gauge Mediated Supersymmetry Breaking}
If at some high energy scale squarks are exactly degenerate and the $A$ terms
proportional to the Yukawa couplings, then the contribution to 
$\varepsilon_K$ comes from RGE and is GIM suppressed, that is
\beq\label{gimsup}
\varepsilon_K\propto\Im[(V_{td}V_{ts}^*)^2]Y_t^4
\left[{\log(\Lambda_S/m_W)\over16\pi^2}\right]^2.
\eeq
This contribution is negligibly small~\cite{Dugan:1985qf}. The contribution 
from genuinely supersymmetric phases ({\it i.e.} the phases in $A_t$ and $\mu$)
is also negligible~\cite{Abel:1997eb,Baek:1999qy}. (This does not necessarily 
mean that there is no supersymmetric effect on $\varepsilon_K$. In some small 
corner of parameter space the supersymmetric contribution from stop-chargino 
diagrams can give up to 20\% of $\varepsilon_K$ 
\cite{Branco:1994eb,Goto:1996zk}.)

In models of Gauge Mediated Supersymmetry Breaking (GMSB) 
\cite{Dine:1995vc,Dine:1996ag}, superpartner masses are generated by the SM 
gauge interactions. These masses are then exactly universal at the scale 
$\Lambda_S$ at which they are generated  (up to tiny high order effects 
associated with Yukawa couplings). Furthermore, $A$ terms are suppressed by 
loop factors. The only contribution to $\varepsilon_K$ is then from the 
running, and since $\Lambda_S$ is low it is highly suppressed.

These models can also readily satisfy the EDM constraints. In most models, the 
$A$ terms and gaugino masses arise from the same supersymmetry breaking 
auxiliary field, that is, they are generated by the same SUSY and $U(1)_R$ 
breaking source. They therefore carry the same phase (up to corrections from 
the Standard Model Yukawa couplings), and $\phi_A$ vanishes to a very good 
approximation:
\beq\label{rada}
\phi_A\propto Y_t^4 Y_c^2 Y_b^2 J_{\rm CKM} 
\left[{\log(\Lambda_S/m_W)\over16\pi^2}\right]^4.
\eeq
The resulting EDM is $d_N\lsim 10^{-31}\ e\ {\rm cm}$. This maximum can be 
reached only for very large $\tan\beta\sim60$ while, for small 
$\tan\beta\sim1$, $d_N$ is about 5 orders of magnitude smaller. This range of 
values for $d_N$ is much below the present ($\sim10^{-25}\ e$ cm) and foreseen
($\sim 10^{-28}\ e$ cm) experimental sensitivities (see {\it e.g.} 
\cite{Romanino:1997cn}).

The value of $\phi_B$ in general depends on the mechanism for generating the 
$\mu$ term. However, running effects can generate an adequate $B$ term at low
scales in these models even if $B(\Lambda_S)=0$.
One then finds \cite{Babu:1996jf}
\beq\label{radb}
{B/\mu} = A_t(\Lambda_S)+ M_2(\Lambda_S)\, (-0.12+0.17\vert Y_t\vert^2)\ ,
\eeq
where $M_2$ is the $SU(2)$ gaugino mass. Since $\phi_A\simeq0$, the resulting 
$\phi_B$ vanishes, again up to corrections involving the Standard Model Yukawa 
couplings~\cite{Dine:1997xk}.

There is therefore no CP problem in simple models of gauge mediation,
even with phases of order one. The supersymmetric contribution to
$D-\bar D$ mixing is similarly small and we expect no observable effects.
As concerns the $B_d$ system, GMSB models predict then a large CP asymmetry
in $B\to\psi K_S$, with small deviations (at most 20\%) from the SM.

More generally, in any supersymmetric model where there are no new flavor
violating sources beyond the Yukawa couplings, CP violation in meson decays is 
hardly modified from the SM predictions \cite{Demir:2000qm}.

\subsubsection{Gravity, Anomaly and Gaugino Mediation}
If different moduli of string theory obtain supersymmetry breaking $F$ terms, 
they would typically induce flavor-dependent soft terms through their 
tree-level couplings to Standard Model fields. There are however various
scenarios in which the leading contribution to the soft terms is flavor
independent. The three most intensively studied frameworks are dilaton
dominance, anomaly madiation and gaugino mediation.

{\bf Dilaton dominance} assumes that the dilaton $F$ term is the dominant one.
Then, at tree level, the resulting soft masses are universal and the $A$ terms 
proportional to the Yukawa couplings. Both universality and proportionality 
are, however, violated by string loop effects. These induce corrections to 
squark masses of order ${\alpha_X\over\pi}m^2_{3/2}$, where 
$\alpha_X=[2\pi(S+S^*)]^{-1}$ is the string coupling. There is no reason why 
these corrections would be flavor blind. However, RGE effects enhance the 
universal part of the squark masses by roughly a factor of five, leaving the 
off-diagonal entries essentially unchanged. The flavor suppression factor is 
then \cite{Louis:1995ht}
\beq\label{dildoe}
\langle{\delta_{12}^d}\rangle\simeq{m^{2\ {\rm one-loop}}_{12}\over 
m^2_{\tilde q}}\simeq{\alpha_X\over\pi}{1\over25}\simeq4\times10^{-4}\ .
\eeq
Dilaton dominance relies on the assumption that loop corrections are small.
This probably presents the most serious theoretical difficulty for this idea, 
because it is hard to see how non-perturbative effects, which are probably 
required to stabilize the dilaton, could do so in a region of weak coupling.  
In the strong coupling regime, these corrections could be much larger.
However, this idea at least gives some plausible theoretical explanation
for how universal masses might emerge in hidden sector models. Given that 
dilaton stabilization might require that non-perturbative effects are 
important, the estimate of flavor suppression~(\ref{dildoe}) might 
well turn out to be an underestimate.

We now turn to the flavor diagonal phases that enter in various EDMs. The phase
$\phi_A$ vanishes at tree-level, so that \cite{Louis:1995ht,Brignole:1994dj}
$\phi_A={\cal O}\left(\alpha_X/\pi\right)$. [The smallness of $\phi_A$ implies 
that there is a suppression of ${\cal O}(\alpha_X/\pi)\sim10^{-2}$ compared 
to~(\ref{naiepe}) and the supersymmetric contribution to 
$\varepsilon^\prime/\varepsilon$ is small.] However, $\phi_B$ is unsuppressed, 
even when $\mu$, and through it $B$, are generated by Kahler potential effects 
through supersymmetry breaking, in which case $B=2 m_{3/2}^* \mu$ 
\cite{Barbieri:1993jk}. While the size of $m_{3/2}$ is determined from the 
requirement that the cosmological constant vanishes, its phase remains 
arbitrary, and in fact depends on the phase of the constant term that is added 
to the superpotential in order to cancel the cosmological constant.

We conclude that the supersymmetric $\varepsilon_K$ problem is solved in these 
models but the EDM problem, in general, is not. For EDM contributions to be 
small in these models, the gravitino mass better give a small physical phase.

{\bf Anomaly mediation} (AMSB) provides another approach to solving the flavor 
problems of supersymmetric theories, as well as to obtaining a predictive 
spectrum. In the presence of some truly `hidden' supersymmetry breaking sector, 
with no couplings to the SM fields (apart from indirect couplings through 
the supergravity multiplet) the conformal anomaly of the Standard Model gives
rise to soft supersymmetry breaking terms for the Standard Model fields 
\cite{Randall:1999uk,Giudice:1998xp}. These terms are generated purely by 
gravitational effects and are given by
\beq\label{anomed}
m^2_0(\mu)=-{1\over 4}{\partial\gamma(\mu)\over\ln\mu}m^2_{3/2},\ \
m_{1/2}(\mu)= {\beta(\mu)\over g(\mu)}m_{3/2},\ \
A(\mu)=-{1\over 2}\gamma(\mu)m_{3/2}\ ,
\eeq
where $\beta$ and $\gamma$ are the appropriate beta function and anomalous
dimension. Thus, apart from the Standard Model gauge and Yukawa couplings, the 
soft terms only involve the parameter $m_{3/2}$.

In general, naturalness  considerations suggest that couplings of hidden and 
visible sectors should appear in the Kahler potential, leading to soft masses 
for scalars already at tree level, and certainly by one loop. As a result, one 
would expect the contributions~(\ref{anomed}) to be irrelevant. However, in 
``sequestered sector models'' \cite{Randall:1999uk}, 
in which the visible sector fields and supersymmetry breaking fields live on 
different branes separated by some distance, the anomaly mediated contribution
(\ref{anomed}) could be the dominant effect. This leads to a predictive picture
for scalar masses. Since the soft terms (\ref{anomed}) are generated by the 
Standard Model gauge and Yukawa couplings, they are universal, up to 
corrections involving the third generation Yukawa couplings. However, the 
resulting slepton masses-squared are negative, so this model requires some 
modification. We will not attempt a complete review of this subject here. Our 
principal concerns are the sources of CP violation, and the extent to which the
AMSB formulae receive corrections, leading to non-degeneracy of the squark 
masses.

For eq. (\ref{anomed}) to correctly give the leading order soft terms, it is 
necessary that all moduli obtain large masses before supersymmetry breaking, 
and that there be no Planck scale VEVs in the supersymmetry breaking 
sector \cite{Bagger:2000rd}. A possible scenario for this to happen is if all 
moduli but the fifth dimensional radius, $R$, sit at an enhanced symmetry 
point, and that $R$ obtains a large mass compared to the supersymmetry-breaking
scale (say, by a racetrack mechanism). Even in this case, however, there is a 
difficulty.  One might expect that some of the moduli have masses well below 
the fundamental scale. If there are light moduli in the bulk, there are 
typically one-loop contributions to scalar masses-squared from exchanges of
bulk fields, proportional to $m_{3/2}^2/R^3$ times a loop factor 
\cite{Randall:1999uk}. If these contributions are non-universal, they may
easily violate the $\Delta m_K$ and $\varepsilon_K$ constraints 
\cite{Dine:2001ne}. 

If there are no light moduli, and if the contributions described above are
adequately suppressed, some modification of the visible sector is required
in order to generate acceptable slepton masses. Different such solutions
have been suggested. In some of these models, there are no new contributions to
CP violation simply because there are few enough new parameters in the theory
that they can all be chosen real by field redefinitions
\cite{Pomarol:1999ie,Katz:1999uw,Chacko:2000wq}. Furthermore, it is possible to
generate the $\mu$ term in these models from AMSB, so that $\phi_B$ vanishes. 
These models are then similar to GMSB models from the point of view of CP 
violation.

We conclude then, that in generic sequestered sector models it is difficult to 
obtain strong degeneracy and a special phase structure is required. It is 
conceivable that there might be theories with a high degree of degeneracy, or 
with no new sources of CP violation. In such theories, the SM predictions for
CP violation are approximately maintained.

{\bf Gaugino mediation} \cite{Kaplan:2000ac,Chacko:2000mi} is in many ways 
similar to anomaly mediation, and poses similar issues. These models also 
suppress dangerous tree level contact terms by invoking extra dimensions, with 
the Standard Model matter fields localized on one brane and the supersymmetry 
breaking sector on another brane. In this case, however, the Standard Model 
gauge fields are in the bulk, so gauginos get masses at tree level, and as a 
result scalar masses are generated by running. Scalar masses are therefore 
universal. Furthermore, the soft terms typically involve only one new 
parameter, namely, the singlet $F$ VEV that gives rise to gaugino masses. 
Therefore, they do not induce any new CP violation. 

Again, however, if there are non-universal tree and one loop contributions
to scalar masses, significant violations of degeneracy and proportionality can 
be expected, and a special structure of CP violating phases is required.

\subsubsection{Supersymmetric Flavor Models}
Various frameworks have been suggested in which flavor symmetries,
designed to explain the hierarchy of the Yukawa couplings, impose at the
same time a special flavor structure on the soft supersymmetry breaking
terms that helps to alleviate the flavor and CP problems. 

In the framework of {\bf alignment}, one does not assume any squark degeneracy.
Instead, flavor violation is suppressed because the squark mass matrices are 
approximately diagonal in the quark mass basis. This is the case in models of 
Abelian flavor symmetries, in which the off-diagonal entries in both the quark
mass matrices and in the squark mass matrices are suppressed by some power of a
small parameter, $\lambda$, that quantifies the breaking of some Abelian flavor
symmetry. A natural choice for the value of $\lambda$ is $\sin\theta_C$, so we 
will take $\lambda\sim0.2$. One would naively expect the first two generation 
squark mixing to be of the order of $\lambda$. However, the $\Delta m_K$ 
constraint is not satisfied with the `naive alignment', $K_{12}^d\sim\lambda$, 
and one has to construct more complicated models to achieve the required 
suppression \cite{Nir:1993mx,Leurer:1994gy}. One can solve the supersymmetric
$\varepsilon_K$ problem by flavor suppression, that is, models with 
$\langle{\delta_{12}^d}\rangle\sim\lambda^5$ \cite{Nir:1996am}. These
models are highly constrained and almost unique. It is simpler to construct 
models where $\langle{\delta_{12}^d}\rangle\sim\lambda^3$ but the CP violating 
phases are also suppressed \cite{Eyal:1998bk}. Such models predict that 
$a_{\psi K_S}\ll1$ and are therefore now excluded. (Models with 
$\langle{\delta_{12}^d}\rangle\sim\lambda^3$ could still be viable with
phases of order one if the RGE contributions enhance squark degeneracy.)

As concerns flavor diagonal phases, the question is more model dependent.
There is however a way to suppress these phases without assuming approximate
CP \cite{Nir:1996am}. The mechanism requires that CP is spontaneously broken by
the same fields that break the flavor symmetry (``flavons"). It is based on the
observation that a Yukawa coupling and the corresponding $A$ term carry the same
horizontal charge and therefore their dependence on the flavon fields is
similar. In particular, if a single flavon dominates a certain coupling,
the CP phase is the same for the Yukawa coupling and for the corresponding $A$
term and the resulting $\phi_A$ vanishes. Similarly, if the $\mu$ term and the
$B$ term depend on one (and the same) flavon, $\phi_B$ is suppressed.

As concerns $\varepsilon^\prime/\varepsilon$, the $\varepsilon_K$ constraint 
requires that the relevant terms are suppressed by at least a factor of 
$\lambda^2$ compared to~(\ref{naiepe}) \cite{Eyal:1999gk} and the contribution
is therefore small.

We conclude that one can construct models in which an Abelian horizontal 
symmetry solves both the $\varepsilon_K$ and the $d_N$ problems. These models 
are however not the generic ones in this framework. 

{\bf Non-Abelian horizontal symmetries} can induce approximate degeneracy 
between the first two squark generations, thus relaxing the flavor and CP  
problems \cite{Dine:1993np}. A review of $\varepsilon_K$ in this class of 
models can be found in \cite{Grossman:1997pa}. Quite generically, the 
supersymmetric contributions to $\varepsilon_K$ are too large and require small
phases (see, for example, the models of ref. \cite{Barbieri:1996uv}). There are
however specific models where the $\varepsilon_K$ problem is solved without the
need for small phases~\cite{Carone:1996nd,Barbieri:1997ww}. Furthermore, 
universal contributions from RGE running might further relax the problem. 

As concerns flavor diagonal phases, it is difficult (though not entirely
impossible) to avoid $\phi_A\gsim\lambda^2\sim0.04$ \cite{Grossman:1997pa}. 
This, however, might be just enough to satisfy the $d_N$ constraint.

With a horizontal U(2) symmetry, the two contributions to 
$\varepsilon^\prime/\varepsilon$ in~(\ref{susyepe}) cancel each other. (More 
generally, this happens for a symmetric $A$ matrix with $A_{11}=0$ 
\cite{Barbieri:2000ax}.)

We conclude that, similar to models of Abelian flavor symmetries, one can 
construct models of non-Abelian symmetries in which the symmetry solves both 
the $\varepsilon_K$ and the $d_N$ problems. These models are however
not the generic ones in this framework.

Finally, one can construct models of {\bf heavy first two generation squarks}.
Here, the basic mechanism to suppress flavor changing processes is actually 
flavor diagonal: $m_{\tilde q_{1,2}}\sim20\ {\rm TeV}$. Naturalness does not 
allow higher masses, but this mass scale is not enough to satisfy even the 
$\Delta m_K$ constraint \cite{Cohen:1997sq}, and one has to invoke alignment, 
$K_{12}^d\sim\lambda$. This is still not enough to satisfy the $\varepsilon_K$ 
constraint of eq.~(\ref{epsKScon}), and a somewhat small phase is required.

Three more comments are in order: First, in this framework, gauginos are 
significantly lighter than the first two generation squarks, and so RGE cannot 
induce degeneracy. Second, the large mass of the squarks is enough to solve the
EDM related problems, and so it is only the $\varepsilon_K$ constraint that 
motivates a special phase structure. Finally, the contribution to 
$\varepsilon^\prime/\varepsilon$ is negligibly small. Instead of 
(\ref{naiepe}), a more likely estimate is~\cite{Eyal:1999gk}
$(\delta_{LR}^d)_{ij}\ \sim\ {m_Z(M_d)_{ij}\over(10\ {\rm TeV})^2}$,
which suppresses the relevant matrix elements by a factor of order $10^4$.

\subsection{(S)Conclusions}
We would like to emphasize the following points:

(i) For supersymmetry to be established, a direct observation of supersymmetric
particles is necessary. Once it is discovered, then measurements of CP 
violating observables will be a very sensitive probe of its flavor structure 
and, consequently, of the mechanism of dynamical supersymmetry breaking.

(ii) It seems possible to distinguish between models of exact universality and 
models with genuine supersymmetric flavor and CP violation. The former tend to
give $d_N\lsim10^{-31}$ e cm while the latter usually predict 
$d_N\gsim10^{-28}$ e cm.

(iii) The proximity of $a_{\psi K_S}$ to the SM predictions is obviously
consistent with models of exact universality. It disfavors models
of heavy squarks such as that of ref. \cite{Cohen:1997sq}. Models of flavor
symmetries allow deviations of order 20\% (or smaller) from the SM predictions.
To be convincingly signalled, an improvement in the theoretical calculations 
that lead to the SM predictions for $a_{\psi K_S}$ will be required 
\cite{Eyal:2000ys}.

(iv) Alternatively, the fact that $K\to\pi\nu\bar\nu$ decays are not affected 
by most supersymmetric flavor models 
\cite{Nir:1998tf,Buras:1998ij,Colangelo:1998pm}
is an advantage here. The Standard Model correlation between 
$a_{\pi\nu\bar\nu}$ and $a_{\psi K_S}$ is a much cleaner test than 
a comparison of $a_{\psi K_S}$ to the CKM constraints.

(v) The neutral $D$ system provides a stringent test of alignment. 
Observation of CP violation in the $D\to K\pi$ decays will make a convincing
case for new physics.

\acknowledgments
I thank Alan Walker for his excellent organization of this school.
I am grateful to the students of the school for their interest and for the
very enjoyable atmosphere.
YN is supported by the Israel Science Foundation founded by the Israel Academy 
of Sciences and Humanities, and by the Minerva Foundation (Munich).

\tighten


\begin{references}

\bibitem{Kobayashi:1973fv} M.~Kobayashi and T.~Maskawa,
Prog.\ Theor.\ Phys.\  {\bf 49}, 652 (1973).

\bibitem{Christenson:1964fg}
J.~H.~Christenson, J.~W.~Cronin, V.~L.~Fitch and R.~Turlay,
Phys.\ Rev.\ Lett.\  {\bf 13}, 138 (1964).

\bibitem{Burkhardt:1988yh} H.~Burkhardt {\it et al.}  [NA31 Collaboration],
Phys.\ Lett.\ B {\bf 206}, 169 (1988).

\bibitem{Barr:1993rx} G.~D.~Barr {\it et al.}  [NA31 Collaboration],
Phys.\ Lett.\ B {\bf 317}, 233 (1993).

\bibitem{Gibbons:1993zq} L.~K.~Gibbons {\it et al.},
Phys.\ Rev.\ Lett.\  {\bf 70}, 1203 (1993).

\bibitem{ktev}
J. Graham, FNAL seminar on {\it `A new measurement of 
$\epsilon^\prime/\epsilon$ from KTeV,'} June 8th, 2001.

\bibitem{nafe}
G. Unal, CERN seminar on `{\it A new Measurement of Direct CP Violation
by NA48},' May 10th, 2001.

\bibitem{Ackerstaff:1998xz} K.~Ackerstaff {\it et al.}  [OPAL collaboration],
Eur.\ Phys.\ J.\ C {\bf 5}, 379 (1998) [hep-ex/9801022].

\bibitem{Affolder:2000gg} T.~Affolder {\it et al.}  [CDF Collaboration],
Phys.\ Rev.\ D {\bf 61}, 072005 (2000) [hep-ex/9909003].

\bibitem{Barate:2000tf} R.~Barate {\it et al.}  [ALEPH Collaboration],
Phys.\ Lett.\ B {\bf 492}, 259 (2000) [hep-ex/0009058].

\bibitem{Aubert:2001nu} B.~Aubert {\it et al.}  [BaBar Collaboration],
Phys.\ Rev.\ Lett.\  {\bf 87}, 091801 (2001) [hep-ex/0107013].

\bibitem{Abe:2001xe} K.~Abe {\it et al.} [Belle Collaboration],
Phys.\ Rev.\ Lett.\  {\bf 87}, 091802 (2001) [hep-ex/0107061].

\bibitem{Wolfenstein:1964ks} L.~Wolfenstein,
Phys.\ Rev.\ Lett.\  {\bf 13}, 562 (1964).

\bibitem{Sakharov:1967dj} A.~D.~Sakharov,
Pisma Zh.\ Eksp.\ Teor.\ Fiz.\  {\bf 5}, 32 (1967)
[JETP Lett.\  {\bf 5}, 24 (1967)].

\bibitem{Burles:2000zk} S.~Burles, K.~M.~Nollett and M.~S.~Turner,
Astrophys.\ J.\  {\bf 552}, L1 (2001) [astro-ph/0010171].

\bibitem{Farrar:1994hn} G.~R.~Farrar and M.~E.~Shaposhnikov,
Phys.\ Rev.\ D {\bf 50}, 774 (1994) [hep-ph/9305275].

\bibitem{Huet:1995jb} P.~Huet and E.~Sather,
Phys.\ Rev.\ D {\bf 51}, 379 (1995) [hep-ph/9404302].

\bibitem{Gavela:1994ds} M.~B.~Gavela, M.~Lozano, J.~Orloff and O.~Pene,
Nucl.\ Phys.\ B {\bf 430}, 345 (1994) [hep-ph/9406288].

\bibitem{Riotto:1999yt} A.~Riotto and M.~Trodden,
Ann.\ Rev.\ Nucl.\ Part.\ Sci.\  {\bf 49}, 35 (1999) [hep-ph/9901362].

\bibitem{Buchmuller:2000as} W.~Buchmuller and M.~Plumacher,
hep-ph/0007176.

\bibitem{Branco:2001pq} 
G.~C.~Branco, T.~Morozumi, B.~M.~Nobre and M.~N.~Rebelo,
hep-ph/0107164.

\bibitem{Cohen:1993nk} A.~G.~Cohen, D.~B.~Kaplan and A.~E.~Nelson,
Ann.\ Rev.\ Nucl.\ Part.\ Sci.\  {\bf 43}, 27 (1993) [hep-ph/9302210].

\bibitem{Crewther:1979pi}
R.~J.~Crewther, P.~Di Vecchia, G.~Veneziano and E.~Witten,
Phys.\ Lett.\ B {\bf 88}, 123 (1979) [Erratum-ibid.\ B {\bf 91}, 487 (1979)].

\bibitem{Dixon:1991cq} L.~J.~Dixon, A.~Langnau, Y.~Nir and B.~Warr,
Phys.\ Lett.\ B {\bf 253}, 459 (1991).

\bibitem{Harris:1999jx} P.~G.~Harris {\it et al.},
Phys.\ Rev.\ Lett.\  {\bf 82}, 904 (1999).

\bibitem{Dine:2000cj} M.~Dine,
hep-ph/0011376.

\bibitem{Banks:1994yg} T.~Banks, Y.~Nir and N.~Seiberg,
hep-ph/9403203.

\bibitem{Peccei:1977hh} R.~D.~Peccei and H.~R.~Quinn,
Phys.\ Rev.\ Lett.\  {\bf 38}, 1440 (1977).

\bibitem{Peccei:1977ur} R.~D.~Peccei and H.~R.~Quinn,
Phys.\ Rev.\ D {\bf 16}, 1791 (1977).

\bibitem{Dine:1992ya} M.~Dine, R.~G.~Leigh and D.~A.~MacIntire,
Phys.\ Rev.\ Lett.\  {\bf 69}, 2030 (1992) [hep-th/9205011].

\bibitem{Choi:1993xp} K.~Choi, D.~B.~Kaplan and A.~E.~Nelson,
Nucl.\ Phys.\ B {\bf 391}, 515 (1993) [hep-ph/9205202].

\bibitem{Worah:1997hk} M.~P.~Worah,
Phys.\ Rev.\ Lett.\  {\bf 79}, 3810 (1997) [hep-ph/9704389].

\bibitem{Worah:1997ni} M.~P.~Worah,
Phys.\ Rev.\ D {\bf 56}, 2010 (1997) [hep-ph/9702423].

\bibitem{Jarlskog:1985ht} C.~Jarlskog,
Phys.\ Rev.\ Lett.\  {\bf 55}, 1039 (1985).

\bibitem{Cabibbo:1963yz} N.~Cabibbo,
Phys.\ Rev.\ Lett.\  {\bf 10}, 531 (1963).

\bibitem{Maki:1962mu} Z.~Maki, M.~Nakagawa and S.~Sakata,
Prog.\ Theor.\ Phys.\  {\bf 28}, 870 (1962).

\bibitem{Chau:1984fp} L.~Chau and W.~Keung,
Phys.\ Rev.\ Lett.\  {\bf 53}, 1802 (1984).

\bibitem{Wolfenstein:1983yz} L.~Wolfenstein,
Phys.\ Rev.\ Lett.\  {\bf 51}, 1945 (1983).

\bibitem{Dib:1990uz} C.~Dib, I.~Dunietz, F.~J.~Gilman and Y.~Nir,
Phys.\ Rev.\ D {\bf 41}, 1522 (1990).

\bibitem{Rosner:1988nx} J.~L.~Rosner, A.~I.~Sanda and M.~P.~Schmidt,
EFI-88-12-CHICAGO {\it Presented at Workshop on High Sensitivity Beauty 
Physics, Batavia, IL, Nov 11-14, 1987}.

\bibitem{Carter:1980hr} A.~B.~Carter and A.~I.~Sanda,
Phys.\ Rev.\ Lett.\  {\bf 45}, 952 (1980).

\bibitem{Carter:1981tk} A.~B.~Carter and A.~I.~Sanda,
Phys.\ Rev.\ D {\bf 23}, 1567 (1981).

\bibitem{Bigi:1981qs} I.~I.~Bigi and A.~I.~Sanda,
Nucl.\ Phys.\ B {\bf 193}, 85 (1981).

\bibitem{Dunietz:1986vi} I.~Dunietz and J.~L.~Rosner,
Phys.\ Rev.\ D {\bf 34}, 1404 (1986).

\bibitem{Bigi:1987vr} I.~I.~Bigi and A.~I.~Sanda,
Nucl.\ Phys.\ B {\bf 281}, 41 (1987).

\bibitem{Harari:1987ex} H.~Harari and Y.~Nir,
Phys.\ Lett.\ B {\bf 195}, 586 (1987).

\bibitem{Hocker:2001xe} A.~Hocker, H.~Lacker, S.~Laplace and F.~Le Diberder,
Eur.\ Phys.\ J.\ C {\bf 21}, 225 (2001) [hep-ph/0104062] (updates on
http://www.slac.stanford.edu/~laplace/ckmfitter.html)

\bibitem{Buchalla:1994tr} G.~Buchalla and A.~J.~Buras,
Phys.\ Lett.\ B {\bf 333}, 221 (1994) [hep-ph/9405259].

\bibitem{Buchalla:1996fp} G.~Buchalla and A.~J.~Buras,
Phys.\ Rev.\ D {\bf 54}, 6782 (1996) [hep-ph/9607447].

\bibitem{Bergmann:2000ak} S.~Bergmann and G.~Perez,
JHEP {\bf 0008}, 034 (2000) [hep-ph/0007170].

\bibitem{Branco:1999fs}
G.~C.~Branco, L.~Lavoura and J.~P.~Silva,
``CP violation,''
{\it  Oxford, UK: Clarendon (1999) 511 p}.

\bibitem{Grossman:1997xn} Y.~Grossman, B.~Kayser and Y.~Nir,
Phys.\ Lett.\ B {\bf 415}, 90 (1997) [hep-ph/9708398].

\bibitem{Buras:2001pn} A.~J.~Buras,
hep-ph/0101336.

\bibitem{Glashow:1970gm} S.~L.~Glashow, J.~Iliopoulos and L.~Maiani,
Phys.\ Rev.\ D {\bf 2}, 1285 (1970).

\bibitem{Littenberg:1989ix} L.~S.~Littenberg,
Phys.\ Rev.\ D {\bf 39}, 3322 (1989).

\bibitem{Buchalla:1998ux} G.~Buchalla and G.~Isidori,
Phys.\ Lett.\ B {\bf 440}, 170 (1998) [hep-ph/9806501].

\bibitem{Perez:1999kw} G.~Perez,
JHEP {\bf 9909}, 019 (1999) [hep-ph/9907205].

\bibitem{Perez:2000qa} G.~Perez,
JHEP {\bf 0002}, 043 (2000) [hep-ph/0001037].

\bibitem{Adler:2000by} S.~Adler {\it et al.}  [E787 Collaboration],
Phys.\ Rev.\ Lett.\  {\bf 84}, 3768 (2000) [hep-ex/0002015].

\bibitem{Alavi-Harati:2000hd}
A.~Alavi-Harati {\it et al.}  [The E799-II/KTeV Collaboration],
Phys.\ Rev.\ D {\bf 61}, 072006 (2000) [hep-ex/9907014].

\bibitem{Grossman:1997sk} Y.~Grossman and Y.~Nir,
Phys.\ Lett.\ B {\bf 398}, 163 (1997) [hep-ph/9701313].

\bibitem{Bergmann:2000id}
S.~Bergmann, Y.~Grossman, Z.~Ligeti, Y.~Nir and A.~A.~Petrov,
Phys.\ Lett.\ B {\bf 486}, 418 (2000) [hep-ph/0005181].

\bibitem{Bergmann:1999pm} S.~Bergmann and Y.~Nir,
JHEP {\bf 9909}, 031 (1999) [hep-ph/9909391].

\bibitem{D'Ambrosio:2001wg} G.~D'Ambrosio and D.~Gao,
Phys.\ Lett.\ B {\bf 513}, 123 (2001) [hep-ph/0105078].

\bibitem{Nir:1999mg} Y.~Nir,
hep-ph/9911321.

\bibitem{Blaylock:1995ay} G.~Blaylock, A.~Seiden and Y.~Nir,
Phys.\ Lett.\ B {\bf 355}, 555 (1995) [hep-ph/9504306].

\bibitem{Wolfenstein:1995kv} L.~Wolfenstein,
Phys.\ Rev.\ Lett.\  {\bf 75}, 2460 (1995) [hep-ph/9505285].

\bibitem{Godang:2000yd} R.~Godang {\it et al.}  [CLEO Collaboration],
Phys.\ Rev.\ Lett.\  {\bf 84}, 5038 (2000) [hep-ex/0001060].

\bibitem{Chau:1994ec} L.~Chau and H.~Cheng,
Phys.\ Lett.\ B {\bf 333}, 514 (1994) [hep-ph/9404207].

\bibitem{Browder:1996ay} T.~E.~Browder and S.~Pakvasa,
Phys.\ Lett.\ B {\bf 383}, 475 (1996) [hep-ph/9508362].

\bibitem{Falk:1999ts} A.~F.~Falk, Y.~Nir and A.~A.~Petrov,
JHEP {\bf 9912}, 019 (1999) [hep-ph/9911369].

\bibitem{Golowich:2001hb} E.~Golowich and S.~Pakvasa,
Phys.\ Lett.\ B {\bf 505}, 94 (2001) [hep-ph/0102068].

\bibitem{Gronau:2001nr} M.~Gronau, Y.~Grossman and J.~L.~Rosner,
Phys.\ Lett.\ B {\bf 508}, 37 (2001) [hep-ph/0103110].

\bibitem{Link:2000cu} J.~M.~Link {\it et al.}  [FOCUS Collaboration],
Phys.\ Lett.\ B {\bf 485}, 62 (2000) [hep-ex/0004034].

\bibitem{Aitala:1999dt} E.~M.~Aitala {\it et al.}  [E791 Collaboration],
Phys.\ Rev.\ Lett.\  {\bf 83}, 32 (1999) [hep-ex/9903012].

\bibitem{Smith:2001ej} A.~B.~Smith  [CLEO Collaboration],
hep-ex/0104008.

\bibitem{Yabsley}
B. Yabsley [Belle Collaboration], talk at the International Europhysics
Conference on High Energy Physics (Budapest, Hungary, July 2001).

\bibitem{Bartelt:1995vr} J.~Bartelt {\it et al.}  [CLEO Collaboration],
Phys.\ Rev.\ D {\bf 52}, 4860 (1995).

\bibitem{Aitala:1998ff} E.~M.~Aitala {\it et al.}  [E791 Collaboration],
Phys.\ Lett.\ B {\bf 421}, 405 (1998) [hep-ex/9711003].

\bibitem{Link:2000aw} J.~M.~Link {\it et al.}  [FOCUS Collaboration],
Phys.\ Lett.\ B {\bf 491}, 232 (2000)
[Erratum-ibid.\ B {\bf 495}, 443 (2000)] [hep-ex/0005037].

\bibitem{Bonvicini:2001qm} G.~Bonvicini {\it et al.}  [CLEO Collaboration],
Phys.\ Rev.\ D {\bf 63}, 071101 (2001) [hep-ex/0012054].

\bibitem{Jaffe:2001hz} D.~E.~Jaffe {\it et al.}  [CLEO Collaboration],
Phys.\ Rev.\ Lett.\  {\bf 86}, 5000 (2001) [hep-ex/0101006].

\bibitem{Abbiendi:2000av} G.~Abbiendi {\it et al.}  [OPAL Collaboration],
Eur.\ Phys.\ J.\ C {\bf 12}, 609 (2000) [hep-ex/9901017].

\bibitem{Barate:2001uk} R.~Barate {\it et al.}  [ALEPH Collaboration],
Eur.\ Phys.\ J.\ C {\bf 20}, 431 (2001).

\bibitem{Aubert:2001xc} B.~Aubert,
hep-ex/0107059.

\bibitem{Bigi:1987in}
I.~I.~Bigi, V.~Khoze, N.~G.~Uraltsev and A.~I.~Sanda,
in {\it CP Violation}, ed. C. Jarlskog (World Scientific, Singapore, 1992).

\bibitem{Aleksan:1993qp}
R.~Aleksan, A.~Le Yaouanc, L.~Oliver, O.~Pene and J.~C.~Raynal,
Phys.\ Lett.\ B {\bf 316}, 567 (1993).

\bibitem{Beneke:1999sy}
M.~Beneke, G.~Buchalla, C.~Greub, A.~Lenz and U.~Nierste,
Phys.\ Lett.\ B {\bf 459}, 631 (1999) [hep-ph/9808385].

\bibitem{Wolfenstein:1998qz} L.~Wolfenstein,
Phys.\ Rev.\ D {\bf 57}, 5453 (1998).

\bibitem{Buchalla:1996vs} G.~Buchalla, A.~J.~Buras and M.~E.~Lautenbacher,
Rev.\ Mod.\ Phys.\  {\bf 68}, 1125 (1996) [hep-ph/9512380].

\bibitem{Hagelin:1981zk} J.~S.~Hagelin,
Nucl.\ Phys.\ B {\bf 193}, 123 (1981).

\bibitem{Buras:1984pq} A.~J.~Buras, W.~Slominski and H.~Steger,
Nucl.\ Phys.\ B {\bf 245}, 369 (1984).

\bibitem{Beneke:1996gn} M.~Beneke, G.~Buchalla and I.~Dunietz,
Phys.\ Rev.\ D {\bf 54}, 4419 (1996) [hep-ph/9605259].

\bibitem{Randall:1999te} L.~Randall and S.~Su,
Nucl.\ Phys.\ B {\bf 540}, 37 (1999) [hep-ph/9807377].

\bibitem{Cahn:1999gx} R.~N.~Cahn and M.~P.~Worah,
Phys.\ Rev.\ D {\bf 60}, 076006 (1999) [hep-ph/9904480].

\bibitem{Barenboim:1999in} G.~Barenboim, G.~Eyal and Y.~Nir,
Phys.\ Rev.\ Lett.\  {\bf 83}, 4486 (1999) [hep-ph/9905397].

\bibitem{Eyal:1999ii} G.~Eyal and Y.~Nir,
JHEP {\bf 9909}, 013 (1999) [hep-ph/9908296].

\bibitem{Charles:1999qx} J.~Charles,
Phys.\ Rev.\ D {\bf 59}, 054007 (1999) [hep-ph/9806468].

\bibitem{Aubert:2001qj} B.~Aubert,
hep-ex/0107074.

\bibitem{Cronin-Hennessy:2000hw}
D.~Cronin-Hennessy {\it et al.}  [CLEO Collaboration],
hep-ex/0001010.

\bibitem{Abe:2001nq} K.~Abe {\it et al.}  [BELLE Collaboration],
Phys.\ Rev.\ Lett.\  {\bf 87}, 101801 (2001) [hep-ex/0104030].

\bibitem{Aubert:2001hs} B.~Aubert {\it et al.}  [BABAR Collaboration],
hep-ex/0105061.

\bibitem{Nir:1991cu} Y.~Nir and H.~R.~Quinn,
Phys.\ Rev.\ Lett.\  {\bf 67}, 541 (1991).

\bibitem{Silva:1994sv} J.~P.~Silva and L.~Wolfenstein,
Phys.\ Rev.\ D {\bf 49}, 1151 (1994) [hep-ph/9309283].

\bibitem{Hernandez:1994rh}
O.~F.~Hernandez, D.~London, M.~Gronau and J.~L.~Rosner,
Phys.\ Lett.\ B {\bf 333}, 500 (1994) [hep-ph/9404281].

\bibitem{Neubert:1998pt} M.~Neubert and J.~L.~Rosner,
Phys.\ Lett.\ B {\bf 441}, 403 (1998) [hep-ph/9808493].

\bibitem{Fleischer:1997bv} R.~Fleischer,
Int.\ J.\ Mod.\ Phys.\ A {\bf 12}, 2459 (1997) [hep-ph/9612446].

\bibitem{Gronau:1990ka} M.~Gronau and D.~London,
Phys.\ Rev.\ Lett.\  {\bf 65}, 3381 (1990).

\bibitem{Grossman:1998jr} Y.~Grossman and H.~R.~Quinn,
Phys.\ Rev.\ D {\bf 58}, 017504 (1998) [hep-ph/9712306].

\bibitem{Gronau:2001ff} M.~Gronau, D.~London, N.~Sinha and R.~Sinha,
Phys.\ Lett.\ B {\bf 514}, 315 (2001) [hep-ph/0105308].

\bibitem{Lipkin:1991st} H.~J.~Lipkin, Y.~Nir, H.~R.~Quinn and A.~Snyder,
Phys.\ Rev.\ D {\bf 44}, 1454 (1991).

\bibitem{Gronau:1991dq} M.~Gronau,
Phys.\ Lett.\ B {\bf 265}, 389 (1991).

\bibitem{Snyder:1993mx} A.~E.~Snyder and H.~R.~Quinn,
Phys.\ Rev.\ D {\bf 48}, 2139 (1993).

\bibitem{Quinn:2000by} H.~R.~Quinn and J.~P.~Silva,
Phys.\ Rev.\ D {\bf 62}, 054002 (2000) [hep-ph/0001290].

\bibitem{Beneke:1999br} M.~Beneke, G.~Buchalla, M.~Neubert and C.~T.~Sachrajda,
Phys.\ Rev.\ Lett.\  {\bf 83}, 1914 (1999) [hep-ph/9905312].

\bibitem{Beneke:2000ry} M.~Beneke, G.~Buchalla, M.~Neubert and C.~T.~Sachrajda,
Nucl.\ Phys.\ B {\bf 591}, 313 (2000) [hep-ph/0006124].

\bibitem{Beneke:2001ev} M.~Beneke, G.~Buchalla, M.~Neubert and C.~T.~Sachrajda, 
Nucl.\ Phys.\ B {\bf 606}, 245 (2001) [hep-ph/0104110].

\bibitem{Keum:2001ph} Y.~Keum, H.~Li and A.~I.~Sanda,
Phys.\ Lett.\ B {\bf 504}, 6 (2001) [hep-ph/0004004].

\bibitem{susbuc}
G. Buchalla, these proceedings.

\bibitem{Grossman:1997dd} Y.~Grossman, Y.~Nir and M.~P.~Worah,
Phys.\ Lett.\ B {\bf 407}, 307 (1997) [hep-ph/9704287].

\bibitem{Georgi:1999wt} H.~Georgi and S.~L.~Glashow,
Phys.\ Lett.\ B {\bf 451}, 372 (1999) [hep-ph/9807399].

\bibitem{Frampton:2001dn} P.~H.~Frampton, S.~L.~Glashow and T.~Yoshikawa,
Phys.\ Rev.\ Lett.\  {\bf 87}, 011801 (2001) [hep-ph/0103022].

\bibitem{Chang:2001uk} D.~Chang, W.~Keung and R.~N.~Mohapatra,
Phys.\ Lett.\ B {\bf 515}, 431 (2001) [hep-ph/0105177].

\bibitem{Ball:2000mb} P.~Ball, J.~M.~Frere and J.~Matias,
Nucl.\ Phys.\ B {\bf 572}, 3 (2000) [hep-ph/9910211].

\bibitem{Bergmann:2001pm} S.~Bergmann and G.~Perez,
hep-ph/0103299.

\bibitem{Pomarol:1993uu} A.~Pomarol,
Phys.\ Rev.\ D {\bf 47}, 273 (1993) [hep-ph/9208205].

\bibitem{Babu:1994ai} K.~S.~Babu and S.~M.~Barr,
Phys.\ Rev.\ Lett.\  {\bf 72}, 2831 (1994) [hep-ph/9309249].

\bibitem{Eyal:1998bk} G.~Eyal and Y.~Nir,
Nucl.\ Phys.\ B {\bf 528}, 21 (1998) [hep-ph/9801411].

\bibitem{Grossman:1997ke} Y.~Grossman and M.~P.~Worah,
Phys.\ Lett.\ B {\bf 395}, 241 (1997) [hep-ph/9612269].

\bibitem{Nir:1990hj} Y.~Nir and D.~J.~Silverman,
Nucl.\ Phys.\ B {\bf 345}, 301 (1990).

\bibitem{Shadmi:2000jy} Y.~Shadmi and Y.~Shirman,
Rev.\ Mod.\ Phys.\  {\bf 72}, 25 (2000) [hep-th/9907225].

\bibitem{Grossman:1997pa} Y.~Grossman, Y.~Nir and R.~Rattazzi,
hep-ph/9701231.

\bibitem{Dine:2001ne} M.~Dine, E.~Kramer, Y.~Nir and Y.~Shadmi,
Phys.\ Rev.\ D {\bf 63}, 116005 (2001) [hep-ph/0101092].

\bibitem{Haber:1998if} H.~E.~Haber,
Nucl.\ Phys.\ Proc.\ Suppl.\  {\bf 62}, 469 (1998) [hep-ph/9709450].

\bibitem{susabe}
S. Abel, these proceedings.

\bibitem{Dugan:1985qf} M.~Dugan, B.~Grinstein and L.~Hall,
Nucl.\ Phys.\ B {\bf 255}, 413 (1985).

\bibitem{Dimopoulos:1996kn} S.~Dimopoulos and S.~Thomas,
Nucl.\ Phys.\ B {\bf 465}, 23 (1996) [hep-ph/9510220].

\bibitem{Buchmuller:1983ye} W.~Buchmuller and D.~Wyler,
Phys.\ Lett.\ B {\bf 121}, 321 (1983).

\bibitem{Polchinski:1983zd} J.~Polchinski and M.~B.~Wise,
Phys.\ Lett.\ B {\bf 125}, 393 (1983).

\bibitem{Fischler:1992ha} W.~Fischler, S.~Paban and S.~Thomas,
Phys.\ Lett.\ B {\bf 289}, 373 (1992) [hep-ph/9205233].

\bibitem{Commins:1994gv} E.~D.~Commins, S.~B.~Ross, D.~DeMille and B.~C.~Regan,
Phys.\ Rev.\ A {\bf 50}, 2960 (1994).

\bibitem{Semertzidis:1999kv} Y.~K.~Semertzidis {\it et al.},
hep-ph/0012087.

\bibitem{Feng:2001sq} J.~L.~Feng, K.~T.~Matchev and Y.~Shadmi,
hep-ph/0107182.

\bibitem{Gabbiani:1996hi}
F.~Gabbiani, E.~Gabrielli, A.~Masiero and L.~Silvestrini,
Nucl.\ Phys.\ B {\bf 477}, 321 (1996) [hep-ph/9604387].

\bibitem{Nir:1986te} Y.~Nir,
Nucl.\ Phys.\ B {\bf 273}, 567 (1986).

\bibitem{Buras:2000da}
A.~J.~Buras, G.~Colangelo, G.~Isidori, A.~Romanino and L.~Silvestrini,
Nucl.\ Phys.\ B {\bf 566}, 3 (2000) [hep-ph/9908371].

\bibitem{Masiero:1999ub} A.~Masiero and H.~Murayama,
Phys.\ Rev.\ Lett.\  {\bf 83}, 907 (1999) [hep-ph/9903363].

\bibitem{Baek:2000jq} S.~Baek, J.~H.~Jang, P.~Ko and J.~H.~Park,
Phys.\ Rev.\ D {\bf 62}, 117701 (2000) [hep-ph/9907572].

\bibitem{Baek:2001kc} S.~Baek, J.~H.~Jang, P.~Ko and J.~H.~Park,
Nucl.\ Phys.\ B {\bf 609}, 442 (2001) [hep-ph/0105028].

\bibitem{Khalil:1999zn} S.~Khalil, T.~Kobayashi and A.~Masiero,
Phys.\ Rev.\ D {\bf 60}, 075003 (1999) [hep-ph/9903544].

\bibitem{Babu:2000xf} K.~S.~Babu, B.~Dutta and R.~N.~Mohapatra,
Phys.\ Rev.\ D {\bf 61}, 091701 (2000) [hep-ph/9905464].

\bibitem{Khalil:1999ym} S.~Khalil and T.~Kobayashi,
Phys.\ Lett.\ B {\bf 460}, 341 (1999) [hep-ph/9906374].

\bibitem{Kagan:1999iq} A.~L.~Kagan and M.~Neubert,
Phys.\ Rev.\ Lett.\  {\bf 83}, 4929 (1999) [hep-ph/9908404].

\bibitem{Khalil:2000ci} S.~Khalil, T.~Kobayashi and O.~Vives,
Nucl.\ Phys.\ B {\bf 580}, 275 (2000) [hep-ph/0003086].

\bibitem{Eyal:1999gk} G.~Eyal, A.~Masiero, Y.~Nir and L.~Silvestrini,
JHEP {\bf 9911}, 032 (1999) [hep-ph/9908382].

\bibitem{Abel:1997eb} S.~A.~Abel and J.~M.~Frere,
Phys.\ Rev.\ D {\bf 55}, 1623 (1997) [hep-ph/9608251].

\bibitem{Baek:1999qy} S.~Baek and P.~Ko,
Phys.\ Lett.\ B {\bf 462}, 95 (1999) [hep-ph/9904283].

\bibitem{Branco:1994eb} G.~C.~Branco, G.~C.~Cho, Y.~Kizukuri and N.~Oshimo,
Phys.\ Lett.\ B {\bf 337}, 316 (1994) [hep-ph/9408229].

\bibitem{Goto:1996zk} T.~Goto, T.~Nihei and Y.~Okada,
Phys.\ Rev.\ D {\bf 53}, 5233 (1996)
[Erratum-ibid.\ D {\bf 54}, 5904 (1996)] [hep-ph/9510286].

\bibitem{Dine:1995vc} M.~Dine, A.~E.~Nelson and Y.~Shirman,
Phys.\ Rev.\ D {\bf 51}, 1362 (1995) [hep-ph/9408384].

\bibitem{Dine:1996ag} M.~Dine, A.~E.~Nelson, Y.~Nir and Y.~Shirman,
Phys.\ Rev.\ D {\bf 53}, 2658 (1996) [hep-ph/9507378].

\bibitem{Romanino:1997cn} A.~Romanino and A.~Strumia,
Nucl.\ Phys.\ B {\bf 490}, 3 (1997) [hep-ph/9610485].

\bibitem{Babu:1996jf} K.~S.~Babu, C.~Kolda and F.~Wilczek,
Phys.\ Rev.\ Lett.\  {\bf 77}, 3070 (1996) [hep-ph/9605408].

\bibitem{Dine:1997xk} M.~Dine, Y.~Nir and Y.~Shirman,
Phys.\ Rev.\ D {\bf 55}, 1501 (1997) [hep-ph/9607397].

\bibitem{Demir:2000qm} D.~A.~Demir, A.~Masiero and O.~Vives,
Phys.\ Lett.\ B {\bf 479}, 230 (2000) [hep-ph/9911337].

\bibitem{Louis:1995ht} J.~Louis and Y.~Nir,
Nucl.\ Phys.\ B {\bf 447}, 18 (1995) [hep-ph/9411429].

\bibitem{Brignole:1994dj} A.~Brignole, L.~E.~Ibanez and C.~Munoz,
Nucl.\ Phys.\ B {\bf 422}, 125 (1994) 
[Erratum-ibid.\ B {\bf 436}, 747 (1994)] [hep-ph/9308271].

\bibitem{Barbieri:1993jk} R.~Barbieri, J.~Louis and M.~Moretti,
Phys.\ Lett.\ B {\bf 312}, 451 (1993)
[Erratum-ibid.\ B {\bf 316}, 632 (1993)] [hep-ph/9305262].

\bibitem{Randall:1999uk} L.~Randall and R.~Sundrum,
Nucl.\ Phys.\ B {\bf 557}, 79 (1999) [hep-th/9810155].

\bibitem{Giudice:1998xp}
G.~F.~Giudice, M.~A.~Luty, H.~Murayama and R.~Rattazzi,
JHEP {\bf 9812}, 027 (1998) [hep-ph/9810442].

\bibitem{Bagger:2000rd} J.~A.~Bagger, T.~Moroi and E.~Poppitz,
JHEP {\bf 0004}, 009 (2000) [hep-th/9911029].

\bibitem{Pomarol:1999ie} A.~Pomarol and R.~Rattazzi,
JHEP {\bf 9905}, 013 (1999) [hep-ph/9903448].

\bibitem{Katz:1999uw} E.~Katz, Y.~Shadmi and Y.~Shirman,
JHEP {\bf 9908}, 015 (1999) [hep-ph/9906296].

\bibitem{Chacko:2000wq}
Z.~Chacko, M.~A.~Luty, E.~Ponton, Y.~Shadmi and Y.~Shirman,
Phys.\ Rev.\ D {\bf 64}, 055009 (2001) [hep-ph/0006047].

\bibitem{Kaplan:2000ac} D.~E.~Kaplan, G.~D.~Kribs and M.~Schmaltz,
Phys.\ Rev.\ D {\bf 62}, 035010 (2000) [hep-ph/9911293].

\bibitem{Chacko:2000mi} Z.~Chacko, M.~A.~Luty, A.~E.~Nelson and E.~Ponton,
JHEP {\bf 0001}, 003 (2000) [hep-ph/9911323].

\bibitem{Nir:1993mx} Y.~Nir and N.~Seiberg,
Phys.\ Lett.\ B {\bf 309}, 337 (1993) [hep-ph/9304307].

\bibitem{Leurer:1994gy} M.~Leurer, Y.~Nir and N.~Seiberg,
Nucl.\ Phys.\ B {\bf 420}, 468 (1994) [hep-ph/9310320].

\bibitem{Nir:1996am} Y.~Nir and R.~Rattazzi,
Phys.\ Lett.\ B {\bf 382}, 363 (1996) [hep-ph/9603233].

\bibitem{Dine:1993np} M.~Dine, R.~Leigh and A.~Kagan,
Phys.\ Rev.\ D {\bf 48}, 4269 (1993) [hep-ph/9304299].

\bibitem{Barbieri:1996uv} R.~Barbieri, G.~Dvali and L.~J.~Hall,
Phys.\ Lett.\ B {\bf 377}, 76 (1996) [hep-ph/9512388].

\bibitem{Carone:1996nd} C.~D.~Carone, L.~J.~Hall and H.~Murayama,
Phys.\ Rev.\ D {\bf 54}, 2328 (1996) [hep-ph/9602364].

\bibitem{Barbieri:1997ww} R.~Barbieri, L.~J.~Hall, S.~Raby and A.~Romanino,
Nucl.\ Phys.\ B {\bf 493}, 3 (1997) [hep-ph/9610449].

\bibitem{Barbieri:2000ax} R.~Barbieri, R.~Contino and A.~Strumia,
Nucl.\ Phys.\ B {\bf 578}, 153 (2000) [hep-ph/9908255].

\bibitem{Cohen:1997sq} A.~G.~Cohen, D.~B.~Kaplan, F.~Lepeintre and A.~E.~Nelson,
Phys.\ Rev.\ Lett.\  {\bf 78}, 2300 (1997) [hep-ph/9610252].

\bibitem{Eyal:2000ys} G.~Eyal, Y.~Nir and G.~Perez,
JHEP {\bf 0008}, 028 (2000) [hep-ph/0008009].

\bibitem{Nir:1998tf} Y.~Nir and M.~P.~Worah,
Phys.\ Lett.\ B {\bf 423}, 319 (1998) [hep-ph/9711215].

\bibitem{Buras:1998ij}
A.~J.~Buras, A.~Romanino and L.~Silvestrini,
Nucl.\ Phys.\ B {\bf 520}, 3 (1998) [hep-ph/9712398].

\bibitem{Colangelo:1998pm} G.~Colangelo and G.~Isidori,
JHEP {\bf 9809}, 009 (1998) [hep-ph/9808487].


\end{references}
\end{document}